\documentclass{article}

\usepackage[T1]{fontenc}
\usepackage{palatino}
\usepackage{fullpage}
\usepackage{graphicx}
\usepackage{amsmath}
\usepackage{amsfonts}
\usepackage{amssymb}
\usepackage[dvipsnames,svgnames,x11names,hyperref]{xcolor}
\definecolor{lkcol}{RGB}{130,70,200}
\usepackage[hidelinks,colorlinks=true,linkcolor=lkcol,citecolor=blue,linktocpage]{hyperref}
\usepackage{cite}
\usepackage{csquotes}
\usepackage{caption}
\usepackage{subcaption}
\usepackage{nicematrix}
\usepackage{mathtools}
\usepackage{tabularray}
\usepackage{float}
\usepackage{amsthm}
\usepackage{dsfont}
\usepackage{calc}
\usepackage{fancyhdr}
\usepackage{titling}

\renewcommand{\mkbegdispquote}[2]{\itshape}

\newcommand{\orcross}{{\includegraphics[width=0.01\textwidth]{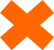}}}

\newcommand{\AND}{\hspace{0.2cm} \text{and} \hspace{0.2cm}}
\newcommand{\xunsp}{w}
\newcommand{\unspmap}{\phi_{\text{unspiral}}}
\newcommand{\net}{\mathcal{W}}

\newcommand{\unspnet}{\widetilde{\net}}
\newcommand{\sqrtbp}{x_{\orcross}}
\newcommand{\sqrtbpw}{w_{\orcross}}
\newcommand{\logbpx}{x_{\bullet}}

\newcommand{\sgn}{\operatorname{sign}}
\newcommand{\incoming}{\text{in}}
\newcommand{\outgoing}{\text{out}}
\newcommand{\curve}{\widetilde{\Sigma}}
\newcommand{\punct}{x_{\bullet}}

\newcommand{\tSigma}{\widetilde{\Sigma}}
\newcommand{\tpi}{\widetilde{\pi}}
\newcommand{\IC}{\mathbb{C}}
\newcommand{\ID}{\mathbb{D}}
\newcommand{\IN}{\mathbb{N}}
\newcommand{\IQ}{\mathbb{Q}}
\newcommand{\IR}{\mathbb{R}}
\newcommand{\IZ}{\mathbb{Z}}
\newcommand{\CC}{\mathcal{C}}
\newcommand{\CE}{\mathcal{E}}
\newcommand{\CS}{\mathcal{S}}

\renewcommand{\Im}{{\rm{Im}}}
\renewcommand{\Re}{{\rm{Re}}}

\def\be{\begin{equation}}
\def\ee{\end{equation}}

\def\IR{{\mathbb{R}}}

\def\IZ{{\mathbb{Z}}}

\def\IC{{\mathbb{C}}}
\def\IN{{\mathbb{N}}}
\def\CI{{\mathcal{I}}}

\def\CF{\mathcal{F}}
\def\CC{\mathcal{C}}
\def\CI{\mathcal{I}}
\def\CL{\mathcal{L}}
\def\CH{\mathcal{H}}
\def\CO{\mathcal{O}}
\def\CM{{\mathcal{M}}}
\def\CN{{\mathcal{N}}}

\def\CW{\mathcal{W}}
\def\tCW{\widetilde{\mathcal{W}}}
\def\Li{\mathrm{Li}}

\def\eff{\mathrm{eff}}

\def\sgn{{\rm{sgn}}}

\def\bQ{{\overline{Q}}}
\def\matter{{\rm{matter}}}
\def\Tr{{\rm{Tr}}}
\def\gauge{{\rm{gauge}}}
\def\sgn{{\rm{sgn}}}
\def\eff{{\rm{eff}}}
\def\bare{{\rm{bare}}}
\def\vortex{{\rm{vortex}}}
\def\disk{{\rm{disk}}}

\def\theory{\mathrm{theory}}
\def\fn{\mathfrak{n}}

\newtheorem{remark}{Remark}
\newtheorem{conjecture}{Conjecture}

\usepackage{authblk}
\title{
Vortices on Cylinders and Warped Exponential Networks
}
\author[1]{Kunal Gupta}
\author[1,2,3]{Pietro Longhi}
\affil[1]{Department of Physics and Astronomy, Uppsala University, Box 516, 751 20 Uppsala, Sweden}
\affil[2]{Department of Mathematics, Uppsala University, Box 480, 751 06 Uppsala, Sweden}
\affil[2]{Centre for Geometry and Physics, Uppsala University, Box 480, 751 20 Uppsala, Sweden}
\date{}                     %% if you don't need date to appear
\fancypagestyle{firstpage}
{
    \fancyhead[L]{}    
    \fancyhead[R]{UUTIP-20/24}
}

\begin{document}

\maketitle
\thispagestyle{firstpage}

\begin{abstract}
We study 3d $\mathcal{N}=2$ $U(1)$ Chern-Simons-matter QFT on a cylinder $C\times\mathbb{R}$.
The topology of $C$ gives rise to BPS sectors of low-energy solitons known as kinky vortices, which interpolate between (possibly) different vacua at the ends of the cylinder and at the same time carry magnetic flux. 

We compute the spectrum of BPS vortices on the cylinder in an isolated Higgs vacuum,
through the framework of \emph{warped} exponential networks, which we introduce.
We then conjecture a relation between these and standard vortices on $\mathbb{R}^2$, which are related to genus-zero open Gromov-Witten invariants of toric branes. 
More specifically, we show that in the limit of large Fayet-Iliopoulos coupling, the spectrum of kinky vortices on $C$ undergoes an infinite sequence of wall-crossing transitions, and eventually stabilizes. 
We then propose an exact relation between a generating series of stabilized CFIV indices and the Gromov-Witten disk potential, and discuss its consequences for the structure of moduli spaces of vortices.

\end{abstract}

\vfill

\newpage

\tableofcontents

\newpage

\section{Introduction}

Within the landscape of three-dimensional Quantum Field Theories with $\CN=2$ supersymmetry, Abelian gauge theories are without doubt among the simplest, and most studied. 
Yet, when considered on certain spacetime backgrounds, even these can feature novel BPS sectors with interesting dynamics.

In this paper we focus on 3d $\CN=2$ $U(1)_\kappa$ Chern-Simons-matter gauge theories on $\IR\times C$, where $\IR$ is regarded as the time direction, and $C=\IR\times S^1$ is a cylinder of radius $R$.
The low energy phases of these theories correspond generically to a finite set of isolated massive vacua. Continuous branches, such as Higgs or Coulomb phases, may also appear for certain values of the moduli, i.e. the Chern-Simons level $\kappa$ and masses.

In the limit of large radius the cylinder $C$ is replaced by $\IR^2$.
It is well known that this background supports a set of isolated vacua $|i\rangle$, characterized by 
a vacuum expectation value of the chiral fields at spatial infinity $\partial \IR^2\simeq S^1$.
It is also well-known that the BPS spectrum of excitations above these vacua includes a spectrum of BPS vortices. 
These well-known features change significantly at finite radius, thanks to the emergence of new BPS sectors and their dynamics.

The first novelty introduced by compactification is that the boundary of $C$ is disconnected $\partial C=S^1\sqcup S^1$. This introduces a new sector in the spectrum of low-energy excitations, corresponding to BPS solitons that interpolate between different vacua $|i\rangle$ and $|j\rangle$ at the two ends of the cylinder.
Another novelty is that the $U(1)$ gauge field can acquire a nontrivial holonomy on the cylinder. 
In a vacuum state $|i\rangle$ the holonomy is fixed to a constant value $e^{i\alpha_i}$ throughout $C$, but in the background of a BPS $(ij)$ soliton the holonomy evolves along the cylinder from $e^{i\alpha_i}$ to $e^{i\alpha_j}$. 
This results in a field configuration with nontrivial magnetic flux through $C$. 
For a fixed pair of vacua $(ij)$ the overall flux depends not only on the labels~$i$ and~$j$, but also on an additional quantum number $n\in \IZ$, which measures the winding of the holonomy across~$C$
\be
	\frac{1}{2\pi}\int_CF = \frac{\alpha_j - \alpha_i}{2\pi} + n\,.
\ee
This kind of BPS solitons, corresponding to kinks that carry magnetic flux, are known as \emph{kinky vortices} \cite{blr2018}.

In this paper we study the spectrum of kinky vortices and its behavior as the moduli of the theory are varied. 
The spectrum is defined by 
viewing 3d $\CN=2$ QFT on $S^1$ as a \emph{Kaluza-Klein} 2d $(2,2)$ QFT, 
whose degrees of freedom are the Fourier modes of 3d quantum fields.  
In the limit $R\to 0$ kinky vortices either become infinitely massive, or descend to BPS solitons of 2d $(2,2)$ quantum field theories. The spectrum of the latter is encoded by the {CFIV} supersymmetric index \cite{Cecotti:1992rm, Cecotti:1992qh}.
At finite radius, the CFIV index counts kinky vortices as BPS states in the framework of 3d $tt^*$ geometry \cite{Cecotti:1991me,Cecotti:2013mba, blr2018}. Following a proposal of \cite{blr2018}, we compute the spectrum of kinky vortices for generic values of the FI parameter through the geometric framework of Exponential Networks \cite{Klemm:1996bj, Gaiotto:2012rg, Eager:2016yxd}.%

The main example considered throughout this work is that of $U(1)_\kappa$ gauge theory with a massless chiral multiplet with unit gauge charge and a Fayet-Ilioupoulous (FI) coupling $\zeta$. More general models will be considered in a follow-up work \cite{GL-to-appear}.
The number of isolated massive vacua is determined by the Chern-Simons level and equals $|\kappa| + \frac{1}{2}$. 
For $|\kappa|=\frac{1}{2}$ (and nonzero FI coupling) the theory has a single isolated vacuum $|i\rangle$, therefore the only kinky vortices are those of type $(ii)$ with arbitrary flux $n$. 
For other values of the coupling there are multiple vacua and the spectrum of kinky vortices features $(ij)$ solitons with arbitrary flux. 
For a fixed choice of $\kappa$ the spectrum is not unique, but depends on the value of the FI coupling. The spectrum jumps by wall-crossing phenomena across walls of marginal stability of real-codimension one in the parameter space of complexified FI couplings.

%%%%

\subsubsection*{Main results}

In this work we focus on a distinguished isolated Higgs vacuum, henceforth denoted by $|i\rangle$, that arises for positive values of the FI parameter.
%For $|\kappa|=\frac{1}{2}$ it is the only vacuum.
Since vacuum $|i\rangle$ exists for any value of the Chern-Simons level,
the BPS spectrum of kinky vortices always includes a universal $(ii)$ sector.

The spectrum of $(ii)$ BPS states depends piecewise continuously on the FI parameter, due to wall-crossing phenomena. 
In Conjecture \ref{conj:disks-from-iin} we propose that the spectrum \emph{stabilizes} for large positive values of the FI coupling.
Let $\mu_{n}$ denote the CFIV index of $(ii)$ kinks with $n$ units of flux on $C$,\footnote{Here we suppress an additional quantum number corresponding to the KK momentum of the vortices. 
See the main text for a discussion.} 
then we conjecture that 
\be\label{eq:intro-conj-1}
	\lim_{\zeta\to+\infty}  \sum_{n\geq 1} \mu_{n}\, x^n
	=
	- \sum_{n\geq 1} \fn_{n} \log(1-x^n) \,,
\ee
where $\fn_{n}\in \IZ$ are integer invariants encoded by the vortex free energy on $\IR^2$ 
\be\label{eq:intro-conj-2}
	W_{\vortex}(x) = - \sum_{n\geq 1} \fn_{n} \Li_2(x^n)\,.
\ee

A physical motivation for our conjecture comes from the field theoretic description of kinky vortices. In particular $(ii)$ BPS states with $n$ units of flux are vortices on the cylinder $C$ with boundary condition given by the same vacuum $|i\rangle$ at both ends. 
As field configurations, these have a characteristic size $R_{{\rm{core}}}$, which is inversely proportional to the mass of the vortex $M \sim \zeta$. As we take the limit $\zeta\to\infty$ the vortex core radius shrinks to zero, and solutions to the BPS equations on $C$ resemble those on $\IR^2$ with fields approaching vacuum $|i\rangle$ at infinity. 
This argument only fully applies in the strict limit, and we expect that the spectrum of $(ii)$ kinky vortices deviates from that of $W_{\vortex}$ for any finite value of $\zeta$, due to finite-size effects. When $R_{\rm{core}}\sim R$, vortices become ``too large'' to fit on $C$ and are expected to disappear altogether.\footnote{A similar jumping behaviour of the spectrum of BPS vortices was observed in the study of 3d $\CN=2$ QFT on compact Riemann surfaces as one varies the volume of $C$ \cite{Bullimore:2018yyb}.}
The physical interpretation in terms of vortices leads to another consequence of our main conjecture, giving a decomposition of the moduli space of vortex configurations with vorticity $n$
\be\label{eq:moduli-space-intro}
	\CM_{n}\approx \bigcup_{d|n} \CI_{n/d} \times \CH_d \,,
\ee
where $\CI_{k}$ is the internal moduli space of a single-center vortex with vorticity $k$ and $\CH_d$ is the Hilbert scheme of $d$ points on $\IC$. Details are given in Section \ref{sec:moduli-space-decomposition}.

We check Conjecture \ref{conj:disks-from-iin} by computing the spectrum of kinky vortices at various points in the moduli space of (complexified) FI couplings, for different values of $\kappa$.
For $\kappa=\frac{1}{2}$ we are able to prove the conjecture.
This case is however very special, due to the absence of wall-crossing.
More generally for $|\kappa|\neq \frac{1}{2}$ we observe that near $\zeta=0$ there are indeed no $(ii)$ kinky vortices, while the spectrum begins to populate gradually as the FI coupling is increased.
One extreme case is when the size of a vortex is comparable to that of the cylinder $R_{\rm{core}}\sim R$, and solutions can decay entirely $\mu_n = 0$. 
As $R_{\rm{core}}$ decreases the spectrum begins to populate, and in the  limit $R_{\rm{core}}\ll R$ all heavier vortices begin to appear. %
As the FI parameter is increased, the spectrum indeed \emph{stabilizes} and matches with \eqref{eq:intro-conj-1} as predicted by independent results on the vortex potential \eqref{eq:intro-conj-2}.

We also provide a nontrivial check of the decomposition \eqref{eq:moduli-space-intro} of the moduli space of vortices, by comparing a prediction based on this picture for the moduli spaces of boosted vortices on the cylinder with the CFIV index for $(ii)$ kinky vortices with nonzero Kaluza-Klein momentum. We find a perfect match, thus supporting the field theoretic interpretation of the moduli space.

%%%%

\subsubsection*{The warped network}

Despite the simplicity of the QFTs that we consider, the wall-crossing behavior of kinky vortices is extremely intricate, and computations by means of exponential networks quickly become too cumbersome to carry out by standard approaches.

We partially tame this problem by observing a novel, universal property of exponential networks concerning $(ii,n)$-type trajectories (see main text for definitions). Thanks to the physical interpretation of $(ii)$ kinky vortices discussed above, it follows that these trajectories must always cross the unit circle on the complexified FI plane at an $n$-th root of unity. Details of the argument can be found in Section \ref{ssec:univ-prop}. 

This observation leads to the surprising consequence that $(ii,n)$ trajectories, those which carry information about $(ii)$ kinky vortices with $n$ units of flux, must follow a predictable pattern, which we call the $n$-th \emph{vortex fan}.
A vortex fan consists of $n$ spirals starting from $n$-th roots of unity on the unit circle and ending at the origin in the complex FI plane, see Figure \ref{fig:vortex-fans}.

We introduce the notion of a \emph{warped exponential network}, defined by a change of coordinate $\unspmap$ that ``straightens out'' all vortex fans. Unlike the original exponential network, whose behavior as a function of the phase appears extremely complex, the warped network is 
essentially constant as we vary the phase, and has a (nearly-)universal structure.
The map $\unspmap$ assigns to a fixed choice of FI parameter an \emph{orbit} in the dual plane. The spectrum of kinky vortices (of all types) is determined by intersections of this orbit with the warped network.\footnote{Another interesting application of the warped network is to compute CFIV indices of kinky vortices of types $ij$ with $i\neq j$ as well, in regions of parameter space with a large $|\zeta|$. 
Recently it was observed how these coincide with counts of framed BPS states with noncompact D4 branes \cite{Banerjee:2024smk}.} 
This significantly simplifies computations, and allows us to exhibit certain features of the spectrum of $(ii)$ kinky vortices in particular, such as their stabilization for large FI parameter.

%%%%

\subsubsection*{Applications to quiver structures of open topological strings}

An interesting application of this correspondence arises by recalling that BPS vortices are related to open Gromov-Witten invariants, namely counts of holomorphic maps with boundaries mapping to a special Lagrangian submanifold \cite{Ooguri:1999bv, Dimofte:2010tz}. 
$\Sigma$ is identified with the moduli space of a certain special Lagrangian $A$-brane, and $W_{\vortex}=W_{\disk}$ with the disk potential \cite{Aganagic:2000gs, Aganagic:2001nx}. 

It is interesting to consider more general QFTs, related to other types of branes such as knot conormals \cite{Ooguri:1999bv, Aganagic:2013jpa}. 
The disk potential encodes genus-zero open Gromov-Witten invariants, which in many cases are encoded by a quiver \cite{Kucharski:2017poe}.

An explanation of this structure was proposed in \cite{Ekholm:2018eee}, which predicted that the quiver would describe the link of boundaries of holomorphic disks.
In light of the correspondence between holomorphic disks (or BPS vortices) and $(ii,n)$ solitons, their realization as $(ii,n)$ BPS states opens up a possibility to compute properties such as linking numbers. 

Indeed we observe that many of the structures predicted in \cite{Ekholm:2018eee} have a realization in terms of data of $(ii)$ BPS solitons encoded by exponential networks.
Details will appear in a companion paper \cite{GL-to-appear}.

\section*{Acknowledgements}
We thank Sibasish Banerjee, Tom Bridgeland, Sergey Cherkis, Fabrizio Del Monte, Tobias Ekholm, Andrea Ferrari, Mauricio Romo and Vivek Shende for discussions.
The work of KG is supported by the Knut and Alice Wallenberg foundation grant KAW 2021.0170, and by the Olle Engkvists Stiftelse Grant 2180108.
The work of PL is supported by the Knut and Alice Wallenberg Foundation grant KAW 2020.0307 and by the VR grant to the Centre of Excellence in Geometry and Physics at Uppsala University.
We also thank the Galileo Galilei Institute for Theoretical Physics for hospitality and INFN for partial support during the completion of this work.
PL gratefully acknowledges support from the Simons Center for Geometry and Physics, Stony Brook University at which some of the research for this paper was performed.

%%%%%%%%%%%%%%%%%%%%%%%%%%%%%%%%%%%%%%%%%%%%%
\section{BPS states of 3d $\CN=2$ QFTs on $S^1\times \IR^2$
}\label{sec:3dQFT}

%%%%%%%%%%%%%%%%%%%%%%%%%%%%%%%%%%%%%%%%%%%%%
\subsection{3d $\CN=2$ Chern-Simons-Matter QFT}

Three dimensional quantum field theories with $\CN=2$ supersymmetry have been studied extensively in past decades, and yet their low energy dynamics continues to be a source of interesting novel phenomena. 
In this work we focus on a class of models which is well studied, namely Chern-Simons-Matter gauge theory with Abelian gauge group. More specifically, we will focus on the case of a $U(1)_\kappa$ gauge symmetry with a massless chiral superfield with electric charge $+1$. 

The 3d $\CN=2$ supersymmetry algebra is\footnote{We follow conventions of \cite{Intriligator:2013lca}. The central charge arises from a compactification of 4d $\CN = 1$ supersymmetry as the momentum component along the circle.} 
\be
	\{Q_\alpha,\bQ_\beta\} = 2\gamma_{\alpha\beta}^\mu \, P_\mu + 2 i \epsilon_{\alpha\beta}Z
\ee 
where $Z$ is a real-valued central extension. 
The field content of the theory includes a vector multiplet and a chiral multiplet. In Wess-Zumino gauge the bosonic components of the former include a real scalar field $\sigma$, a vector field $A_\mu$ (with $\mu=0,1,2$) and an auxiliary scalar $D$, while the latter contains instead a complex scalar $\phi$ and an auxiliary scalar $F$.

The classical Lagrangian features the following terms involving $A_\mu$ and $D$
\be\label{eq:Lagrangian}
\begin{split}
	\CL & = -\frac{1}{4 e^2} F_{\mu\nu} F^{\mu\nu} + \frac{\kappa}{4\pi} \epsilon^{\mu\nu\rho} A_\mu\partial_\nu A_\rho + A_\mu j^\mu_{\matter}
	\\
	& + \frac{1}{2e^2} D^2 +\frac{D}{2} \left( |\phi|^2 - \frac{\kappa}{2\pi} \sigma - \frac{\zeta}{2\pi}\right) + \dots
\end{split}
\ee
where $e$ is the bare gauge coupling, $\kappa$ is the bare Chern-Simons level, $\zeta$ is the real-valued bare Fayet-Ilioupoulos coupling, and we denote by $\phi$ the complex scalar in the charged chiral multiplet.

After integrating out the auxiliary scalar field $D$, we obtain the classical potential
\be
	V_{\rm{cl}} = \frac{e^2}{32\pi^2} \left(2\pi |\phi|^2 - \zeta - \kappa\sigma\right)^2 + \sigma^2 |\phi|^2\,.
\ee
Minimizing this potential gives the classical space of vacua of the theory. However, $\kappa,\zeta$ and $e$ can receive quantum corrections that partially lift the above classical vacua. In the theory we are considering, with a single massless chiral ($m_{\bare} = 0$) with electric charge $+1$, we have 
\be
	\zeta_{\eff} = \zeta\,,\qquad
	\kappa_{\eff} = \kappa+\frac{1}{2}\sgn(\sigma)\,.
\ee
The quantum correction to $\kappa$ comes from 1-loop effects due to integrating out the chiral multiplet, and the discontinuity at $\sigma=0$ corresponds to the vanishing of the effective mass $m_{\eff} = \sigma+m_{\bare}$ of the chiral. Quantisation condition of the effective Chern-Simons level then gives us the following  condition on the bare Chern-Simons level 
\be
	\kappa  \in \IZ+\frac{1}{2}
\ee
Details of running of the renormalized (field-dependent) gauge coupling $e_\eff$ will not be relevant here. Then, we have the semi-classical potential of the theory (including only perturbative corrections) as
\be\label{eq:V-sc}
	V_{\rm{s.c.}} = \frac{e_\eff^2}{32\pi^2} \left(2\pi |\phi|^2 - \zeta_\eff -\left (\kappa + \frac{1}{2} \sgn(\sigma)\right)\sigma\right)^2 + \sigma^2 |\phi|^2\,.
\ee
Next we consider the equations of motion for $A_\mu$ deduced from the semiclassical Lagrangian obtained from \eqref{eq:Lagrangian} after the substitutions $e\to e_{\eff}, \zeta\to\zeta_\eff$ and $\kappa\to\kappa_\eff$. These are the Maxwell equations modified by the Chern-Simons term and by the matter current
\be\label{eq:Amu-eom}
	-\frac{1}{e_\eff^2} \partial_\nu F^{\nu\mu} = j_{\matter}^\mu +\kappa_\eff j_J^\mu\,,
\ee
where $j_J$ is the topological conserved current 
\be
	j_J^\mu := \frac{1}{2\pi} \epsilon^{\mu\nu\rho}\partial_\nu A_\rho
\ee
whose charge generates a topological $U(1)_J$ symmetry dual to the global $U(1)$ subgroup of $U(1)_{\gauge}$.
In particular, note that $j_J^0 = \frac{F_{12}}{2\pi} = \frac{B}{2\pi}$ measures the magnetic flux on a space-like hypersurface, so that the associated charge $q_J$ corresponds to the magnetic charge of a field configuration at a given time.
From \eqref{eq:Amu-eom} it follows that particles with magnetic charge also get dressed with electric charge if $\kappa_\eff\neq 0$ 
\be\label{eq:witten-effect}
	q_{{\rm Gauss}} = q_{\gauge} + \kappa_\eff q_J\,.
\ee

We will now discuss the structure of the semi-classical moduli space of vacua obtained by minimizing the semi-classical potential \eqref{eq:V-sc}. Although our ultimate goal will be the low energy dynamics on $S^1\times \IR^2$, it will be useful to begin with a review of known results on $\IR^3$.

%%%%%%%%%%%%%%%%%%%%%%%%%%%%%%%%
\subsubsection{Low energy phases in Euclidean space}

The semiclassical space of vacua, i.e. the constant expectation values of $\sigma$ and $\phi$ that minimize $V_{\rm{s.c.}}$, is determined by the couplings $\zeta$ and $\kappa$.

\paragraph{Higgs vacua.}
The vanishing of the term $\sigma^2|\phi|^2$ requires either $\sigma=0$ or $\phi=0$.
If $\sigma=0$ then the remaining terms in the potential vanish if an only if
\be
	|\phi|^2 = \frac{\zeta}{2\pi}\,,\qquad \sigma=0\,.
\ee
Clearly this requires $\zeta>0$.
Giving a v.e.v. to $\phi$ gives mass to the gauge boson, and therefore this is a Higgs vacuum. This is a class of isolated vacua (the phase of $\phi$ is unphysical by gauge symmetry) that exists for any value of $\kappa$.

\paragraph{Topological vacua.}
If $\phi=0$, minimization of $V_{s.c.}$ requires 
\be
	\sigma = -\frac{\zeta}{\kappa + \frac{1}{2} \sgn(\sigma)}
 \label{eq:topological-vacua}
\ee
As long as $\kappa \pm \frac{1}{2} \neq 0$, this gives an isolated vacuum for each value of $\zeta \in \IR$. Note that this include vacua both for $\zeta>0$ and for $\zeta<0$.  These vacua are often referred to as the topological vacua.

Combining topological vacua with the Higgs vacua, we see that at the semiclassical level and for generic $\kappa$, there is a single isolated vacuum for $\zeta<0$ and a pair of isolated vacua for $\zeta>0$.

\paragraph{Coulomb branch.}
In Eq \eqref{eq:topological-vacua}, when $\kappa = \pm \frac{1}{2}$ and $\zeta = 0$, there are branches where $\sigma$ can take arbitrary negative/positive values ie.
\be
	\kappa = \pm \frac{1}{2} \,,\qquad
	\zeta = \phi = 0, \qquad \sigma\in \IR_{\mp}\,.
\ee
This characterises the Coulomb branch.

\begin{figure}[h!]
\begin{center}
\includegraphics[width=1.0\textwidth]{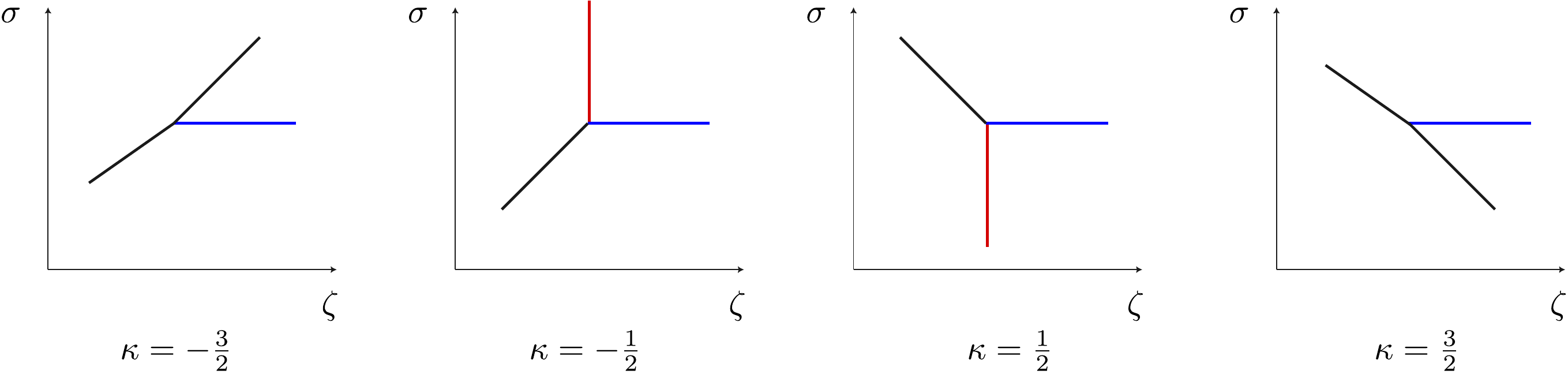}
\caption{Vacua on $\IR^3$ for various values of $\kappa$. Higgs vacua are colored in blue, topological vacua in black, and Coulomb branches (when they exist) are in red.}
\label{fig:R3-vacua}
\end{center}
\end{figure}

We summarize the structure of the space of vacua with a diagram in the $(\zeta,\sigma)$ plane shown in Figure \ref{fig:R3-vacua} for various values of $\kappa$.

\bigskip

\paragraph{Witten index.}
The number of semiclassical vacua of the theory on $\IR^3$ appears to jump with the choice of $\zeta$. However, the Witten index, computed as the trace $\Tr_{\CH}(-1)^F$ over the Hilbert space of the quantum mechanics obtained by reducing the QFT on $\IR\times T^2$, is constant in $\zeta$ \cite{Intriligator:2013lca}. To see this, we recall that 
while the Higgs vacuum contributes $1$ to the Witten index, the topological vacua at $(\sigma_{\rm Top},\zeta)$ contribute $|\kappa_{\eff}(\sigma_{\text{Top}})| = |\kappa + \frac{1}{2}\sgn(\sigma_{\text{Top}})|$ to the index, corresponding to the dimension of the space of $U(1)_{\kappa_\eff}$ conformal blocks on $T^2$.
Therefore, for $\kappa \neq \pm \frac{1}{2}$, we have the following values
\be\label{eq:Witten-index}
	\begin{array}{c||c|c||c|c|}
		& \multicolumn{2}{c||}{\kappa\leq -3/2} & \multicolumn{2}{c|}{\kappa\geq 3/2} \\
		\hline
		& \zeta<0 & \zeta > 0 &\zeta<0 & \zeta > 0\\
		\hline
		\sgn(\sigma_{\rm Top}) & - & + & + & - \\
		\hline
		\Tr_{\text{Top}}(-1)^F & |\kappa - \frac{1}{2}|& |\kappa + \frac{1}{2}| & |\kappa + \frac{1}{2}| & |\kappa - \frac{1}{2}| \\
		\Tr_{\text{Higgs}}(-1)^F & 0 & 1 & 0 & 1 \\
		\hline
		\Tr_{\CH}(-1)^F & \multicolumn{2}{c||}{|\kappa|+ \frac{1}{2}}  &  \multicolumn{2}{c|}{|\kappa| + \frac{1}{2}}  \\
	\end{array}
\ee
which are consistent with the Witten index being constant in $\zeta$. Note  that in the case of $\kappa=\pm\frac{1}{2}$, the Witten index is ill-defined due to the presence of non-compact Coulomb branches.

%%%%%%%%%%%%%%%%%%%%%%%%%%%%%%%%%%%
\subsubsection{Vortices and monopoles in Euclidean space}
Introducing combinations of supercharges\footnote{This slightly unusual convention is due to \cite{Intriligator:2013lca}, leading to the complex conjugate pairs being $(Q_-, \overline{Q}_+)$ and $(Q_+, \overline{Q}_-)$} $Q_\pm := \frac{1}{2} (Q_1\pm i Q_2)$ and $\overline{Q}_{\pm} := \frac{1}{2}(\overline{Q}_1 \pm i\overline{Q}_2)$, one finds that 
\be
	\{Q_\pm,\overline Q_\mp\} = P^0 \pm Z\,.
\ee
On a massive particle state in the rest frame, the right hand side evaluates to $P^0 \pm Z=m\pm Z$, where $m>0$ is the mass of the particle. The two cases are related by CPT symmetry, therefore we discuss one of them.
If $Z|\psi\rangle=m|\psi\rangle$ then $Q_-|\psi\rangle = \overline Q_+|\psi\rangle=0$ vanish on the particle state. The resulting multiplet is therefore BPS since it is generated by the remaining half of the supercharges $\overline Q_-$ and $Q_+$. This generates a doublet of states $|a\rangle$ and $|b\rangle$ defined by
\be\label{eq:BPS-doublet}
	\overline Q_{-} |a\rangle = 0\,,
	\qquad
	Q_{+} |a\rangle = |b\rangle \,.
\ee
These are one particle states with a time-like worldline and $U(1)$ spin in the space-like plane $\IR^2$ given by 
\be\label{eq:doublet-spins}
	J_3 | a\rangle = s|a\rangle\,,\qquad
	J_3 | b\rangle = \left(s+\frac{1}{2}\right)|b\rangle\,.
\ee

In the case of 3d $\CN=2$ QFTs, the central charge $Z$ gets contributions only from global symmetries (and not from gauge symmetries, in contrast to the case of 4d $\CN=2$ QFTs). For the class of models that we are considering, we have only $U(1)_J$ and hence
\be\label{eq:Zqjzeta}
	Z = q_J\, \zeta\,.
\ee
Therefore, BPS states with positive magnetic charge and with $Z>0$ can only exist in isolated topological vacua or Higgs vacua, not on the Coulomb branches. 
Conversely, BPS states with negative magnetic charge and $Z>0$ can only exist if $\zeta<0$, and hence only in isolated topological vacua.

The BPS field equations are found by studying the constraints imposed by invariance under $Q_-$ and $\overline Q_+$ (see \cite[Appendix A]{Intriligator:2013lca} and references therein for details)  
\be\label{eq:vortex-equations}
	\partial_t \sigma = 0\,,\qquad
	F_{zt}+\partial_z\sigma = 0\,,\qquad
	(\partial_t - i (\sigma+A_0))\phi = 0\,,\qquad
	F_{z\bar z} - i D = 0\,,\qquad
	D_zQ = F=0,
\ee
where $z=x_1+i x_2$ is a complex coordinate on the spatial plane and $t$ is the time coordinate.
A vortex corresponds to a static configuration with $A_0=-\sigma$, and has a magnetic charge 
\be
	2\pi q_J = -i \int d^2z \, F_{z\bar z} = \int d^2z \, D = -\frac{e^2}{2} \int d^2z \, \left(|\phi|^2  - \frac{\zeta_\eff}{2\pi}  - \frac{\kappa_\eff}{2\pi}\sigma\right) =: \frac{\zeta e_\eff^2}{4\pi}\, R^2_{\rm{core}}
\ee
Here $D$ is determined by solving the Lagrangian constraints, and the fields $\phi,\sigma$ must asymptote to a vacuum at spatial infinity but can deviate from it at finite distance. The above integral defines a length scale $R_{\rm{core}}$ which represents the characteristic size of the vortex 
\be\label{eq:vortex-size}
	R_{\rm{core}}^2 = \frac{q_J}{\zeta}\, \frac{8\pi^2}{e_\eff^2}  \,.
\ee

Recall that $\zeta\neq 0$ parametrizes either Higgs or topological vacua.
Thus we learn that BPS vortices are very thin when $\zeta$ is large, such as in asymptotic regions of the Higgs vacua or of the topological branches, but become fuzzy as $\zeta$ goes to zero, where several vacua meet. An important consequence of this that we will use later on is that for large $\zeta$, these very thin vortices will not feel the effect of the topology of the surface they are put on.

In the Higgs vacuum the vortex field configuration has large $|z|$ asymptotics 
\be
	\lim_{z\to\infty} \phi = \sqrt{\frac{\zeta}{2\pi}}\, e^{i q_J\varphi} + \dots
	\qquad
	A_\varphi = q_J +\dots
\ee
where $z = |z|e^{i\varphi}$. The vorticity coincides with the magnetic charge for the well-known ANO vortices.

From \eqref{eq:witten-effect} it follows that any field configuration with magnetic charge must also carry electric charge. This is what happens for vortices. In three dimensions, magnetic and electric charges are computed by 
\be
	q_J = \int_\Sigma \frac{F_{12}}{2\pi} = \int_\Sigma \frac{B}{2\pi}\,, \qquad 
	q_{\rm{Gauss}} = -\frac{1}{e^2}\oint_{\gamma} \star F = \frac{1}{e^2}\oint_{\gamma} \vec E\cdot d\ell\,,
\ee
along a surface $\Sigma$ encircling the magnetic charge, or along a loop $\gamma$ encircling the electric charge.
In the case of vortices $\Sigma\simeq \IR^2$ is the spatial plane, and $\gamma = \partial \Sigma$ is a circle at spatial infinity. 
The magnetic charge $q_J$ is exactly the Aharonov-Bohm phase around the vortex worldline
\be
	q_J 
	= \frac{1}{2\pi}\oint_\gamma A\,.
\ee

\begin{figure}[h!]
\begin{center}
\includegraphics[width=0.4\textwidth]{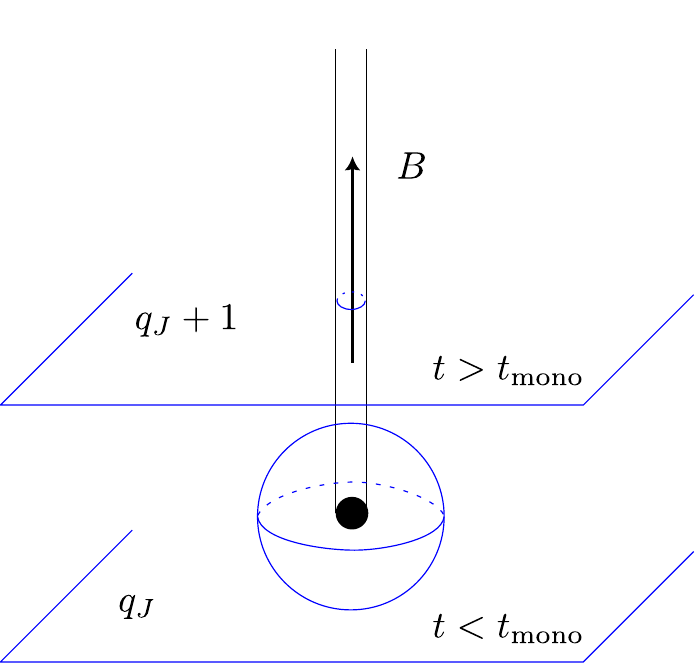}
\caption{Creation of a vortex BPS state at late times by a monopole operator insertion at early times. Schematically this is depicted as a vortex ending on a monopole.}
\label{fig:vortex-creation}
\end{center}
\end{figure}

Another way to think about $\Sigma$ is to view the spatial plane as an $S^2$ (punctured at infinity) surrounding a monopole of charge $q_J$ which lives at $t\to -\infty$.
Vortex states can be created by monopole operators, and correspond to the lowest energy state defined by late-time evolution \cite{Intriligator:2013lca}
\be\label{eq:vortex-monopole-creation}
	|v\rangle = \lim_{t\to\infty} e^{-(H-Z)t} \, \CO_{{\rm monopole}} |0\rangle\,.
\ee
In a Higgsed or topological phase the gauge field is massive and the flux of a monopole operator becomes confined to a flux tube, corresponding to the core of the vortex. See Figure \ref{fig:vortex-creation}.

We recall that vortices can carry spin, which is given by
\footnote{Chern-Simons vortices have a structure, in particular while the electric field is radial on the spatial plane generated from the point like charge $q_\gauge$, the magnetic fields come out of the space plane on a ring around this point like charge. Due to the simultaneous presence of transverse magnetic field $B$ and electric field $\vec E$, there is a nontrivial Poynting vector which rotates around the vortex worldline giving rise to a contribution to the spin of the vortex due to the electromagnetic field. 
For a comprehensive review see for example \cite{Dunne:1998qy}.}
\be\label{eq:em-spin}
	s = \frac{1}{2} q_{\gauge}q_J\,.
\ee
In the Higgs phase, which will be of particular interest to us, one has $\vec E / e_\eff^2 \to 0$, leading to the vanishing of the screened charge, ie. $q_{\rm{Gauss}}=0$, implying $q_{\gauge} = -\kappa_\eff q_J$. Thus, the spin of this vortex becomes
\be\label{eq:vortex-spin}
	s = -\frac{1}{2} \kappa_\eff q_J^2\,.
\ee
Note an important subtlety: $\kappa_\eff$ is ill-defined in the Higgs phase since $\sigma=0$. 
If we consider the case of a vortex with $q_J=1$ in the Higgs phase, it has a unique bosonic zero mode which parametrizes the position of the vortex in the plane. The fermionic partner completes the BPS doublet. These two modes are $|a\rangle$ and $|b\rangle$ defined earlier via Eq \eqref{eq:BPS-doublet}, and their spins, via Eq \eqref{eq:doublet-spins} are
\be\label{eq:ab-spin}
	J_3 |a\rangle =\left(-\frac{\kappa}{2} - \frac{1}{4}\right) |a\rangle\,,
	\qquad
	J_3|b\rangle = \left( -\frac{\kappa}{2} + \frac{1}{4}\right) |b\rangle\,.
\ee
Therefore the spin of $|a\rangle$ corresponds to \eqref{eq:vortex-spin} with $\kappa_\eff$ evaluated for $\sigma=0^+$ while for $|b\rangle$ it is evaluated at $\sigma=0^-$, compatibly with \eqref{eq:doublet-spins}.
In fact, exactly for values $\kappa=\pm \frac{1}{2}$ for which there is a Coulomb branch, one of these two states has spin zero. 

This is because the Coulomb branch is parameterized by monopole operators $X_\pm$.
It is argued in \cite{Intriligator:2013lca}, by matching quantum numbers, that in the case of $\kappa = -\frac{1}{2}$ the monopole operator is $(X_+)^{\dagger}$ creating the state $|a\rangle$, while in the case $\kappa = \frac{1}{2}$ the monopole operator is $(X_-)^{\dagger}$ creating the state~$|b\rangle$. We summarise this in the following table-
\be
	\begin{array}{c|c|c}
	\kappa & \CO_{\rm monopole}   & \text{BPS vortex } |v\rangle \\
	\hline
	- \frac{1}{2} & X_{+}\sim e^{2\pi\sigma / e^2_{\eff} + i \tilde A} & |a\rangle\\
	\frac{1}{2} & X_{-}\sim e^{-(2\pi\sigma / e^2_{\eff} + i \tilde A)} & |b\rangle \\
	\end{array}
\ee
where $\tilde A$ is the magnetic scalar potential $F = \frac{e_\eff^2}{2\pi}\star d\tilde A$. The other state in the multiplet is generated by acting with $\overline Q_{-}$ or $Q_{+}$ as in \eqref{eq:BPS-doublet}.
In both cases, vortex state with $q_J=1$ is created by the monopole operator that has spin $s=0$.
In fact, it is precisely for $|\kappa|=\frac{1}{2}$ that the monopole operators (and the vortices created by them) have zero electric charge and zero spin. Correspondingly monopole operators can acquire a v.e.v. and generate a Coulomb branch only for these values of the Chern-Simons level.

More generally, in the Higgs phase the Clifford vacuum $|v\rangle$ with $q_J=1$ is $|a\rangle$ for $\kappa\leq -1/2$, but it switches to $|b\rangle$ for $\kappa\geq \frac{1}{2}$.
Parametrizing the bare Chern-Simons level by an integer called framing
\be\label{eq:f-kappa}
	f := \kappa  + \frac{1}{2}\in \IZ,
\ee 
we have the following spin for Clifford vacua
\be
	\begin{array}{c||cc|c|c|cc}
		\text{Framing } f & \cdots &-1 & 0 & 1 & 2&\cdots\\
		\hline\hline
		2s_{|a\rangle} & \cdots& 1 & 0 & -1 & -2& \cdots\\
		2s_{|b\rangle} & \cdots& 2 & 1 & 0 & -1 &\cdots\\
		\hline\hline
		\text{Clifford Vacuum } |v\rangle & \multicolumn{3}{c|}{|a\rangle}  & \multicolumn{3}{c}{|b\rangle}  \\
		\hline
		2s_{|v\rangle} & \cdots &1 & 0 & 0 & -1&\cdots
	\end{array}
\ee
As a function of $f$, the spin of the vortex $q_J=1$ in the Higgs phase $\zeta>0$ is 
\be
	2s_{|v\rangle} = \left\{
	\begin{array}{lr}
	-f \qquad& f\leq 0\\
	-f+1 \qquad& f\geq 1
	\end{array}\right.
\ee
When $|\kappa|\neq \frac{1}{2}$, the spin is nonzero, and vortex condensation is forbidden by rotational symmetry. This is in line with the expectation that monopole operators carry the same quantum numbers as vortices, and hence they lift the Coulomb branch.

%%%%%%%%%%%%%%%%%%%%%%%%%%%%%%%%%%%%%%%%
\subsection{Circle compactification}

Next we consider the theory discussed so far on $ \IR^2\times S^1$, where we compactify $x^2\in \IR/2\pi R\IZ$. 
Both the space of vacua and that of BPS states change significantly.
Fourier modes of 3d fields on the circle give rise to towers of massive fields on the plane. 
Viewing the 3d theory as a 2d $\CN=(2,2)$ theory on $\IR^2$ with towers of Kaluza-Klein fields allows us to invoke results about the dynamics of 2d $(2,2)$ QFTs and their BPS states. 

With the introduction of the circle, we can express the component of the gauge potential along it as 
\be\label{eq:A-theta}
	A_\theta\, d\theta = \left(\sum_{n\in \IZ} \alpha_n \cos(n\theta) + \beta_n\sin(n \theta) \,\right) d\theta \, ,
\ee
where $\theta = x^2/R$ and by construction the engineering dimension of $\alpha_n$ and $\beta_n$ is $0$. A gauge transformation $A\to A+dh$ allows us to set to zero all $\alpha_n,\beta_n$ except for the zero mode coefficient $\alpha_0$, which measures the holonomy on $S^1$.
This becomes a scalar field on $\IR^2$, complexifying the zero mode of the vectormultiplet scalar $\sigma$ as\footnote{Compared to \eqref{eq:vortex-equations} we have a rotation exchanging $x^0$ and $x^2$ as the time direction.}
\be
	2 \pi R \sigma  \to 2 \pi R \sigma +i\oint_{S^1} A_\theta\, d\theta = 2\pi R \sigma_0 + \sum_{n\geq 1 } \left[
	\delta_{n} \, \sin(n\theta)
	+ \epsilon_{n} \, \cos(n\theta)\right]  +2\pi i \alpha_0\,.
\ee
where $R$ is the radius of the circle and $\delta_n, \epsilon_n$ are the non-trivial Fourier components of $\sigma$. 
Any vacuum field configuration must minimize the kinetic term, and therefore these non-trivial Fourier modes must vanish, ie. $\delta_n=\epsilon_n=0$. 

The above quantities have the engineering dimension $0$, as expected for a scalar field in two dimensions. This allows us to define the v.e.v. of the effective complex scalar field in two dimensions as
\be\label{eq:Y-def}
	Y = -2\pi \langle R  \sigma_0 + \, i \alpha_0\rangle,
\ee
where the minus sign is taken in order to ensure consistency with \cite{blr2018}. Large gauge transformations $A_\theta\to A_\theta+ \partial_\theta \chi$ for some function $\chi$ on the compactified circle lead to the identification 
\be\label{eq:shift-sym}
	Y \sim Y+2\pi i\,.
\ee
The space of physical vacua of the compactified theory is therefore parameterized by the exponentiated variable $y=e^Y$. 

In a similar way, the Fayet-Iliopoulos coupling $\zeta$ also pairs up with the holonomy $\tilde\alpha_0$ of the connection associated to the topological $U(1)_J$ symmetry. This allows us to introduce the complexified FI coupling
\be\label{eq:X-def}
	X = -2\pi  \left(R \zeta +i\tilde \alpha_0\right)\,,
\ee
and its exponentiated version $x = e^X$.

The 3d chiral multiplet descends to a tower of 2d chiral fields with mass given by the Kaluza-Klein momentum. In the deep infrared these modes are non-dynamical and their contribution to the low energy effective action consists of 1-loop corrections to the effective twisted superpotential \cite{Dimofte:2011jd}
\be\label{eq:C3-tCW}
	2\pi R \tCW =  \Li_2(e^Y) + \frac{f}{2} \, (Y+\pi i )^2 +  YX\,.
\ee
where we defined framing $f$ from the bare Chern-Simons coupling in \eqref{eq:f-kappa}.
The vacuum manifold defined by $\exp( 2\pi R\, { \partial \tCW}/{\partial Y}) = 1$ 
is the algebraic curve $\Sigma\subset (\IC^*)^2$
\be\label{twistedchiralringequation}
	1-y-x (-y)^f=0\,.
\ee
Viewed as a ramified covering over $\IC^*_x$, the curve $\Sigma$ has $f$ sheets  if $f>0$, and it is $|f|+1$ if $f\leq 0$. The number of sheets is therefore
\be
	\text{\# sheets}  = \left|f-\frac{1}{2}\right| + \frac{1}{2} = |\kappa| + \frac{1}{2} = \Tr_{\CH}(-1)^F
\ee 
matching the number of vacua computed by the Witten index \eqref{eq:Witten-index}. 
For a given choice of complexified FI coupling $x$, the points 
\be\label{eq:y-i}
	y_i(x)\,, \qquad i = 1\dots |\kappa|+\frac{1}{2}
\ee
correspond to isolated massive vacua of the 3d theory on a circle.
Note that $\tCW$ is not single valued on $\Sigma$ due to the presence of a dilogarithm in \eqref{eq:C3-tCW}. While an overall constant shift of the vacuum energy does not affect equations of motion, multivaluedness can play an important role when studying transitions between vacua.

%%%%%%%%%%%%%%%%%%%%%%%%%%%%%%%%
\subsection{BPS sectors of the circle compactification}

The theory on a circle has two types of BPS states that will be of interest to us. 
Both are related to the BPS vortices of the 3d theory in Euclidean space.

%%%%%%%%%%%%%%%%%%%%%%%%%%%%%%%
\subsubsection{BPS vortices} 

If the circle is regarded as the time dimension, the BPS vortices of the 3d theory wrap $S^1$ and descend to BPS vortices of the 2d $(2,2)$ theory of Kaluza-Klein modes on space-like $\IR^2$. 
Far out in the Higgs phase, near $x\to 0$, this BPS sector is captured by the vortex partition function $Z_{\vortex}$. For the $U(1)_\kappa$ theory with a charged chiral multiplet that we consider, the partition function is related to the half-index or holomorphic block \cite{Yoshida:2014ssa, Beem:2012mb}
\be\label{eq:Z-vortex}
	Z_{\vortex} = \sum_{n\geq 0} q^{f \, n^2} \frac{((-q)^{-f} x)^n}{(q^2;q^2)_n}\,.
\ee
Here $q$ is a parameter for the $\Omega$ deformation on $S^1\times \IR^2$ introduced to regularize integration over noncompact moduli spaces of vortices.

Three dimensional theories with $\CN=2$ supersymmetry arise in the context of the 3d-3d correspondence \cite{Dimofte:2011ju,Terashima:2011qi}. In particular, abelian gauge theories like the one considered here arises from a twisted compactification of the abelian 6d $(2,0)$ theory on a toric Lagrangian in a toric Calabi-Yau threefold \cite{Aganagic:2000gs, Aganagic:2001nx, Dimofte:2010tz}.
Through this correspondence, the partition function of vortices is identified with the spectrum of open worldsheet instantons, which is described mathematically by open Gromov-Witten invariants.
In this context $q=e^{g_s}$ is related to the topological string coupling and the free energy $\log Z_{\rm vortex}$ is organized by a topological expansion in $g_s$. The leading order contribution comes from holomorphic disks, and takes the form 
\be\label{eq:W-vortex}
	W_{\vortex} :=
	\lim_{g_s\to 0} \, 2g_s \log Z_{\rm vortex} = \sum_{k\geq 0} \fn_{k}\Li_2(x^k)
\ee
where $\fn_k\in \IZ$ are integers whose interpretation can be given in terms of M-theory as Euler characteristic of moduli spaces of M2 branes wrapping holomorphic disks \cite{Ooguri:1999bv}. 
The values of $\fn_k$ depend on the Chern-Simons coupling via $f$, which is identified with the framing parameter for the toric brane \cite{Aganagic:2001nx}. The first few $\fn_k$ for different framings read 
\be\label{eq:AKV-invariants}
\begin{array}{c|cc|c|c|cc}
	& \cdots & f=-1 & f=0 & f=1 & f=2 &\cdots\\
	\hline
	k=1 & \cdots & 1 & -1 & 1 & -1 & \cdots \\
	k=2 & \cdots & -1 & 0 & 0 & 1 & \cdots \\
	k=3 & \cdots & 1 & 0 & 0 & -1 & \cdots \\
	k=4 & \cdots & -2 & 0 & 0 & 2 & \cdots \\
	k=5 & \cdots & 5 & 0 & 0 & -5 & \cdots \\
\end{array}
\ee

%%%%%%%%%%%%%%%%%%%%%%%%%%%%%%%%%%%%%%%%%%
\subsubsection{Kinky vortices} 

If $\IR\subset \IR^2$ is instead regarded as the time direction, a space-like slice becomes a cylinder $S^1\times \IR$. 
An important novelty is that the region at spatial infinity is disconnected 
\be
	\partial (S^1\times\IR)  = S^1\sqcup S^1\,.
\ee
If follows that this geometry supports BPS soliton field configurations, with $\sigma$ interpolating between different vacua at the two ends. 
As remarked above, only the zero mode of $\sigma$ is involved in the selection of a vacuum state and it is complexified by the gauge holonomy into a 2d complex scalar $Y(x_1,x_2)$. 

Soliton configurations are time independent $\partial_{x_2}Y=0$, and must approach vacua at both ends
\be\label{eq:Y-bc}
	\lim_{x_1\to-\infty}Y(x_1) = \log y_i + 2\pi i \, N\,,
	\qquad
	\lim_{x_1\to+\infty}Y(x_1) = \log y_j + 2\pi i \, M\,.
\ee
The BPS central charge of the soliton is 
\be\label{eq:Z-general}
	Z_{ij,n,k}(x) =\tCW(Y_j + 2\pi i M) -\tCW(Y_i + 2\pi i N) + \frac{2\pi}{R}k\,,
\ee
where $Y_i(x)$ and $Y_j(x)$ are vacua, and $M,N$ label branches of the logarithm.\footnote{Although the logarithmic index is unphysical in the definition of individual vacua, it affects physical properties of BPS solitons.}
A soliton can interpolate between a fixed pair of vacua in different ways, due to the multivalued nature of $\tCW$ remarked below \eqref{eq:C3-tCW}.
The monodromy group of $\tCW$ is generated by that of the logarithm $Y=\log y$, with branching at $y=0,\infty$, and by that of the the dilogarithm $\Li_2(y)$, with branching at $0,1,\infty$.
The index $n$ keeps track of the winding number $n$ of $y(x_1)$ around $\IC^*_y$, as $x_1$ goes $-\infty$ to $+\infty$
\be
	n=M-N\,.
\ee
The additional index $k$ instead keeps track of the monodromy $\Li_2(y)\to \Li_2(y)+4\pi^2 k$ arising from combined loops around $0,1,\infty$, see \cite{blr2018} for details. 
Physically, the BPS central charge only depends on the logarithmic branches via their difference 
and not on the overall logarithmic branch, as a consequence of the shift symmetry \eqref{eq:shift-sym}. 
Moreover, the contribution of $k$ to the BPS central charge is quantized in units of the Kaluza-Klein momentum $2\pi/R$ along $S^1$.

Solitons with $n=k=0$ correspond to a trivial circle uplift of the 2d BPS states familiar from the context of 2d $tt^*$ geometry \cite{Cecotti:1991me, Cecotti:1992rm}, whose BPS degeneracy is captured by the CFIV index \cite{Cecotti:1992qh}. When $n\neq 0$, the 2d soliton carries a magnetic flux on the cylinder
\be\label{eq:cylinder-flux}
	\int_{S^1\times \IR} \frac{F}{2\pi} = \frac{1}{2\pi} \oint_{S^1} \left[A_\theta(x_1=+\infty)-A_\theta(x_1=-\infty)\right] = \frac{\Im(Y_j) - \Im(Y_i)}{2\pi}+ n\,,
\ee
where we made use of \eqref{eq:Y-def}. 
A kinky vortex owes its name (coined in \cite{blr2018}) to the resemblance to a boundstate of a kink interpolating between different 2d vacua $\sigma_i\sim \Re(Y_i)$ and a vortex with magnetic flux $2\pi n$ and localized at a point on $S^1\times \IR$. Thanks to the fact that the vortex is localized on $S^1$, it can acquire nonzero KK momentum, which corresponds to $k\neq 0$.

The spectrum of BPS kinky vortices depends on $x$. It is piecewise constant within certain chambers, but jumps by wall-crossing at walls of marginal stability that separate chambers.
Even in the simplest models, such as the one considered in this work, the spectrum is very rich compared to the case of 2d $(2,2)$ BPS solitons. A systematic approach to computing the CFIV index of kinky vortices, which we will use below, is the framework of nonabelianization \cite{blr2018}.

While the spectrum of BPS vortices is related to the open Gromov-Witten invariants of a toric brane in $\IC^3$, that of kinky vortices is related to another kind of enumerative invariants of the same Calabi-Yau. As discussed at length in \cite{Klemm:1996bj, Eager:2016yxd, Gaiotto:2009hg, blr2018} exponential networks capture the spectrum of D-branes on Calabi-Yau threefolds, which mathematically is related to rank-zero (generalized) Donaldson-Thomas invariants \cite{Kontsevich:2008fj}.

%%%%%%%%%%%%%%%%%%%%%%%%%%%%%%%%%%%%%%%%%%
\section{From Exponential Networks to holomorphic disks}

Our main tool in the study of different BPS sectors will be the nonabelianization map for Exponential Networks introduced in \cite{blr2018}. 
This section begins with a brief review of this framework. We then introduce a generalization of nonabelianization, that extends the domain of application of the original construction to cover a class of geometries relevant for this work. 
The section ends with the formulation of a conjecture that relates certain BPS states computed by Exponential Networks ($(ii,n)$ kinky vortices) and BPS vortices.

%%%%%%%%%%%%%%%%%%%%%%%%%%%%%%%%%%%%%%%%%%%%
\subsection{Overview of counting 3d BPS kinky vortices with exponential networks}

Let $\Sigma$ be an algebraic curve described by a polynomial equation $F(x,y)=0$ with $(x,y) \in \IC^*\times \IC^*$. Viewing $\Sigma$ as a ramified covering of the $x$-coordinate plane, we fix a system of branch cuts and denote by $y_i(x)$ the sheets. 
An exponential network $\CW(\vartheta)$ for the covering $\Sigma\to\IC^*$ is a network of trajectories on $\IC^*_x$, known as $\CE$-walls.\footnote{The letter $\CW$ used for networks is similar to the notation $\tCW$ for the twisted superpotential \eqref{eq:C3-tCW}. The two should not be confused.}

Each $\CE$-wall is labeled by an ordered pair of sheets $i,j$ and by an integer $n\in \IZ$. 
The shape of an $\CE$-wall of type $(ij,n)$ is determined by the differential equation
\be\label{eq:E-wall-eq}
	(\log y_j - \log y_i + 2\pi i \, n ) \frac{d\log x}{dt} = e^{i\vartheta}
\ee
where $t$ is a local coordinate along the trajectory. 
To make sense of this equation it is essential to introduce a trivialization for $\log y$, which requires choosing a branch cut for the logarithmic map $\log :\IC^*\to \IC$ associating $y\mapsto \log y+2\pi i N$. In $\IC^*\times \IC^*$ the branch cut has real dimension three and intersects $\Sigma$ along a system of arcs connecting the punctures where $y\to 0,\infty$, also known as \emph{logarithmic punctures}. We define a $\IZ$-covering $\tilde \pi: \ \tSigma\to \Sigma$ ramified at these arcs, and then functions $\log y_i$ are defined on $\Sigma$ on the complement of the branch locus for $\tpi$.

A solution $x(t)$ to \eqref{eq:E-wall-eq} is determined by a choice of basepoint $x_0 = x(0)$. 
There are two types of basepoints: if $x_0$ is a branch point the corresponding $\CE$-wall is said to be primary, while if $x_0$ is an intersection of $\CE$-walls it is the starting point for a descendant trajectory.\footnote{We assume a generic choice of moduli, so that all branch points are quadratic. In this case there are three emanating trajectories. In general there can also be higher order polynomial branch points, and in that case more than three lines would be generated.}
Primary walls always have $n=0$, but as they evolve on the $x$-plane they may cross a logarithmic branch cut either for $\log y_j$ or for $\log y_i$. When this happens the label of the $\CE$-wall shifts from $(ij,0)$ to $(ij,n)$ with $n\neq 0$.

In addition each $\CE$-wall carries certain \emph{soliton data}. 
Given a generic point $x$ along the $\CE$-wall, let $a$ be the relative homology class of a path that runs from $\log y_i(x)+2\pi i N$ to $\log y_j(x)+2\pi i (N+n)$ on $\tSigma$. We further denote by $\Gamma_{ij,N,N+n}$ the charge lattice of such relative homology classes.
After choosing a trivialization of the charge lattice\footnote{This may involve both a choice of trivialization for $\tSigma\to\IC^*$ and a choice of trivialization of the moduli space of $\Sigma$, see \cite{blr2018}.}, we can identify each relative homology class with a triple of indices $(n,k,\beta)\in \IZ^{2+r}$ that label charges of BPS kinky vortices in the topological sector $(ij)$
\be\label{eq:a-n-k-n-beta}
	a  \ \ \leftrightarrow\ \ (n,k,\beta)\,.
\ee
Here $(n,k)$ correspond to the charges in \eqref{eq:Z-general} arising for the theory with vacuum manifold \eqref{twistedchiralringequation}, while $\beta$ accounts for the possibility of twisted masses in more general models.

The soliton data of the $\CE$-wall is a map 
\be\label{eq:soliton-data}
	\mu_\CE:\ \bigcup_{N\in \IZ}\Gamma_{ij,N,N+n}\to \mathbb{Q}
\ee 
determined entirely by the global topology of the network $\CW(\vartheta)$
through the nonabelianization map. 
A certain shift symmetry of the latter ensures that $\mu$ is (in a suitable sense) independent of $N$ \cite{blr2018}.
From a field theory perspective the soliton data of $\CE$-walls coincides with indices of BPS states of the 3d $\CN=2$ QFT on $S^1\times\IR^2$. 
More precisely, $\mu(a)$ is the CFIV index for kinky vortices interpolating between vacua $(i,N)$ and $(j,N+n)$. The CFIV index of \cite{Cecotti:1992qh} can be extended to count BPS kinks in 2d-4d systems \cite{Gaiotto:2011tf}. In the context of exponential networks we have a 3d-5d system engineered by an M5 brane wrapping a toric Lagrangian in a toric Calabi-Yau threefold. When spacetime is $S^1\times \IR^4$ with the M5 supported on $S^1\times \IR^2$ this can be viewed as a 2d-4d system of the Kaluza-Klein modes, thus extending the definition of CFIV index to our setting. 

By construction, an exponential network $\CW(\vartheta)$ encodes data of BPS kinky vortices with central charges of phase $\vartheta$. 
Fixing a choice of $x\in \IC^*$ corresponds to fixing a choice of 3d theory, i.e. to a choice of complexified FI coupling, see \eqref{eq:X-def}. 
Then the entire spectrum of BPS states is obtained by varying $\vartheta$ and studying which trajectories sweep across the theory point $x$.
Each sweeping trajectory carries information about a part of the BPS spectrum, namely the kinky vortices with central charge \eqref{eq:Z-general} of phase $\vartheta$
\be\label{eq:phase-condition}
	\arg Z_{a}(x) = \vartheta \,.
\ee
This procedure gives the full spectrum of BPS kinky vortices for the theory at $x$.

In general, the BPS spectrum is only piecewise continuous in $x$, due to jumps that occur at \emph{walls of marginal stability}. 
There is a decomposition into chambers
\be
\label{eq:chamber-structure}
	\IC^* = \bigcup_\alpha \CC_\alpha\,,
\ee
delimited by walls of marginal stability where the soliton data can jump.
Given a chamber $\CC$, there is a map
\be
	\mu_\CC : \bigcup_{(ij,n)}\bigcup_{N\in \IZ}\Gamma_{ij,N,N+n}\to \mathbb{Q}
\ee
defined by soliton data \eqref{eq:soliton-data} of $\CE$-walls that sweep through any point $x\in \CC$
as we vary $\vartheta\in[0,2\pi)$.
This map is defined on the entire charge lattice (of BPS kinky vortices) of the theory whose FI coupling corresponds to $x$. 
This map jumps at MS walls $\mu_\CC\neq \mu_{\CC'}$. These come in two types in the framework of exponential networks.
The first one corresponds to real-codimension one loci in $\IC^*_x$ where two (or more) trajectories intersect transversely for some values of $\vartheta$. 
The second type of MS wall corresponds to the appearance of \emph{generalized saddles} of the network. 
A generalized saddle consists of a system of $\CE$-walls that begins and ends on branch points, and can only exist for specific values of $\vartheta$, see \cite{Klemm:1996bj, Eager:2016yxd, blr2018} for examples. 
For the class of theories we consider in this paper there is only one phase where generalized saddles appear, which is $\vartheta=0$ mod $\pi$. The saddle corresponds to the tower of D0 branes in $\IC^3$.\footnote{More precisely, this is only the fixed point in a moduli space of compact leaves of the foliation associated to the network, see \cite{Banerjee:2022oed}.}

%%%%%%%%%%%%%%%%%%%%%%%%%%%%%%%%%%%%%%%%%%
\subsection{Including logarithmic punctures at finite distance}\label{sec:log-punctures-exp-net}

The definition of exponential networks reviewed above hinges on the assumption that there is at least one branch point, where primary $\CE$-walls are sourced. 
If the covering $\Sigma\to \IC^*$ has only one sheet, the definition does not apply since there are no branch points and therefore no $\CE$-walls.
Moreover in the construction of \cite{blr2018} it was assumed that there are no logarithmic singularities (i.e. points where $y\to 0$ or $\infty$) at finite values of $x$. 
We will now introduce a generalization of exponential networks that includes both of these possibilities.

Suppose that one of the sheets of the covering $\Sigma\to \IC^*_x$, labeled $y_i$, has a logarithmic puncture at $x=\logbpx$, with $|\log \logbpx|<\infty$.
The exponential network should then include additional $\CE$-walls of types $(ii,n)$ for $n\in \IZ$ that emanate from $\logbpx$, as first pointed out in \cite{Grassi:2022zuk, Alim:2022oll}.

Our task will be to determine the soliton data associated with these trajectories. 
For this purpose we introduce a transition function for each $\CE$-wall, and demand that the parallel transport defined by these has trivial monodromy around any contractible path.
This flatness constraint of the parallel transport will end up determining the exact form of the transition function, after taking into account certain signs coming from twisting \cite{Gaiotto:2012rg,blr2018}.

\begin{figure}[h!]
   \centering
   \begin{subfigure}{.5\textwidth}
       \centering
       \includegraphics[width=0.7\textwidth]{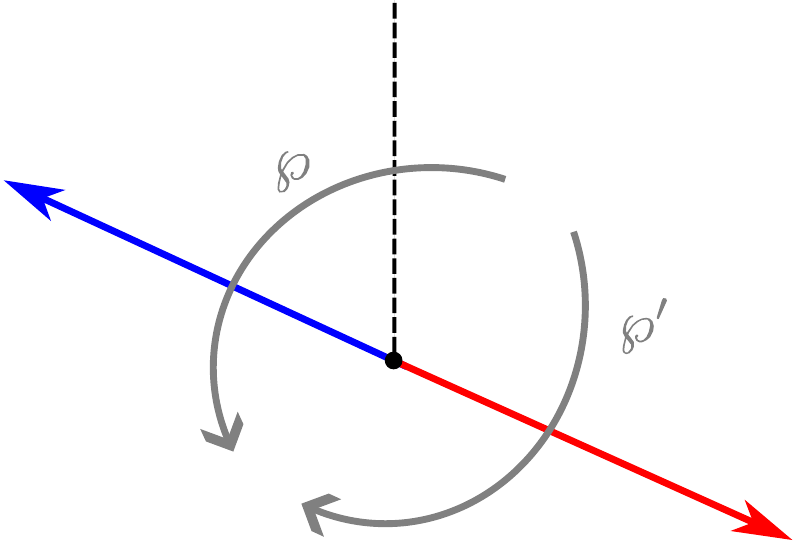}
       \caption{}
       \label{trivhomotopypath1}
   \end{subfigure}%
   \begin{subfigure}{.5\textwidth}
       \centering
       \includegraphics[width=0.7\textwidth]{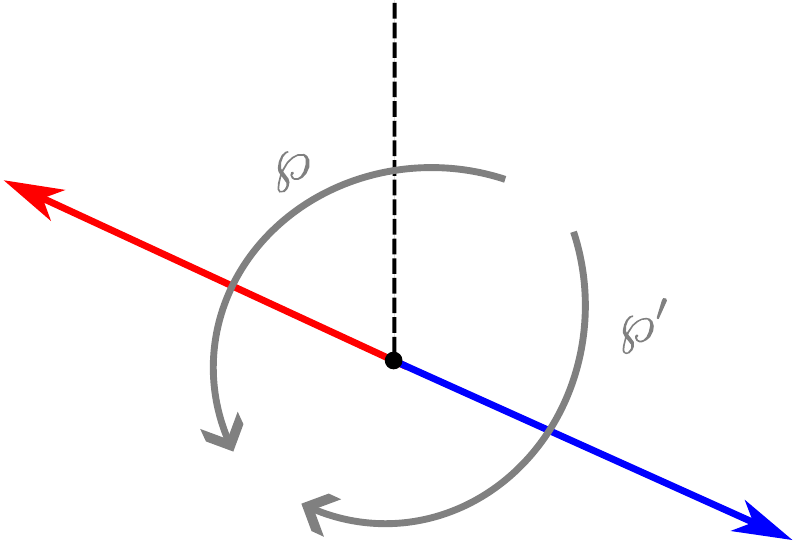}
       \caption{}
       \label{trivhomotopypath2}
   \end{subfigure}
   \caption{Homotopy paths for the two topologies around a logarithmic puncture at finite $x$. Note that trajectories are not emanated from logarithmic punctures at $0$ or $\infty$. 
}
   \label{trivhomtoppath}
\end{figure}
We consider a pair of paths $\wp,\wp'$ with matching endpoints and tangent vectors and passing on either side of $x=\logbpx$, as shown in Figure \ref{trivhomotopypath2}.
Each path crosses a families of $\CE$-walls, whose geometry is determined by \eqref{eq:E-wall-eq} specialized to $i=j$
\be
	x(t) = \logbpx \,\exp\left(\frac{t\, e^{i\vartheta}}{2\pi i n}\right)\,.
\ee 
$\CE$-walls with $n>0$ are here denoted in red, while those with $n<0$ are denoted in blue. There are two distinct cyclic orderings of the $\CE$-walls and the logarithmic cut, shown in Figure \ref{trivhomtoppath}.

An $\CE$-wall of type $(ii,n)$ carries solitons whose topological charges belong to the affine lattice $\Gamma_{ii,N,N+n}$, as defined above \eqref{eq:soliton-data}. 
Without making any assumptions on the global geometry of the curve, there is only one relative homology class (for each $N$) that we can define in the local geometry in the neighborhood of $\logbpx$. 
Let $p$ denote the segment of the $(ii,n)$ $\CE$-wall between $\logbpx$ and $x$, oriented towards $x$. 
Denoting by $p^{(M)}$ the lift of $p$ to sheet $(i,M)$ with the same orientation, we define
\be
	a_n^{(N)} := p^{(N+n)} - p^{(N)} \,.
\ee
This definition applies to all $\CE$-walls, with both signs of $n$.
Note that we have a relation
\be\label{eq:a-concat}
	a_n^{(N)} \approx  a_k^{(N)}\circ a_{n-k}^{(N+k)} \,,\qquad 1\leq k\leq N-1\,.
\ee
Here $\approx$ signals that the equivalence involves the concatenation of paths, which involves a choice of how to match tangent directions at endpoints. 

As in \cite{blr2018}, we introduce formal variables $X_{a_n^{N}}$ such that their multiplication rule respects \eqref{eq:a-concat}, and the product is zero if paths do not concatenate 
\be\label{eq:X-algebra}
	 X_{a_n^{(N)}} \, X_{a_{m}^{(M)}}= \delta_{M,N+n}X_{a_{n+m}^{(N)}} \,.
\ee
The transition function associated to an $\CE$-wall with label $(ii,n)$ is
\be
	S_n := \exp\left( \sum_{N\in \IZ} \mu(a_n^{(N)}) \,X_{a_n^{(N)}} \right)\,.
\ee
We will also need to introduce the (counterclockwise) monodromy transition function associated to the logarithmic cut, which we denote by $M_{\logbpx}$. 
The action of this matrix is to shift the logarithmic index of soliton data as
\be
	\left[M_{\logbpx}\right]_{(j,N),({k,M})} = [\mathds{1}]_{j,k} \otimes 
	[\mathds{1}]_{N+\ell \delta_{i,j},M}
\ee 
where $i$ denotes the sheet on which the logarithmic puncture lies, and $\ell$ is the order of divergence of $y\sim x^\ell$.
In other words, the logarithmic cut shifts the logarithmic label of a path by $\ell$ if an only if that path lies on a lift to $\tSigma$ of the $i$-th sheet. Since in this paper we will only need $\ell=1$ we restrict to this case for simplicity, although the following discussion has a straightforward generalization to arbitrary $\ell\in \IZ$.
The flatness equation for the configuration shown in Figure \ref{trivhomotopypath1} can then be expressed as follows 
\be
	F(\wp) \equiv  M_{\logbpx}\cdot \left(\prod_{n\geq 1}  S_{-n}\right) 
	= \left(\prod_{n\geq 1} S_{n}\right)^{-1} \equiv F(\wp')\,.
\ee
with the extra minus sign coming due to the twisted homotopy condition \cite{Gaiotto:2012rg,blr2018}. Solving this equation uniquely determines the soliton data $\mu$.

To illustrate the solution we take a slight shortcut: we observe that by large gauge transformations the soliton data should be independent of $N$, namely $\mu(a_n^{(N)}) = \mu_n$ for all $N\in \IZ$.\footnote{As in \cite{blr2018}, this can be verified a posteriori, as a shift symmetry of the soliton data.}
This allows us to rewrite 
$
	S_n = \exp\left( \mu_n \Xi_n \right)
$ with
$
\Xi_n:=\sum_{N\in \IZ}  X_{a_n^{(N)}} 
$.
The concatenation algebra of formal variables \eqref{eq:X-algebra} implies that $\Xi_n^k = \Xi_k$ behaves as a simple complex number, which we denote by $\xi$. 
This allows us to work with the \emph{reduced} vector space of finite dimension $d$ equal to the degree of the covering map $\pi:\Sigma\to \IC^*$. On this vector space the transition matrix and monodromy matrix are diagonal with entries\footnote{The sign in the monodromy matrix around the logarithmic branch point arises as a consequence of our choice of conventions for the twisted flat connection in the nonabelianization construction of \cite{Gaiotto:2012rg, blr2018}.}
\be
	[S_n]_{jj} = \left\{
	\begin{array}{lr}
	1 & (j\neq i)\\
	\exp\left( \mu_n \xi^n \right) \quad& (j=i)
	\end{array}\right.\,,
	\qquad
	[M_{\logbpx}]_{jj} = \left\{
	\begin{array}{lr}
	1 & (j\neq i)\\
	-\xi  \quad& (j=i)
	\end{array}\right.
\ee
having set $\ell=1$ as explained above.
The flatness constraint is then trivial for matrix elements $j\neq i$, while for $j=i$ it boils down to the identity 
\be
	-\xi\cdot
	\exp\left(\sum_{n\geq 1} \mu_{-n} \xi^{-n}\right)
	=
	\exp\left(-\sum_{n\geq 1} \mu_{n} \xi^n\right)\,.
\ee
It is important that the left hand side has a series expansion in $\xi^{-1}$ while the right hand side has a series expansion in $\xi$.
The solution is given by
\be
	\exp\left(\sum_{n\geq 1} \mu_{-n} \xi^{-n}\right) = 1-\xi^{-1}\,,
	\qquad
	\exp\left(\sum_{n\geq 1} \mu_{n} \xi^n\right) = \frac{1}{1-\xi}\,,
\ee
since 
\be
	-\xi\cdot (1-\xi^{-1})  = {1-\xi}\,.
\ee
The soliton data is therefore 
\begin{equation}\label{eq:log-p-soliton-data}
    \mu_n = \frac{1}{n} \qquad  (n \in \mathbb{Z}^*)
\end{equation}
This is the main result of this subsection, and it concludes the analysis for the configuration of Figure \ref{trivhomotopypath1} in the case $\ell=1$. Switching to general $\ell$ simply amounts to replacing $\xi\to \xi^\ell$ in all expressions, so that $\mu_n$ is only nontrivial for integer multiples of~$\ell$.
For the configuration of Figure \ref{trivhomotopypath2} the story is very similar. The flatness equation reads now
\be
	-\exp\left(\sum_{n\geq 1} \mu_{n} \xi^{n}\right)\cdot 
	\xi
	=
	\exp\left(-\sum_{n\geq 1} \mu_{-n} \xi^{-n}\right)\,.
\ee
which is again solved by the same soliton data, since
\be
	- \frac{1}{1-\xi} \cdot \xi = \frac{1}{1-\xi^{-1}}\,.
\ee
Therefore once again the soliton data is given by \eqref{eq:log-p-soliton-data}.

%%%%%%%%%%%%%%%%%%%%%%%%%%%%%%%%%%%%%%%%%%
\subsection{Main conjecture: Open Gromov-Witten invariants from $(ii,n)$ kinky vortices}\label{eq:Open-GW-from ii}

We will now formulate a conjecture that relates two different BPS sectors of 3d $\CN=2$ QFTs on $S^1\times \IR^2$. On one side there are BPS vortices located at a point in $\IR^2$ and wrapping the $S^1$, while on the other side there are the kinky vortices located at a point in $S^1\times \IR$ and wrapping the transverse $\IR$.

The spectrum of BPS vortices is well understood thanks to results based on localization \cite{Yoshida:2014ssa}, which allow for the systematic computation of vortex partition functions such as \eqref{eq:Z-vortex}. Recall the definition of $W_{\vortex}$ from \eqref{eq:W-vortex}. As pointed out in \cite{Aganagic:2000gs,Aganagic:2001nx,Ooguri:1999bv,Ekholm:2018eee}, the vortex superpotential can be expressed in terms of the vacuum manifold $\Sigma$ as follows
\be\label{eq:W-vortex-log-y}
	W_{\vortex} = \int^x \log y_i\, d\log x\,,
\ee
where $y_i$ is the sheet of $\Sigma\to \IC^*_x$ with the property that $\lim_{x\to 0}y_i=1$, i.e. it corresponds to the isolated Higgs vacuum of the theory in Euclidean space, see Figure \ref{fig:R3-vacua}. 

While the vortex superpotential can be easily obtained from the geometry of $\Sigma$, much less is known about the spectrum of kinky vortices. 
In part this is due to the fact that it is unclear how to approach the computation of the CFIV index with localization techniques. Another difficulty is that the spectrum of BPS kinky vortices jumps by wall-crossing in the moduli space of complexified FI couplings $x$, see \eqref{eq:chamber-structure}.

Our conjecture involves specifically kinky vortices of type $(ii,n)$ which, according to \eqref{eq:Y-def}, correspond to vortices on $S^1\times \IR^2$.
More precisely, these vortices wrap a noncompact worldline $\IR$ and correspond to solutions of the vortex equations on the cylinder $S^1\times \IR^2$.
Boundary conditions for the BPS vortex equations are given by the \emph{same} vacuum $\sigma_i$ at both endpoints, but with different winding number for the gauge holonomy $\oint A$. 

The kinky vortices of type $(ii,n)$ therefore resemble closely standard vortices on $\IR^3$ encoded by $W_{\vortex}$, where the vacuum at infinity is $\sigma_i$ and with magnetic flux measured by the winding number $n$, see \eqref{eq:cylinder-flux}.
A crucial difference is however given by the periodic boundary conditions introduced by the circle compactification.
The moduli space of solutions of vortex equations with different boundary conditions are different, and for this reason the BPS spectrum of kinky vortices cannot, in general, be related to $W_{\vortex}$.

Nevertheless, the two can be expected to coincide when the radius of the circle is infinitely large $R\to \infty$ since in that limit $\IR^2\times S^1\to \IR^3$. 
To make this more precise, we introduce the generating series of $(ii)$ kinky vortices 
\be\label{eq:ii-n-gen-series}
	\Xi(x,Q) = \sum_{n\geq 1}\sum_{\beta} \mu_{n,0,\beta}\, x^n\, Q^{\beta}\,,
\ee
where $i$ is the label of the isolated Higgs vacuum for $\zeta>0$.
Here $(n,k=0,\beta)$ denotes the charge of a kinky vortex of type $(ii,n)$, where $n$ is the vorticity, $k=0$ the KK momentum, and $\beta$ takes into account the possibility of additional flavor charges, see \eqref{eq:a-n-k-n-beta}. The coefficient $\mu_{n,k,\beta}(x)\in \IQ$ is the CFIV index of the kinky vortex computed in the theory at $x\in \IC^*$. 
Due to wall-crossing $\Xi(x,Q)$ is only piecewise continuous in $(x,Q)$.

From the viewpoint of QFT, the fugacities $x$ and $Q$ correspond to exponentiated central charges. 
Comparing \eqref{eq:X-def} with \eqref{eq:Zqjzeta} gives in fact
\be\label{eq:vortex-charge-and-Z}
	x = e^{-2\pi R\, Z_c}\,,\qquad n=q_J\,,
\ee 
where $Z_c = \zeta+\frac{i}{R}\tilde\alpha_0$ is the complexified 3d central charge at finite $R$.
For the class of theories considered in this paper the flavor symmetry is actually trivial (hence $\beta=0$), but solitons can always carry nontrivial Kaluza-Klein charge \cite{blr2018}. 
By definition we fix the KK momentum to zero and do \emph{not} sum over KK modes in~\eqref{eq:ii-n-gen-series}. If the KK momentum is nonzero, the CFIV index in general changes $\mu_{n,k,\beta}\neq \mu_{n,0,\beta}$, we return to this point in Section \ref{sec:higher-KK-modes}.

The magnitude of central charges is important because it provides a scale against which the limit $R\to\infty$ can be meaningfully studied.
In the isolated Higgs vacuum $(i,N)$ the characteristic size of $(ii,n)$ kinky vortices is controlled by the FI coupling through \eqref{eq:vortex-size}.
Therefore taking $x\to 0$ shrinks vortices to vanishing size compared to the $S^1$ radius
\be\label{eq:vortex-scale}
	\frac{R_{\rm{core}}}{R}\to 0\,.
\ee
This is the limit in which the spectrum of $(ii,n)$ kinky vortices can be expected to agree with the spectrum of BPS vortices computed by \eqref{eq:Z-vortex}.

\begin{conjecture}\label{conj:disks-from-iin}
In the limit $x\to 0$ 
%the CFIV index becomes independent of $k$, i.e. $\mu_{n,k,\beta} = \mu_{n,\beta}$ for all $k\in \IZ$, and 
the generating series \eqref{eq:ii-n-gen-series} stabilizes to 
\be\label{eq:Xi-stable}
	\lim_{x\to 0} \Xi(x,Q) 
	= 
	- \sum_{n\geq 1}\sum_{\beta} \fn_{n,\beta} \log(1-x^n Q^\beta) 
\ee
where $\fn_{n,\beta}\in \IZ$ count single-center BPS vortices on $\IR^2$ in vacuum $i$
\be\label{eq:W-vortex-Li2-n}
	W_{\vortex}(x,Q) = - \sum_{n\geq 1}\sum_{\beta} \fn_{n,\beta} \Li_2(x^n Q^\beta)\,.
\ee
\end{conjecture}

The integers $\fn_{n,\beta}$ are LMOV invariants of a special Lagrangian brane in a Calabi-Yau threefold. These give a reorganization of the open string free energy based on counts of M2 branes \cite{Ooguri:1999bv, Labastida:2000zp, Labastida:2000yw}.
The appearance of $\Li_1$ in \eqref{eq:Xi-stable} should be contrasted with that of $\Li_2$ in \eqref{eq:W-vortex-Li2-n}. The relation can be understood in terms of \emph{framed 2d-4d wall-crossing} (lifted to 3d-5d) where multiplicative ``halo factors'' of the form $(1-x^nQ^\beta)$ describing jumps of generating series of framed BPS states arise from conjugation by quantum dilogarithms \cite{Gaiotto:2011tf,Galakhov:2014xba}, or equivalently as the statement that $\Li_2$ describes the generating function for a Kontsevich-Soibelman canonical transformation (symplectomorphism) \cite{Coman:2020qgf}.

%%%%%%%%%%%%%%%%%%%%%
\subsection{Moduli space decomposition}\label{sec:moduli-space-decomposition}

Conjecture \ref{conj:disks-from-iin} has an interesting consequence for the structure of moduli spaces of vortices.

An $n$-vortex state is generically composed of a collection of single-center vortices with vorticities $n_i$, such that $n = \sum_{i} n_i$. See e.g. \cite{Dunne:1998qy} and references therein.
We may label such a vortex configuration by a partition $\lambda = (n_1,\dots, n_\ell)$ with $n_i\leq n_{i+1}$ corresponding to vorticities of single centers, whose overall number is $|\lambda|=\ell$.
Our conjecture implies certain constraints on the kind of partitions that can appear.

Using \eqref{eq:ii-n-gen-series} into \eqref{eq:Xi-stable} (with $\beta=0$ for simplicity) gives $\mu_{n,0}$ in terms of genus zero LMOV invariants as follows 
\be\label{eq:mu-0-from-n}
	\mu_{n,0} = \sum_{d|n} \frac{1}{d} \, \fn_{n/d}
\ee
This formula can be understood as follows
\begin{itemize}
\item
On the one hand, $\fn_{n/d}$ is the LMOV invariant, a (possibly weighted) Euler characteristic of the moduli space of sheaves on embedded disks with boundary on a toric Lagrangian in $\mathbb{C}^3$. 
In the 3d $\CN=2$ QFT setting we may regard $\fn_{n/d}$ as the (again weighted) Euler characteristic of the moduli space $\CI_{n/d}$ of \emph{internal} degrees of freedom of a \emph{single-center} vortex configuration with vorticity $n/d$. This will clearly depend on the choice of Chern-Simons coupling~$\kappa$, as LMOV invariants are known to do.
\item
On the other hand $\frac{1}{d}$ corresponds to the Euler characteristic of the configuration space of $d$ points in the plane. These correspond to the centers of identical single-center vortices, regardless of their vorticities. The moduli space is the Hilbert scheme of $d$ points in $\IC$, denoted $\CH_d$.
\end{itemize}
This interpretation leads to the conclusion that an $n$-vortex configuration only gets contributions from
$d$ copies of single-center vortices with vorticity $n_1=\dots=n_d = n/d$.
This implies the following decomposition of the moduli space of vortices with $k=0$ KK momentum
\be\label{eq:mod-space-decomp}
	\CM_{n,0}\approx \bigcup_{d|n} \CI_{n/d} \times \CH_d \,.
\ee
Taking the Euler characteristic recovers \eqref{eq:mu-0-from-n}.

%%%%%%%%%%%%%%%%%%%%%%%%

\section{Linear Framing}\label{sec:f0}
When $f=0$ or $1$, the curve \eqref{twistedchiralringequation} is linear in $y$, and we refer to this as a linear framing. As it will turn out, linear framing is much easier to analyse than ``non-linear'' framings, hence they provide a useful entry point to illustrate our conjecture and related computations, before moving on to the nonlinear case.

\subsection{Geometry of vacua and soliton charges}

Setting $f=0$ in \eqref{twistedchiralringequation} gives the curve
\begin{equation}
    y = 1-x\,,
\end{equation}
For each value of the (exponentiated) Fayet-Iliopoulos coupling $x$, there is a single vacuum $y$.
Geometrically, the curve $\Sigma$ is a 1:1 covering of $\IC^*_x$ with punctures at $x=0,1,\infty$.

\begin{figure}[h!]
   \includegraphics[width=0.9\linewidth]{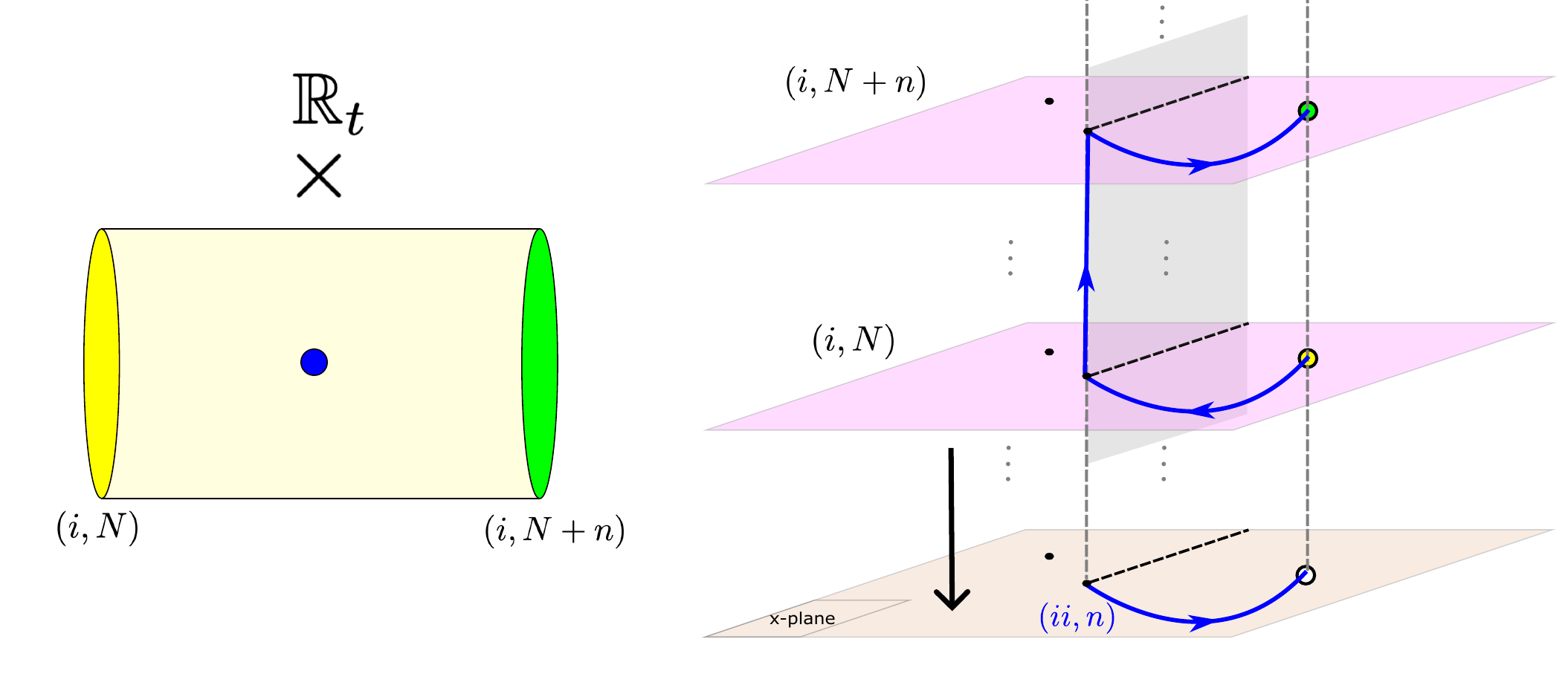}
   \centering
   \caption{
   On the left hand side, we show a BPS soliton extrapolating between the two vacua as shown in yellow and green, with the jump in the middle in blue denoting the vortex solution on the cylinder. On the right hand side, we show the corresponding BPS string wrapping $\curve$ along with its projection on the $x$-plane, forming a part of the exponential network.
}
   \label{trivsoliton}
\end{figure}

Since the field space is not simply connected, it will be convenient to pass to its universal covering, with local coordinate $Y =\log y$. We thus introduce the $\IZ$-covering 
\be
	\curve\to\Sigma
\ee
with logarithmic branching at points where $y=0$ and $y=\infty$, which happens at $x=1,\infty$, see Figure \ref{trivsoliton}.
The unique vacuum \eqref{eq:Y-def} then lifts to a whole tower 
\begin{equation}
	    Y_N = \log (1-x) + 2\pi i N \,,
\end{equation}
with $N$ labeling the choice of logarithmic branch involved in the lift.
The Seiberg-Witten differential is single-valued on the universal covering curve $\curve$, and given by
\begin{equation}
    \lambda_N =\frac{Y_N(x)}{2\pi R}\, \frac{dx}{x} 
    \label{trivswdiff}
\end{equation}
Using this, we can construct BPS strings and hence the BPS solitons, and we show this in Fig \ref{trivsoliton}. Since we only have one polynomial branch, all solitons in the $f=0$ case are $ii$ solitons.
We fix a choice of trivialization for $\log y$ by choosing a branch cut that runs from $x=1$ to $\infty$ along a straight line, as shown in Figure \ref{trivsoliton}.

The topological charges of $ii$ kinky vortices include the logarithmic shift $n$, which appears in the defining equation \eqref{eq:E-wall-eq}, as well as their relative homology class $a$.
Let us fix a generic choice of $x_\theory\in \IC^*$ corresponding to a choice of coupling for the theory. Then the (affine) lattice of relative homology 1-chains with endpoints $\log y_i$ and $\log y_i+2\pi i n$ can be trivialized by writing 
\be\label{eq:triv-f=0}
	a = a_0 + \gamma
\ee
where $a_0$ is the charge of the lightest BPS state (the one with lowest value of $|Z|$) and $\gamma\in H_1(\tSigma,\IZ)/{\rm ker} \, Z\approx \IZ$ belongs to the homology lattice of $\tilde \Sigma$.\footnote{For a discussion of the quotient by the kernel of central charge and other technical details on the lattice of charges, see \cite{blr2018}.}
For the theory we are studying, the latter is one dimensional, and generated by the D0 brane charge $\gamma_{D0}$.

The logarithmic index $n$ is identified with the \emph{vorticity} of the kinky vortex, since the central charge is
\be
	Z_{a_0} = \frac{i}{R} n \log x_\theory\,.
\ee
It follows from \eqref{eq:E-wall-eq} that BPS states with logarithmic index $n$ can only pick up shifts $\gamma$ corresponding to multiples of $n \gamma_{D0}$. Therefore, after fixing a choice of trivialization \eqref{eq:triv-f=0} for the charge lattice, the topological charge of a $(ii,n)$ kinky vortex is labeled by two integers
\be\label{eq:topological-charges}
	a : \ (n,k)\in\IZ^2\,.
\ee
The central charge is then
\be\label{eq:central-charge-framing-zero}
	Z_a = \int_a\lambda =  \frac{i}{R} \, n  \left(\log x_\theory  + 2\pi i\, k \right) \,,
\ee
where $R$ is the $S^1$ compactification radius. In our conventions, which follow \cite{blr2018} the D0 central charge is $Z_{\gamma_{D0}} =\frac{2\pi}{R}$. 

The central charge of $Z_a$ with charge $(n,k)=(1,0)$ coincides with that of a vortex with unit vorticity \eqref{eq:Zqjzeta}, provided the identifications
\be\label{eq:vortex-soliton-id}
	q_J = n\,,\qquad
	\zeta =  \frac{i}{R}  \, \log x_\theory \,,
\ee
in line with \eqref{eq:X-def}.
The trivialization of the charge lattice therefore corresponds to a choice of trivialization for $\log x$.

As topological charges on $\tilde \Sigma$, the integer $n$ tracks the logarithmic jump between sheets $(i,N)$ and $(i,N+n)$ as shown in Figure \ref{trivsoliton}, while the integer $k$ tracks the winding number around $\IC^*_x$ of the trajectory connecting $x$ to $x_0=1$. Each unit of winding shifts $Z_a$ by $- \frac{2\pi n}{R}$, which can be attributed to the fact that we have to give a KK-boost to a system with $n$ vorticity \footnote{\label{foot:n-k-relation}A nice physical interpretation for this is to consider the solution of vorticity $n$ to rather be a solution which contains $n$ vortices of unit charge. Then going up the KK-ladder is equivalent to giving a KK-momentum to the solution along the circle, which requires giving a KK-momentum to all $n$ vortices of unit charge.}.
%

%%%%%%%%%%%%%%%%%%%%%%%%%%
\subsection{Exponential network}

\begin{figure}[h!]
   \centering
   \begin{subfigure}{.5\textwidth}
       \centering
       \includegraphics[width=0.7\linewidth]{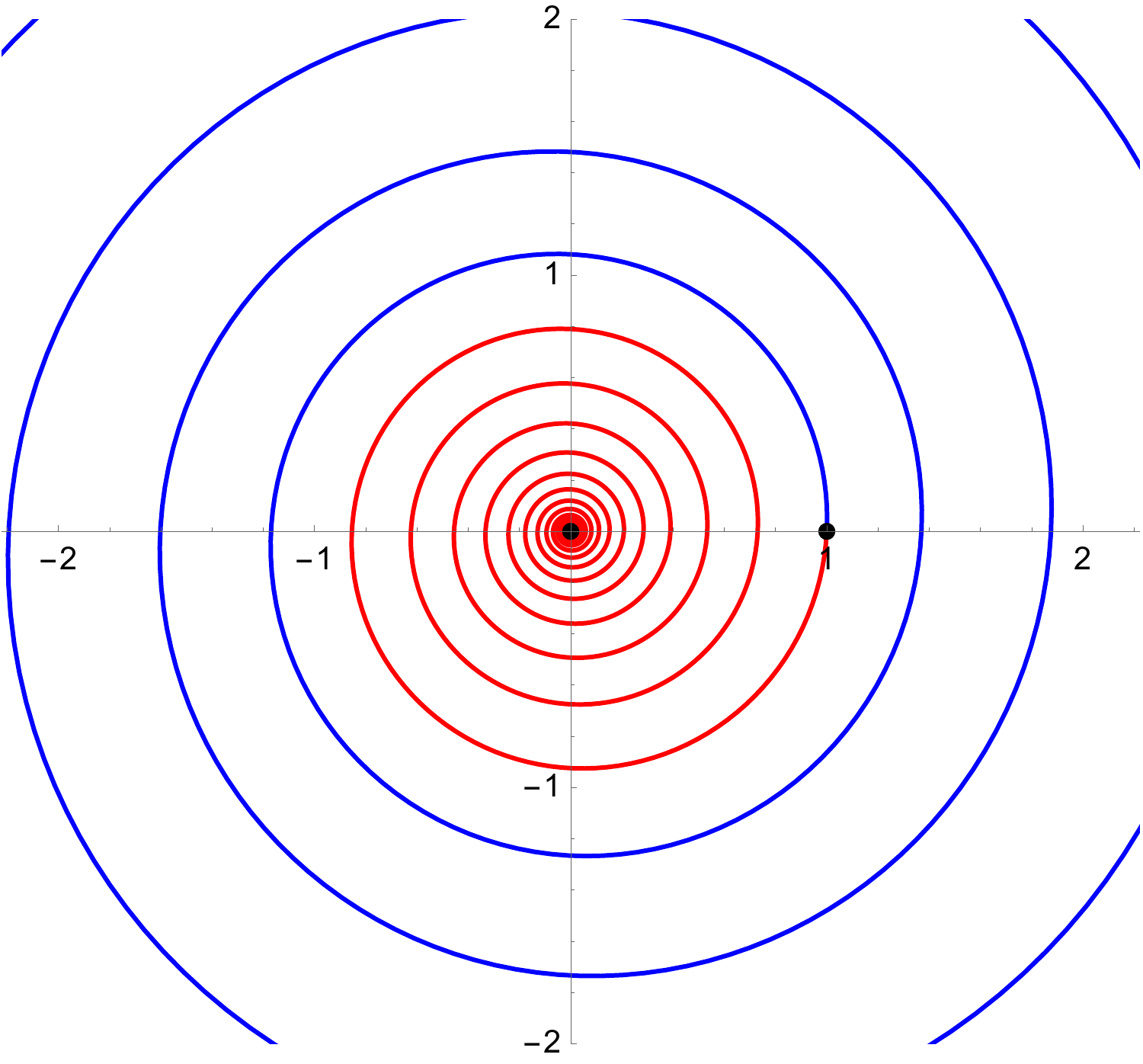}
       \caption{$-\pi<\vartheta < 0$}
       \label{triv1}
   \end{subfigure}%
   \begin{subfigure}{.5\textwidth}
       \centering
       \includegraphics[width=0.7\linewidth]{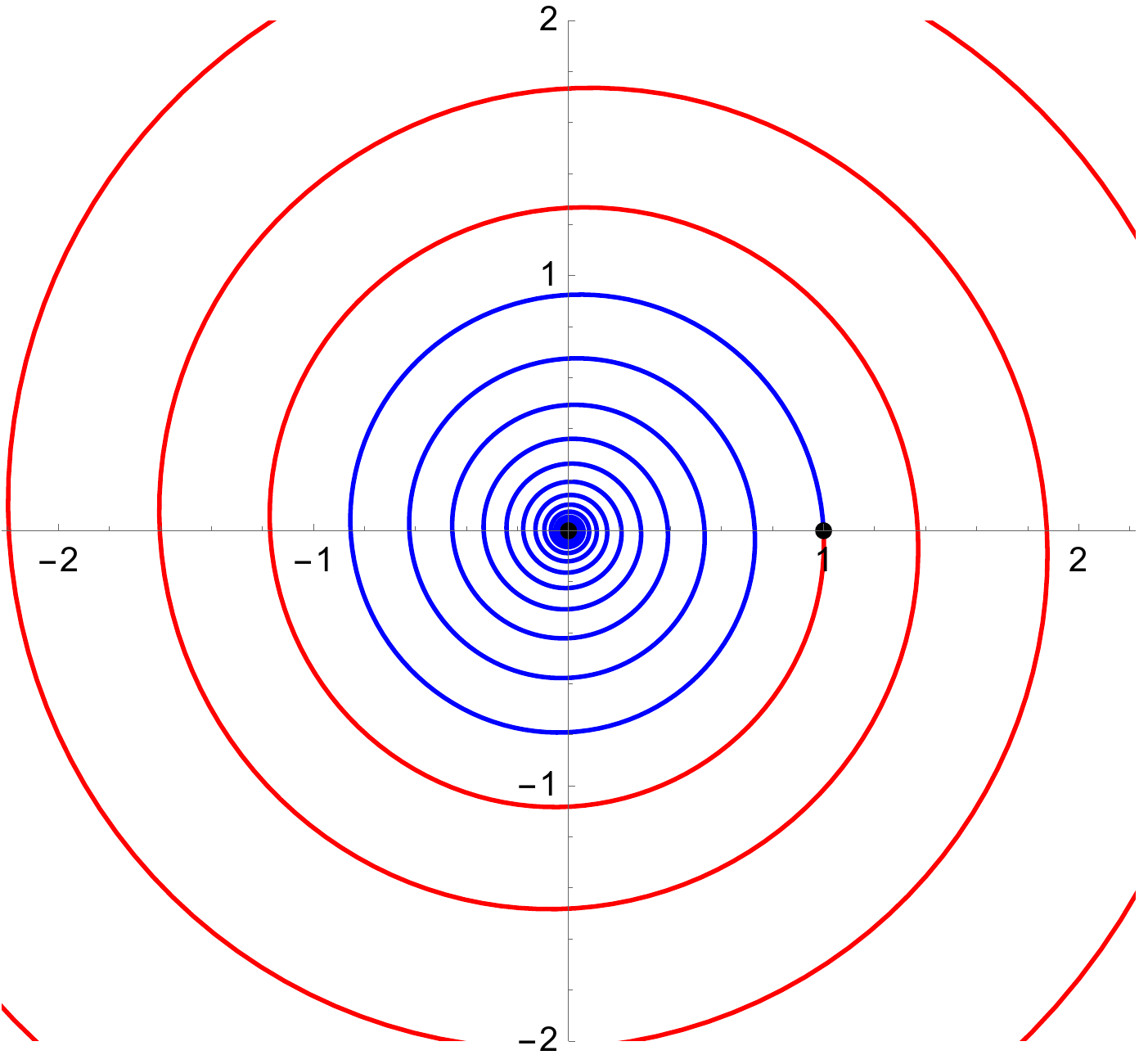}
       \caption{$0<\vartheta<\pi$}
       \label{triv2}
   \end{subfigure}
   \caption{$\net_\vartheta$ for $f=0$. The red trajectory corresponds to $n>0$ and the blue trajectory corresponds to $n<0$. 
   The logarithmic branch cut (not shown) runs between $x=1$ and infinity. }
   \label{trivtrajs}
\end{figure}

In linear framing there are no polynomial branch points, only a logarithmic branch point located at $\logbpx=1$ on the single sheet $i$ of the covering.
As explained in Section \ref{sec:log-punctures-exp-net} there are two families of $\CE$-walls that emanate from $\logbpx$. 
These have the following spiralling shapes
\be\label{trivtrajectory}
	\CE_n\,:\quad x(t) = \exp \left( \sgn(n) \, \frac{t e^{i \vartheta}}{2 \pi i}\right)\,,
\ee
for $t \geq 0$. 
One family of spirals is determined by $n > 0$ and another for $n < 0$, see Figure \ref{trivtrajs}. 
When $0<\vartheta<\pi$ it is trajectories with $n<0$ that spiral towards $x=0$, while those with $n>0$ spiral towards infinity.

It it important to observe that, although at $\logbpx=1$ the value of $\log y$ diverges, the central charge of BPS solitons associated to the $\CE$-wall originating from this puncture is nevertheless finite.
In fact, the central charge of an $(ii,n)$ BPS soliton supported along $\CE_n$ is obtained by plugging \eqref{trivtrajectory} into \eqref{eq:central-charge-framing-zero} which gives\footnote{The fact that $Z$ does not depend on the sign of $n$ might seems strange. However observe that this is the central charge for BPS states supported at a point at time $t$ along $\CE_n$. If $x_n(t,\vartheta)$ for $t>0$ denotes a point on $\CE_n$ for fixed $\vartheta$, the corresponding point on $\CE_{-n}$, namely $x_{-n}(t,\vartheta)$ is the inverse $x_n(t,\vartheta)^{-1}$. Therefore $n$ and $-n$ concern theories at different points. Moreover $x_{n}(t,\vartheta) = x_{-n}(t,-\vartheta) $. Therefore $Z_n$ and $Z_{-n}$ evaluated at the \emph{same point} indeed have opposite signs.}
\be
	Z_n(t) = |n|  \frac{t e^{i \vartheta}}{2 \pi R}\,.
\ee

The spectrum of BPS solitons is encoded by the supersymmetric CFIV index \cite{Cecotti:1992qh}.
For a fixed value of the (exponentiated) FI coupling $x$, the CFIV index of solitons with charge $a\in \Gamma_{ii,N,N+n}$ is given by \eqref{eq:log-p-soliton-data}.
In general the index is piecewise constant along $\CE_n$, but it could jump at walls of marginal stability. 
In this case there are no intersections among $\CE$-walls, therefore the only MS wall corresponds to the generalized saddle appearing at $\vartheta=0,\pi$.
The saddle is a circular trajectory beginning and ending at $x=1$, which divides $\IC^*_x$ into two regions: $|x|<1$ and $|x|>1$ with different yet related BPS spectra - the spectrum at $x$ and $1/x$ is same up to the sign change of vorticity: $n \leftrightarrow -n$, compatibly with \eqref{eq:log-p-soliton-data}.

%%%%%%%%%%%%%%%%%%%%%%%%%%%%%%
\subsection{BPS states near $x=0$ and test of Conjecture \ref{conj:disks-from-iin}}\label{sec:BPS-spectrum-from-exp-net-fr-0}

As explained above there is only one wall of marginal stability, corresponding to the saddle at $\vartheta=0,\pi$ which appears on the unit circle.
We fix a choice of theory point $x_\theory$ in the chamber $|x_\theory|<1$, and determine the BPS spectrum there.\footnote{The spectrum in the chamber $|x|>1$ is essentially identical, and can be obtained in a similar way.}

To probe the entire spectrum of $(ii,n)$ kinky vortices we vary the phase $\vartheta$ of the exponential network over a half-circle, since by CPT symmetry the spectrum of anti-particles can be obtained from that of the particles.  Without loss of generality we shall restrict to $0< \vartheta<\pi$.
Whenever an $\CE$-wall passes through $x$ we record the spectrum of solitons carried by that walls, and include them as part of the spectrum.

\begin{figure}[h!]
\begin{center}
\includegraphics[width=0.8\textwidth]{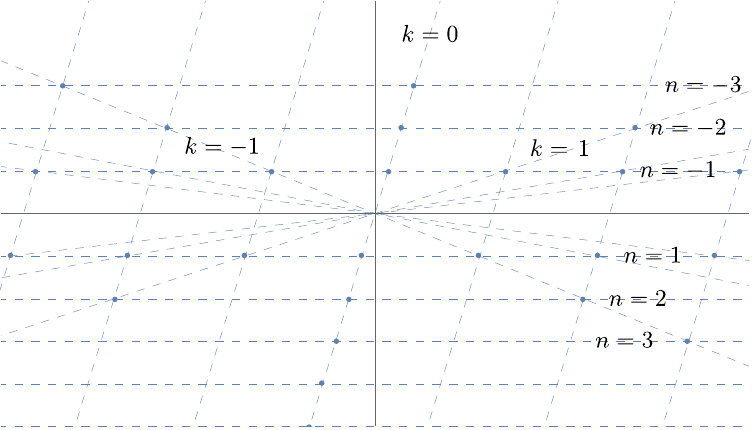}
\caption{The spectrum of $(ii,n)$ BPS states with charges $(n,k)$ in the $Z$-plane, for $|x|<1$ and $f=0$.}
\label{fig:Znk-grid}
\end{center}
\end{figure}

From the exact formula for trajectories \eqref{trivtrajectory} it follows immediately that there will be BPS states whenever
\be\label{eq:theta-k-def}
	\vartheta_{k} = 
	\arg  \left[i\,\sgn(n) \left(\log x_\theory + 2\pi i \, k\right) \right]\,,
\ee
for all $k\in \IZ$. 
Precisely at this phase, a family of BPS states with charges \eqref{eq:topological-charges} given by $(n,k)$ is detected at $x_\theory$.
Since we are working in the chamber $|x_\theory|<1$, we detect BPS states with all values of $n<0$ in the phase range $0<\vartheta<\pi$ of choice. 
Then, for each $\vartheta_k$ there is a whole tower of states with different values of $n$, lying along the radial dashed lines in Figure \ref{fig:Znk-grid}.

BPS states with charge $(n,k)$ 
have central charges given by \eqref{eq:central-charge-framing-zero}, namely
\be\label{eq:Z-n-k-def}
	Z_{n,nk} = n\, Z_{1,0} - nk\, Z_{D0}
\ee
with
\be
	Z_{1,0} = \frac{i}{R}\log x_\theory\,,\qquad
	Z_{D0} = \frac{2\pi}{R}\,.
\ee
The CFIV index is moreover independent of $k$
\be\label{eq:r-n-k-f0}
	r_{n,nk} = \mu_n
\ee
for all $k\in \IZ$, with $\mu_n$ given by \eqref{eq:log-p-soliton-data}. This symmetry arises geometrically because different values of $k$ correspond to different phases for which the same family of spiraling $\CE$-walls \eqref{trivtrajectory} goes through $x$. Since the invariants $\mu_n$ are constsnt throughout the spiral $\CE_n$, the soliton spectrum is the same for all $k$. 
As $k$ increases, so does the length of the spiral between $\logbpx=1$ and $x$, as measured by $t$. This gives solitons with higher mass, organized in Kaluza-Klein towers with step $n\, Z_{D0}$. 
The reason why the Kaluza-Klein momentum is quantized in multiples of $n$ is that 
when boosting an entire $n$-vortex solution along the circle, its momentum shifts linearly with mass which is proportional to $n$, see also footnote \ref{foot:n-k-relation}.
In conclusion, the 3d BPS spectrum of $(ii,n)$ BPS states for a theory at $|x_{\theory}|<1$ is given by $(r_{n,nk},Z_{n,nk})$ as in \eqref{eq:Z-n-k-def} and \eqref{eq:r-n-k-f0}.

The generating series \eqref{eq:ii-n-gen-series} of $(ii,n)$ vortices for the theory at $|x|<1$ is therefore
\be
	\Xi(x) = \sum_{n\geq 1}  \frac{1}{n}\, x^n = -\log(1-x) \,,
\ee
where we excluded summation over $k\in \IZ$, as by definition. Indeed Kaluza-Klein shift symmetry is manifest in this example for any value of $x$, not just near $x=0$.
The expression for $\Xi(x)$ in terms of $\Li_1$ agrees with the decomposition of the vortex potential in terms of $\Li_2$
\be\label{eq:W-vortex-f=0}
	W_\vortex = \int^x \log(1-x)\,d\log x = -\Li_2(x)
\ee
since in both cases $\fn_\ell = \delta_{\ell,0}$.
We find an exact check of Conjecture \ref{conj:disks-from-iin} for $f=0$.

%%%%%%%%%%%%%%%%%%%%%%%%%%%%%%%%%%%%%%%%%%%%%%%%
\subsection{Unspiralling coordinates}

We now describe an alternative way to obtain the BPS spectrum for a theory whose FI coupling corresponds to a given $x\in \IC^*$. 
Although we already determined the spectrum in Section \ref{sec:BPS-spectrum-from-exp-net-fr-0} for the theory currently under study, it will be useful to develop an alternative approach when considering other choices of $f$ (nonlinear framings), and for richer models which will be considered in a separate publication \cite{GL-to-appear}.

The main issue is that in general the exponential networks analysis becomes much more involved. 
Here we introduce a change of coordinates, which we call the \emph{unspiralling map}
\be\label{eq:unspiralling-map}
	\unspmap^{\vartheta} : \ \ID^* \to \IC 
\ee 
that significantly simplifies the analysis.
Here $\ID$ is the open unit disk $\ID=\{|x|<1\}$ and $\ID^*$ denotes the punctured unit disk obtained by removing $x=0$. Restriction to the unit disk is natural, since we are interested in the isolated Higgs vacua that lie in this region of the moduli space, see Figure \ref{fig:R3-vacua}. The parameter $\vartheta$ takes values in the interval $(0,\pi)$ and determines the map as follows
\begin{equation}\label{unspiraltransformation}
    \xunsp = \unspmap^\vartheta(x) = \frac{\log |x|}{\sin\vartheta}\exp\left(i\arg x + \frac{i\log|x|}{\tan \vartheta}\right),
\end{equation} 
The inverse transformation is 
\begin{equation}
    x = (\unspmap^\vartheta)^{-1}(w) = \exp\left(i\arg \xunsp - i |\xunsp| e^{i\vartheta}\right)\,.
\end{equation}

The unspiralling map inverts the $x$-plane by mapping the unit circle boundary of $\ID$ to the origin (hence introducing a real blowdown), and maps the origin in $\ID^*$ to infinity
\begin{align*}
	\lim_{|x|\to 1}\unspmap^{\vartheta}(x)= 0, \qquad
	\lim_{x\to 0}\unspmap^{\vartheta}(x) = \infty\,.
\end{align*}
Moreover this map acts on trajectories of type $(ii,n)$, given by \eqref{trivtrajectory}, as follows.
For any $\vartheta$ the map $\unspmap^{\vartheta}$ takes a generic point $x\in \CE_{n}$ 
to
\be\label{eq:unspiralled-spiral-f0}
	\xunsp = \unspmap^{\vartheta}(x) 
	= 
	- \frac{t}{2 \pi} \,.
\ee
Here we used $\sgn(n)=-1$ since we are in the region $|x|<1$ where only trajectories with $n<0$ appear, for $\vartheta\in (0,\pi)$. 
Trajectories that start at $x=1$ therefore begin at $w=0$ in unspiralling coordinates $w$, and evolve along a straight line along $\IR_{<0}$ towards $-\infty$. 
This is the main property of $\unspmap$: it maps $(ii,n)$ trajectories, which are $\vartheta$-dependent spirals, to straight lines that are \emph{independent} of $\vartheta$.

\begin{figure}[h!]
\begin{center}
\includegraphics[width=0.4\textwidth]{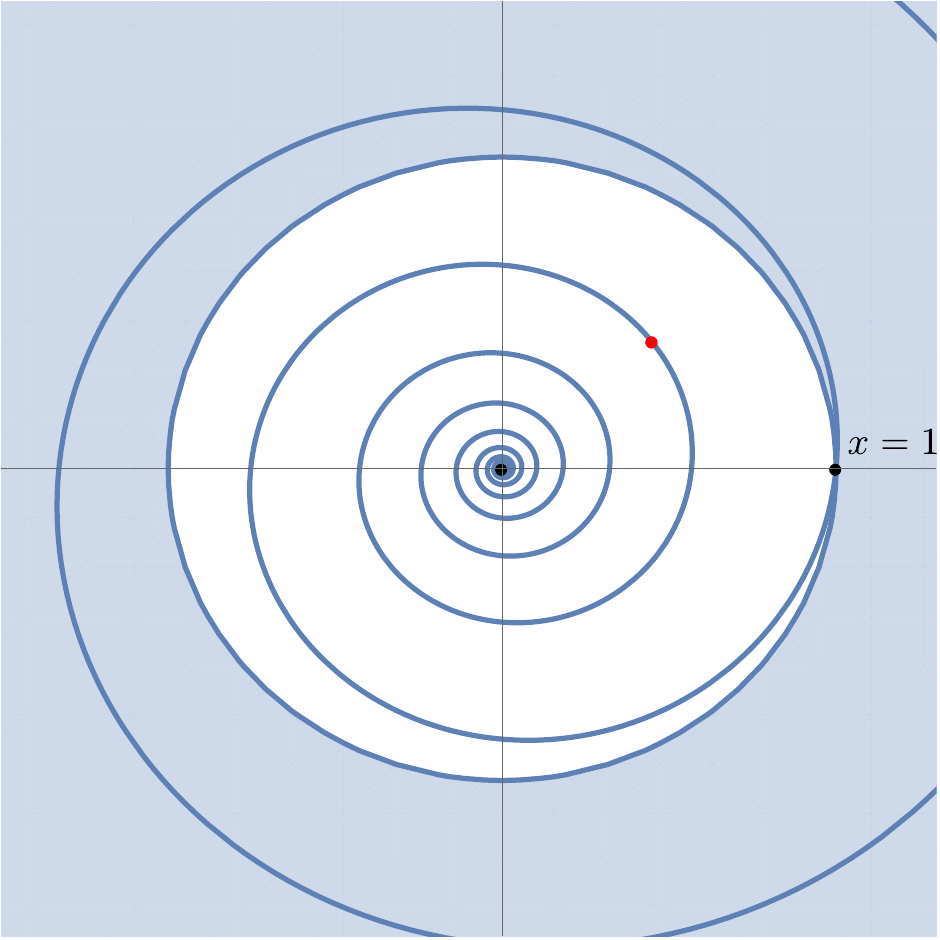}
\qquad
\includegraphics[width=0.4\textwidth]{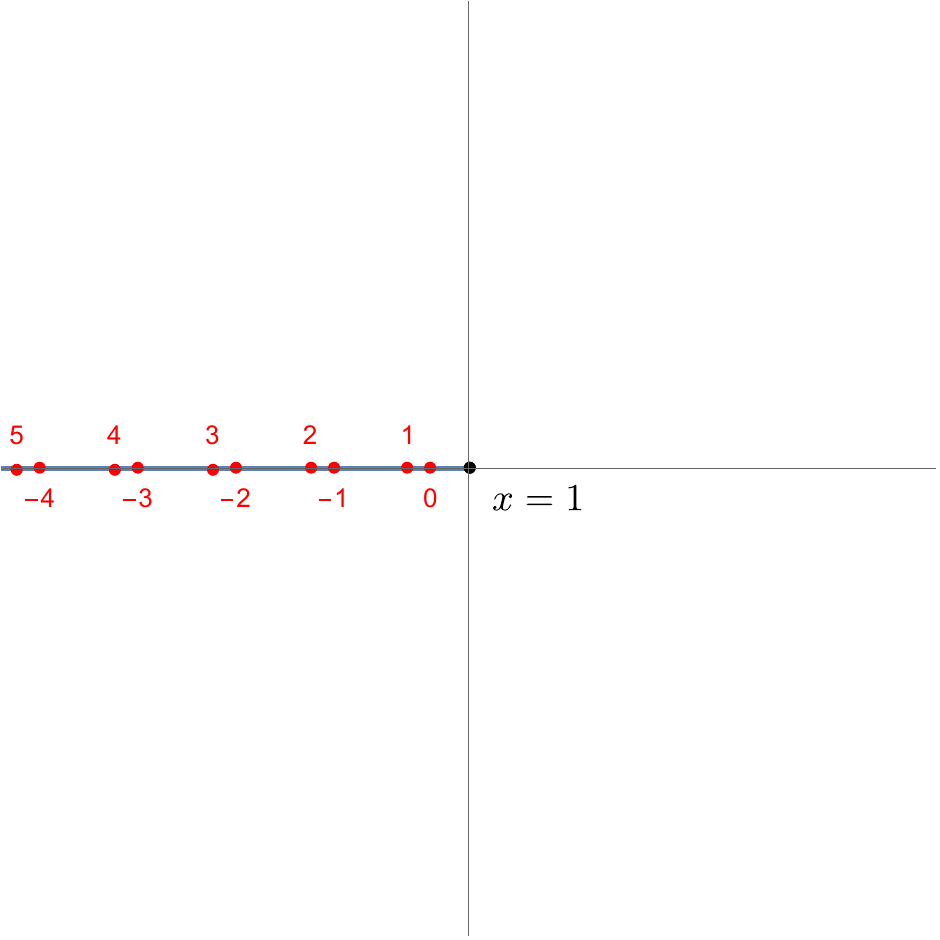}
\caption{
Left frame: The unit punctured disk $\ID^*\subset \IC^*_x$ with the theory point $x_{\theory}$ marked by a red dot, and the family $\CE_n$ for a value of the phase $\vartheta_k$ ($k=0$ in this case) for which the trajectory goes through $x_{\theory}$. 
Right frame: The plane of the unspiralled coordinate $\xunsp$. The family of $\CE_n$ maps to the same straight line for all values of $\vartheta$. 
The theory point evolves along a line as $\vartheta$ varies, and whenever $\vartheta=\vartheta_k$ it lies on $\CE_n$ (red dots marked by $k$).}
\label{fig:unspiralling-C3}
\end{center}
\end{figure}

On the other hand, the theory point defined by a given choice of $x$ traces a curve in the $\xunsp$ plane, parameterized by $\vartheta\in(0,\pi)$. The shape of the curve is obtained by fixing $x$ and letting $\vartheta$ vary in \eqref{unspiraltransformation}.
This curve will intersect the unspiralled $\CE$-wall \eqref{eq:unspiralled-spiral-f0} whenever $\vartheta=\vartheta_k$ as defined in \eqref{eq:theta-k-def} since these are the values of the phase for which $\CE$-walls run through the theory point in the original coordinate $x$.

Therefore, in unspiralled coordinates the BPS spectrum of the theory at $x_{\theory}$ is encoded by a set of points
\be
	\xunsp_k:= \unspmap^{\vartheta_k}(x_{\theory})\,,
\ee
on the straight line \eqref{eq:unspiralled-spiral-f0}, see Figure \ref{fig:unspiralling-C3}.
Each point encodes a whole tower of states with $n=-1,-2,\dots$ and CFIV index given by \eqref{eq:r-n-k-f0}. Different points correspond to different values of $k$, which labels the Kaluza-Klen modes of the $n$-vortex states.

%%%%%%%%%%%%%%%%%%%%%%%%%%%%%%%%
\section{Nonlinear framing}

In this section, we consider the theory with Chern-Simons level $\kappa$ corresponding to framing $f\neq 0,1$ through \eqref{eq:f-kappa}.
When this is the case, the curve $\Sigma$ in \eqref{twistedchiralringequation} has more than one sheet as a covering of $\IC^*_x$.
This corresponds to the presence of multiple vacua, and gives rise to interesting new sectors of the BPS spectrum, including kinky vortices of types $(ij,n)$ with $i\neq j$ that interpolate between different vacua. Interactions of $(ij,n)$ and $(ji,m)$ can result in new contributions to the spectrum of $(ii,k(n+m))$ BPS states via wall-crossing, therefore the analysis of the latter becomes more involved.

For concreteness we focus on the case $f=-1$, although the ideas underlying the forthcoming discussion generalize to arbitrary $f$.

%%%%%%%%%%%%%%%%
\subsection{Geometry of vacua and soliton charges}

\begin{figure}[h!]
   \includegraphics[width=0.5\textwidth]{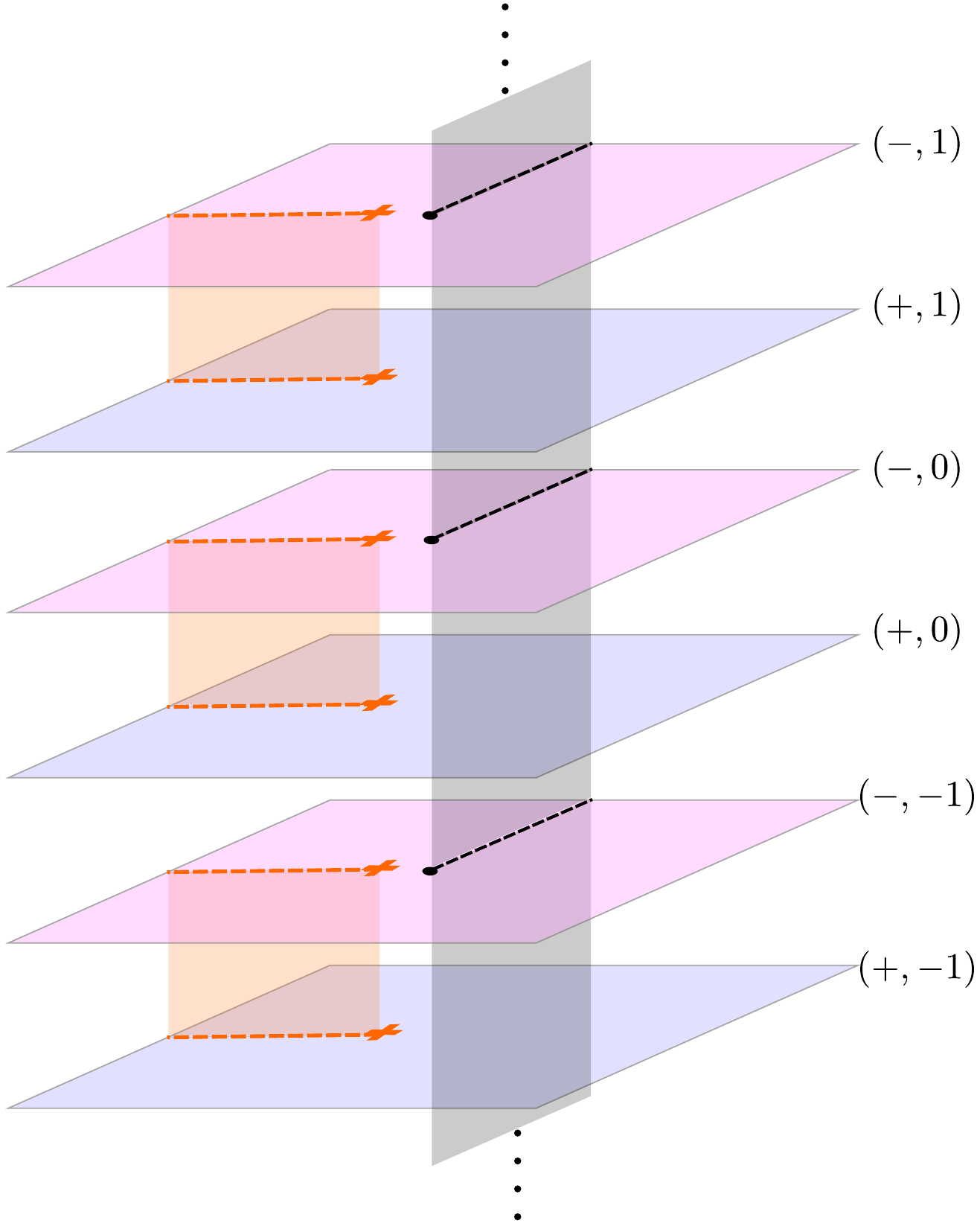}
   \centering
   \caption{
   Various branches $(i,n)$ of $\curve$ in coordinates $(x,Y)$. $\sqrtbp = -1/4$ is the orange cross, $\logbpx = 0$ is the black dot, square-root branch cut is the dashed orange line, and logarithmic branch cut is the dashed black line. 
  }
   \label{swcurve}
\end{figure}

Substituting $f=-1$ in Eq \eqref{twistedchiralringequation} gives the curve 
\begin{equation}\label{eq:Sigma-f1}
    y^2-y-x = 0 \,.
\end{equation}
As a covering of $\IC^*_x$ this has two sheets $y_\pm = \frac{1 \pm \sqrt{1 + 4x}}{2}$
exchanged by monodromy around a square-root branch points located at $\sqrtbp = -1/4$ and at the puncture $x=\infty$. We trivialize $\pi:\Sigma\to\IC^*_x$ by a square-root branch cut running along the negative real axis starting between the two branch points, see Figure \ref{swcurve}.
The two sheets $y_\pm$ lift to two towers of sheets on $\tSigma$
\be \label{VacuumFieldConfigurations}
	Y_{\pm,N}
	= -\frac{1}{2\pi R}\left(\log \frac{1 \pm \sqrt{1 +4x}}{2} + 2\pi i N\right)\,.
\ee
The covering $\tpi:\tSigma\to \Sigma$ is branched at two logarithmic branch points, one located on sheet $y_-$ at $\punct = 0$ ($\lim_{x\to 0}y_-=0$), and the other located at $\punct=\infty$ where both sheets converge ($\lim_{x\to\infty}y_+=\lim_{x\to\infty} y_-=\infty$). 
We fix a choice of trivialization for $\log y$
 by choosing a branch cut that runs from $0$ to $\infty$ along a straight line on sheet $y_-$, whose exponentiated version we show in Figure \ref{swcurve}.
The Seiberg-Witten differential is then globally defined on $\tSigma$ as follows
\be\label{eq:lambda-f1}
    \lambda_{(\pm,N)} =\frac{Y_{\pm,N}(x)}{2\pi R}\, \frac{dx}{x}\,.
\ee

The main novelty of nonlinear framing is the presence of multiple polynomial branches, labeled in this case by $\pm$. This leads to new BPS soliton sectors.
%%%%%
Topological charges of $ij$ kinky vortices take values in the (affine) lattice $\Gamma_{ij,N,N+n}$ defined above \eqref{eq:soliton-data}.
The labels $i,j\in\{\pm\}$ and $N,N+n$ determine boundary conditions for the fields at the two ends of the space-like cylinder, see \eqref{eq:Y-bc}.\footnote{Only the difference $n$ is physically meaningful, due to invariance under large gauge transformations.}
For a given charge sector, the topological charge is determined by the relative homology class $a$ of a soliton path interpolating between the two vacua (two points on $\tSigma$).
As in the case of linear framing, it will be useful to fix a trivialization for the affine lattices of charges.
Let us fix a generic choice of $x\in \IC^*$. Then the (affine) lattice of relative homology 1-chains with endpoints $\log y_i$ and $\log y_j+2\pi i n$ can be trivialized by writing 
$
	a = a_0 + \gamma
$
where $a_0$ is the charge of the lightest BPS state (the one with lowest value of $|Z|$) in sector $(ij,n)$, and $\gamma\in H_1(\tSigma,\IZ)/{\rm ker} \, Z\approx \IZ$ belongs to the homology lattice of $\tilde \Sigma$.\footnote{For a discussion of the quotient by the kernel of central charge and other technical details on the lattice of charges, see \cite{blr2018}.}
As in the case of linear framing, the latter is one dimensional and generated by the (mirror) D0 brane charge. 
Therefore one may trivialize the affine charge lattice $\Gamma_{ij,N,N+n}\simeq \IZ$ by decomposing $a=a_0+k\gamma_{D0}$. Using this, all information about a BPS soliton charge $a$ can be specified by a triple which specifices the charge sector and the shift relative to $a_0$
\be\label{eq:vortex-charge-triv}
	a \,  : \ \ (ij,n,k)\,.
\ee

\begin{figure}[h!]
   \includegraphics[width=0.5\textwidth]{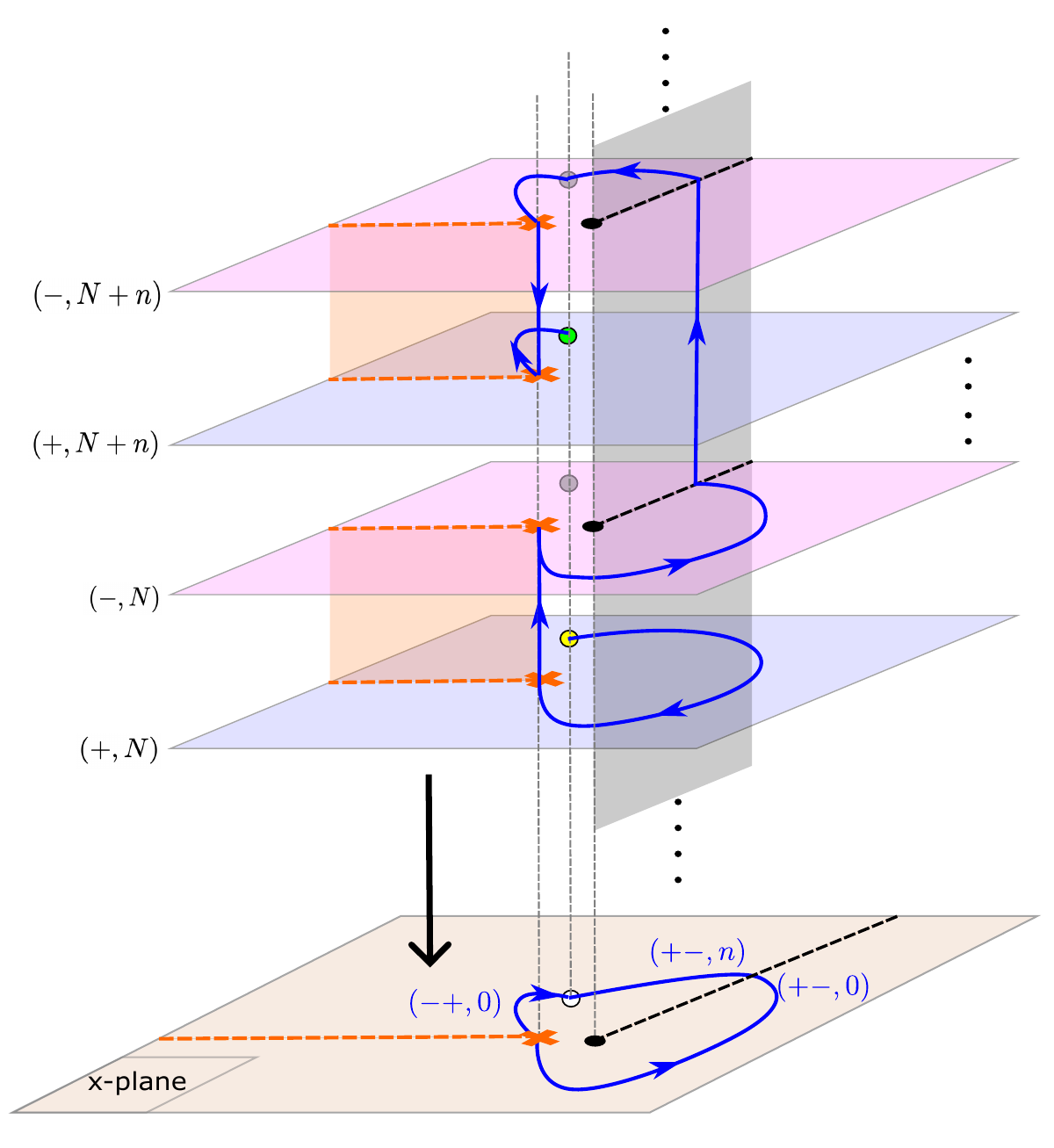}
   \centering
   \caption{An example of a string (blue) wrapping on $\curve$ leading to a transition from one vacuum $(+,N)$ to the other $(+,N+n)$ for some $n > 0$. We also show the projection of the string on the $x$-plane, giving us a part of the exponential network. 
   }
   \label{soliton}
\end{figure}

\begin{figure}[h!]
\begin{center}
\includegraphics[width=0.8\textwidth]{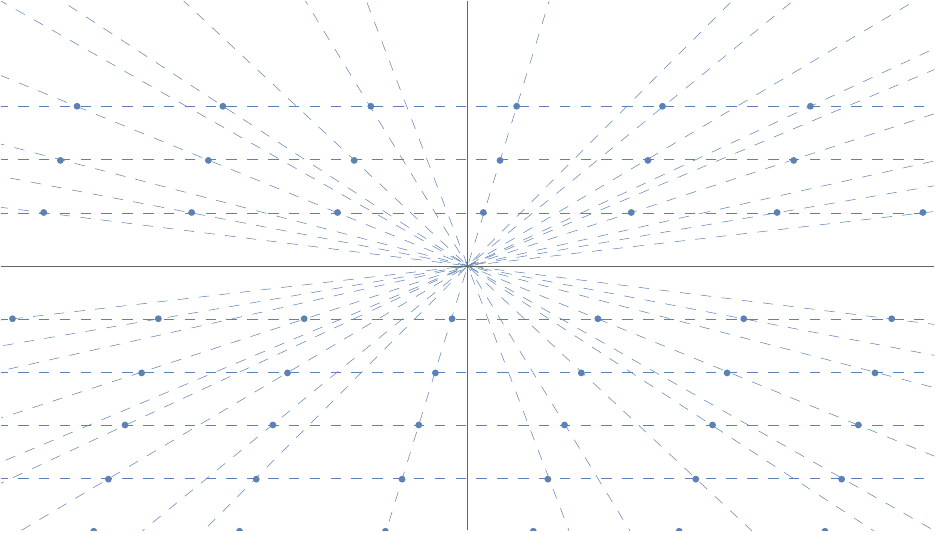}
\caption{The spectrum of $(ii,n)$ BPS states with charges $(n,k)$ in the $Z$-plane, for $|x|<1$ and $f=-1$.}
\label{fig:Znk-grid-f1}
\end{center}
\end{figure}

We now restrict to the case $i=j=+$. 
Since $\lim_{x\to 0}y_+=1$ this sheet of $\Sigma$ corresponds to the circle compactification of the isolated Higgs vacuum, see Figure \ref{fig:R3-vacua}.
BPS states of $(++,n)$ can therefore be regarded as \emph{vortices on a cylinder} in the Higgs vacuum $y_+$, and the logarithmic index $n$ is identified with the {vorticity}.
If we choose the convention that the BPS central charge of the lightest $1$-vortex is 
$
	Z_{a_0} = \frac{i}{R} \log x
$
the BPS central charge of a generic $n$-vortex with charge \eqref{eq:vortex-charge-triv} is then
\be\label{eq:central-charge-nonlinear-framing}
	Z_{n,k} 
	:=
	\int_a\lambda 
	=  \frac{i}{R} \,   \left(n \log x  + 2\pi i\, k \right)  
	= n \, Z_{1,0} - k \, Z_{D0} 
	\,.
\ee
The lattice of charges is shown in the central charge plane in Figure \ref{fig:Znk-grid-f1}.

\begin{remark}[Boosted vortices]\label{rmk:vortex-partitions}
Note an important difference with the case of linear framing \eqref{eq:Z-n-k-def}.
In that case $k$ was multiplied by $n$, but here it is not. This formula is more general and encompasses the $f=0$ case (by restricting $k$ to multiples of $n$).
Physically this generalization is needed because in general an $n$-vortex state can be composed of a collection of single-center vortices labeled by a partition $\lambda = (\lambda_1,\dots, \lambda_\ell)$, see Section \ref{sec:moduli-space-decomposition}.
When a multi-center configuration $\lambda$ is boosted along $S^1$ by $m$ units of KK momentum,
each center acquires $k$ units of momentum, therefore overall momentum of a configuration $\lambda$ gets boosted in units of $|\lambda|$ units of KK momentum
\be\label{eq:vortex-partition}
	\sum_{i=1}^{|\lambda|}  (n_i \, Z_{1,0} + m Z_{D0})
	=
	n \, Z_{1,0} + m |\lambda| Z_{D0}\,.
\ee
Comparing with \eqref{eq:central-charge-nonlinear-framing}, we see that $k=m|\lambda|$ is quantized in units of $\lambda$ for a vortex configuration labeled by the partition $\lambda$.
In the case with only single centers of vorticity $n_i=1$ we have $\lambda = (1,1\dots,1)$ and $|\lambda|=n$, therefore recovering \eqref{eq:Z-n-k-def}.
 
A crucial point is that, depending on the value of $f$, not all the values of $n_i$ are allowed.
In the case $f=0$ the only allowed value for $n_i$ is $1$, while for $f\neq 0,1$ all values of $n_i$ are allowed.
\end{remark}

%%%%%%%%%%%%%%%%%%%%%%%%%%%%%%%%
\subsection{Exponential Networks and 3d BPS wall-crossing}

The exponential networks of the curve \eqref{eq:Sigma-f1} were studied extensively in \cite{blr2018}.
For generic $\vartheta$, the network consists of infinitely many trajectories.
There are three \emph{primary} $\CE$-walls sourced at the square-root branch point $\sqrtbp = -1/4$, which evolve and generate new trajectories through mutual and self-intersections. The descendant trajectories further intersect among themselves and the primary ones, creating another generation of trajectories, and the pattern repeats forever.

For the purpose of studying the spectrum of $(ii,n)$ BPS states of a theory with coupling $x_\theory$, we need to keep track of the $\CE$-walls of type $(ii,n)$ that pass through $x_\theory$ for different values of $\vartheta$, and of the soliton data $\mu(a)$ that each trajectory carries.
There are two aspects of the network's geometry as a function of $\vartheta$ that will play a relevant role. We discuss each in turn.

\begin{figure}[h!]
\begin{center}
\includegraphics[width=0.8\textwidth]{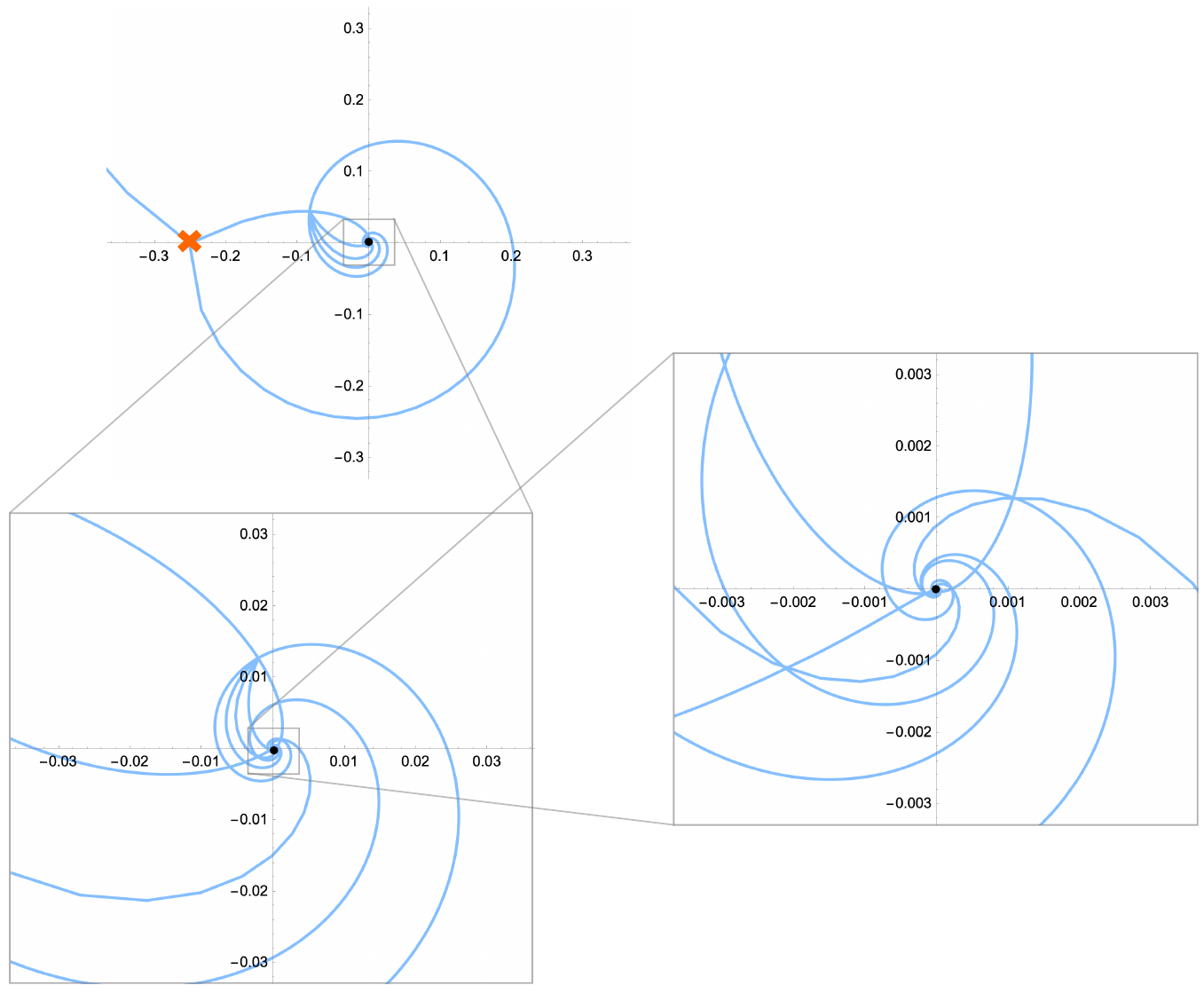}
\caption{Structure of Exponential Networks near the puncture $x=0$.
}
\label{fig:networks-example-f1}
\end{center}
\end{figure}

While the geometry of the network is quite involved, we will be interested in a specific sector of the space of BPS states, namely those with charges $(++,n)$.
Unlike in the case of linear framing studied in Section \ref{sec:f0}, $\CE$-walls of types $(ii,n)$ cannot be generated from logarithmic punctures because now these lie at zero and at infinity.\footnote{The masses of corresponding BPS states would be infinite, as a result.}
Instead $(ii,n)$ $\CE$-walls must be generated at intersections of $(ij,m)$ and $(ji,n-m)$ $\CE$-walls, see \cite{blr2018}.
The generic exponential network is shown in Figure \ref{fig:networks-example-f1}. 
There are infinitely many intersections between walls of type $(ij,m)$ and $(ji,n-m)$. Each of these generates new walls of type $(ii,n)$ and $(jj,n)$.

\begin{figure}[h!]
\begin{center}
\includegraphics[width=0.5\textwidth]{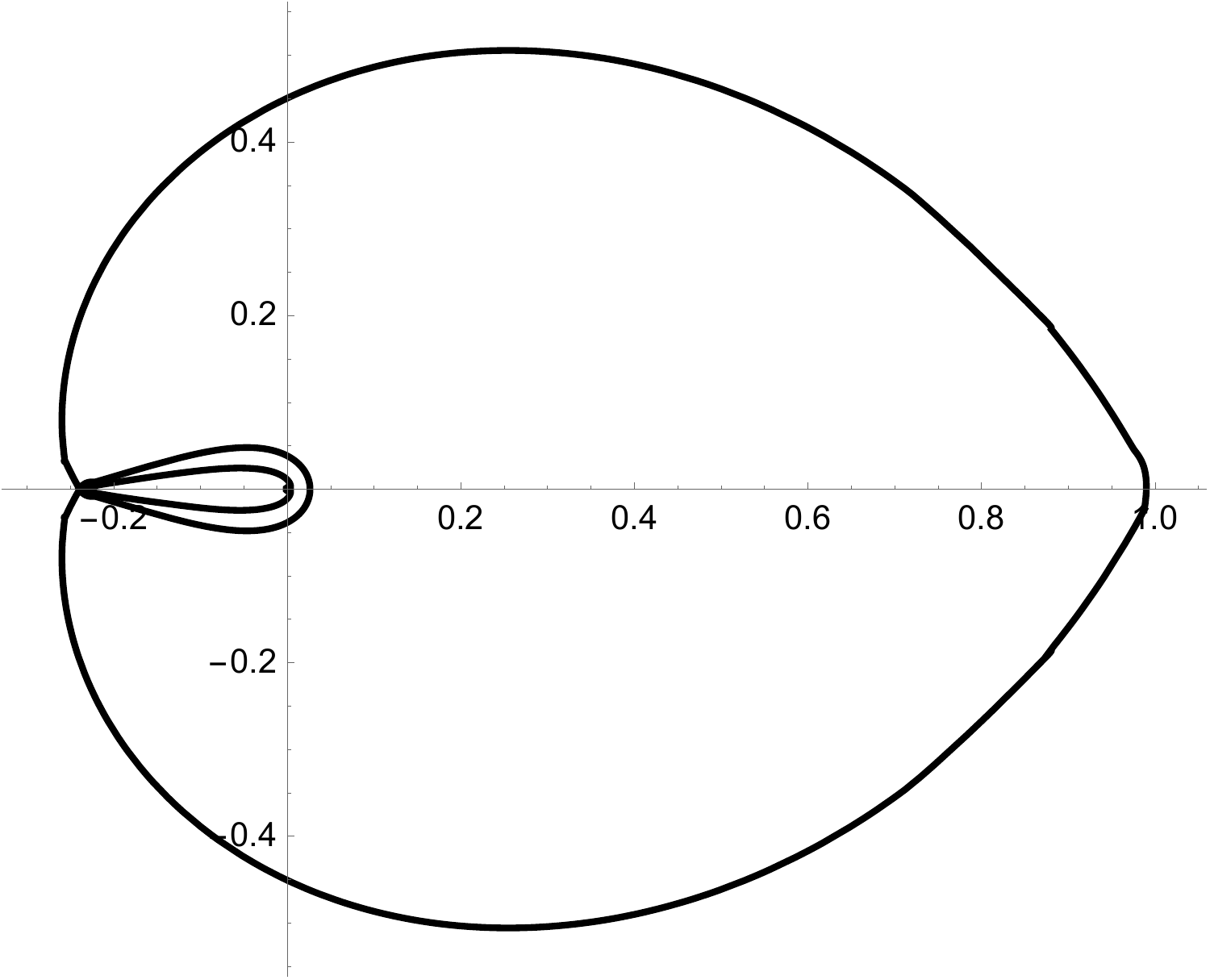}
\caption{Accumulation of MS walls around $x=0$}
\label{fig:MSWalls}
\end{center}
\end{figure}

Tracing the position of intersections as we vary $\vartheta$ gives the location on $\IC^*_x$ where the BPS spectrum jumps, namely the walls of marginal stability.
As shown in Figure \ref{fig:MSWalls}, 
MS walls accumulate towards $x=0$, and the BPS spectrum undergoes \emph{infinitely many} jumps in the limit $x\to 0$. Therefore if we wish to study the spectrum of $(++,n)$ BPS states, the answer will depend on which chamber we choose for the theory parameter $x_{\theory}$. 
Even in the simple model we are considering, the structure of BPS chambers turns out to be rather complicated. We will describe below how to approach the computation of BPS states with networks in a systematic way.

%%%%%%%%%%%%%%%%%%%%%%%%%%%%%%%%%%%%%%%%%%%%%
\subsection{A universal property of exponential networks}\label{ssec:univ-prop}

While networks are rather complicated in this example as compared to $f=0$, physical reasoning based on the interpretation of $(++,n)$ solitons as BPS vortices on a cylinder provides useful guidance.
Let us fix a choice of $x_{\theory}$ somewhere within the unit disk.
The BPS central charge $Z_{n,k}$ of a state with vorticity $n$ and KK momentum $kn$ is given by \eqref{eq:central-charge-nonlinear-framing} with $x=x_{\theory}$.
Thus we expect that $\CE$-walls of type $(++,n)$ will pass through $x_{\theory}$ precisely when $\vartheta = \arg Z_{n,k}$. 
A difference with the case $f=0$ studied above is the occurrence of wall-crossing. There may be multiple (or none) trajectories passing through $x_{\theory}$ at each of these phases, depending on which BPS chamber $x_\theory$ belongs to.

A universal property of $(ii,n)$ walls is that they all have the same shape: solutions of \eqref{eq:E-wall-eq} always have the shape of spirals given by
\footnote{We rescaled $t$ by $|n|$ to uniformize the treatment for all $n$.}
\be\label{eq:generic-spirals}
	\CE_n\,:\quad x(t) = x(0)\, \exp \left( \sgn(n)\, \frac{t \, e^{i \vartheta}}{2 \pi i}\right)\,.
\ee
The starting point $x(0)$ lies on a wall of marginal stability, and coincides with the intersection point of $(ij,m)$ and $(ji,n-m)$ trajectories that create $(ii,n)$. 
A closed formula for $x(0)$ is not known, however recall that if there is a BPS state with charge $(++,n,k)$ then we expect $\CE$-walls of type $(++,n)$ to go through $x_{\theory}$ 
when the phase of the network coincides with 
\be\label{eq:theta-n-k-def}
	\vartheta_{k,n} = \arg Z_{n,k} 
	=\arg \left[i\,\sgn(n)   \left(\log x_\theory  + 2\pi i\, \frac{k}{n} \right)\right] \,.
\ee

\begin{remark}[Phase stabilization]\label{rmk:phase-stab}
In the limit $x_\theory\to 0$ the phases of BPS states accumulate to
\be
	\vartheta_{k,n} \approx \arg (i \, \log |x_\theory| ) \approx \left\{
	\begin{array}{lr}
	\pi/2 & \qquad (n<0) \\
	-\pi/2 & \qquad (n>0)
	\end{array}\right.
\ee
for finite values of $k\in \IZ$. 
Phases of $(++,n,k)$ BPS states, corresponding to slopes of rays
rays in the central charge plane of Figure \ref{fig:Znk-grid-f1},
 get more and more collimated around the imaginary axis.
\end{remark}

Evaluating $\CE_n$ at phases $\vartheta_{k,n}$ gives a family 
of trajectories that pass through $x_{\theory}$ 
\be
	\CS_{k/n}: = \CE_{n}\Big|_{x(0)=x_\theory, \vartheta=\vartheta_{n,k}}\, .
\ee
Up to a positive rescaling of $t$ these are spirals of the following form
\be\label{eq:trajec-th-kn}
	\CS_{k/n} \, : \ \ 
	x(t,\vartheta) = x_\theory^{1+t}\, \exp \left(\frac{2\pi i \,k}{n} \, t\right)\,.
\ee
Recall that the boundary of the unit disk has a special role in the unspiralling map \eqref{eq:unspiralling-map}. 
Also observe that $\CS_{k/n}$ intersects the unit circle precisely for $t=-1$, and it does so at the point
\be\label{eq:unit-circle-point}
	\CS_{n,k}\cap \{|x|=1\} = e^{-\frac{2\pi i \,k}{n}}\,.
\ee

\begin{remark}[A universal property of exponential networks]
A key property of this point is that it is \emph{independent} of $x_\theory$. 
This condition is strikingly strong.
Since $\vartheta_{k,n}$ can be continuously deformed (by varying $x_\theory$), but \eqref{eq:unit-circle-point} is fixed, it follows that not only for $\vartheta_{k,n}$ but in fact for \emph{any} choice of $\vartheta$ trajectories of type $(ii,n)$ must pass through the unit circle on one of the points \eqref{eq:unit-circle-point}.
In other words, in any exponential network $\CW(\vartheta)$, $\CE$-walls of type $(ii,n)$ can cross the unit circle $|x|=1$ only on one of the $n$-th roots of unity, for any $\vartheta$.
\end{remark}

Clearly the situation encountered in linear framing $f=0$ is a special case of this, since the trajectories originated from the puncture located at $x=1$.
Since $\CS_{k/n}$ is an $(ii,n)$ trajectory of the form \eqref{eq:generic-spirals} with basepoint \eqref{eq:unit-circle-point}, it can be expressed in a way that makes no reference to $x_\theory$
\be\label{eq:vortex-line-kn}
	\CS_{k/n}\, : \ \ 
	x(t,\vartheta) = e^{-\frac{2\pi i \,k}{n} + t \frac{e^{i\vartheta}}{2\pi i n}}\,.
\ee
This property of exponential networks, which follows from the physical interpretation of $(ii,n)$ kinky vortices as BPS states with central charge of phase \eqref{eq:theta-n-k-def}, is surprising from a purely  mathematical point of view. 
This property will also be extremely useful in the study of $(ii,n)$ soliton data later on, playing a central role in testing our main conjecture.\footnote{We slightly abuse notation here. The spirals $\CS_{k/n}$ do not just depend on $k/n$ but also on the sign of $n$. Since we will be working with $|x|<1$ and $\vartheta\in(0,\pi)$ we always restrict to $n<0$ unless otherwise stated.}

%%%%%%%%%%%%%%%%%%%%%%%%%%%%%%%%%%%%%%%%%%%%%%%%%
\subsection{Vortex fans and orbits of $x_\theory$}

\begin{figure}[h!]
\begin{center}
\includegraphics[width=0.30\textwidth]{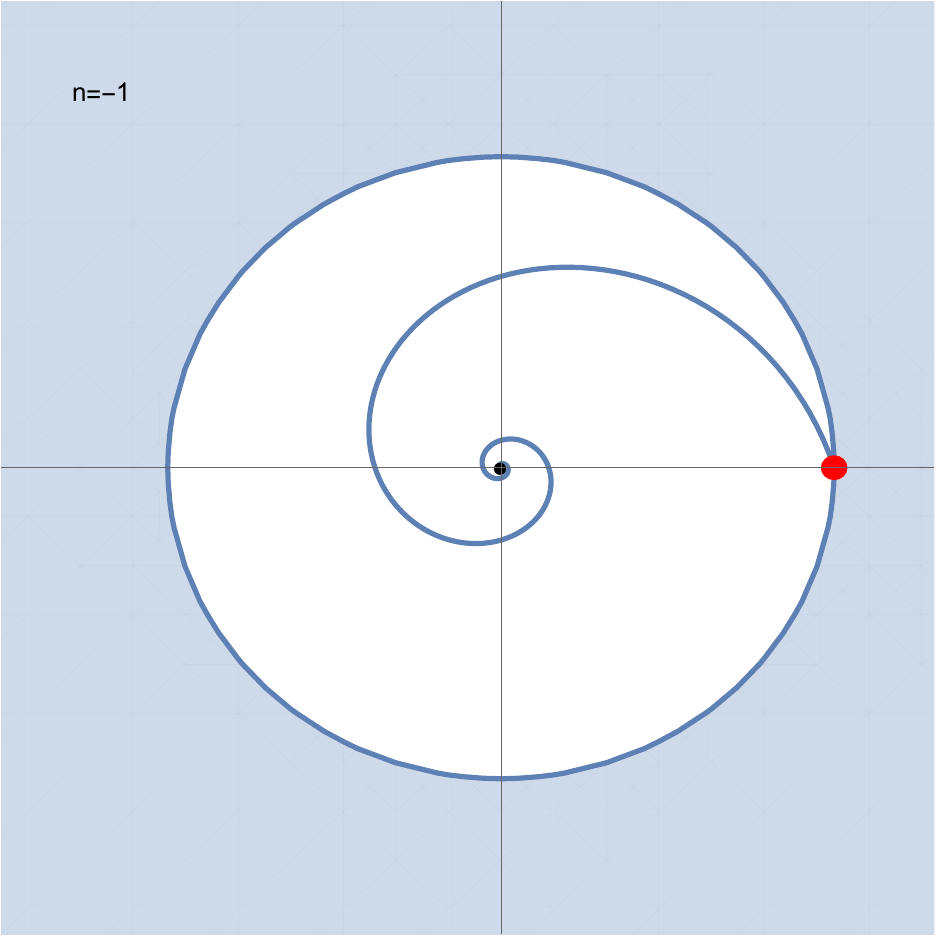}
\includegraphics[width=0.30\textwidth]{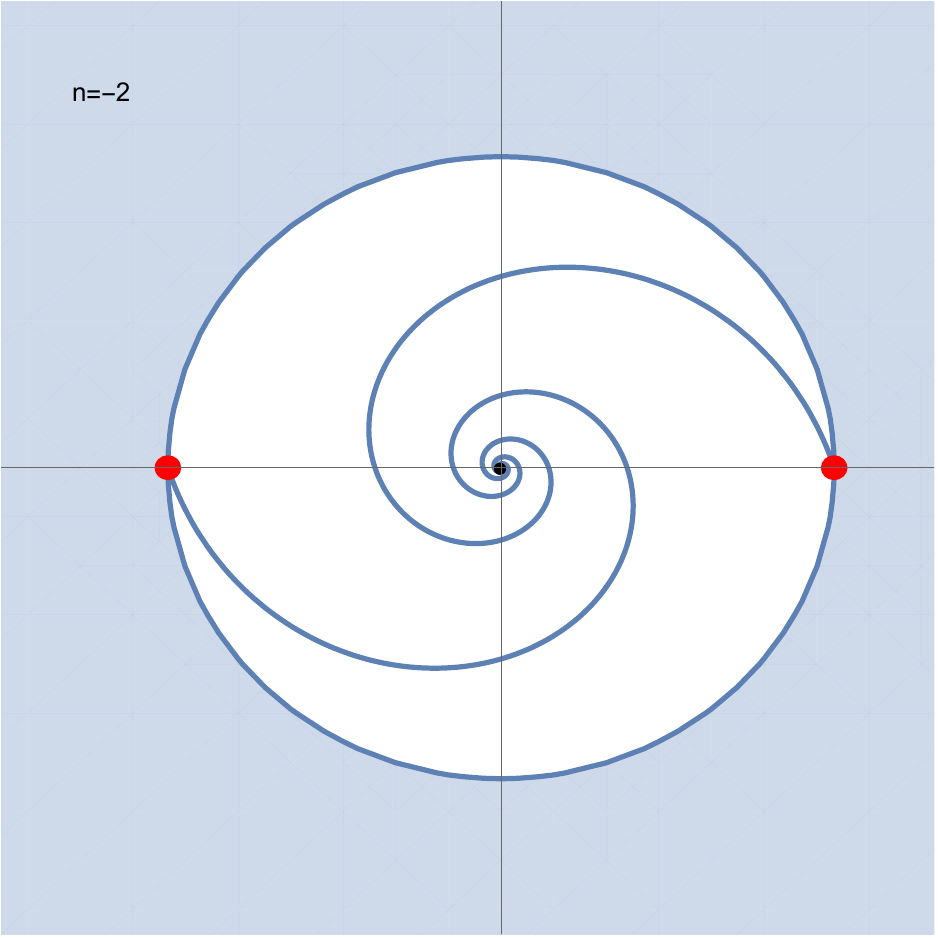}
\includegraphics[width=0.30\textwidth]{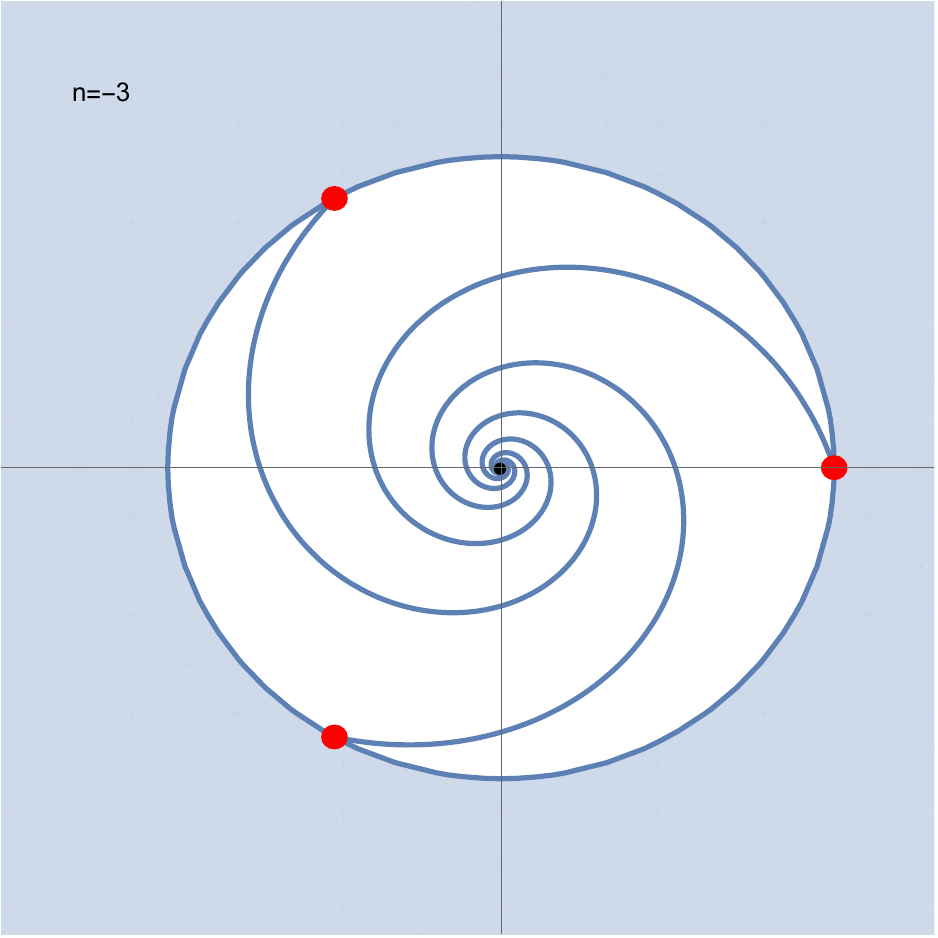}\\
\includegraphics[width=0.30\textwidth]{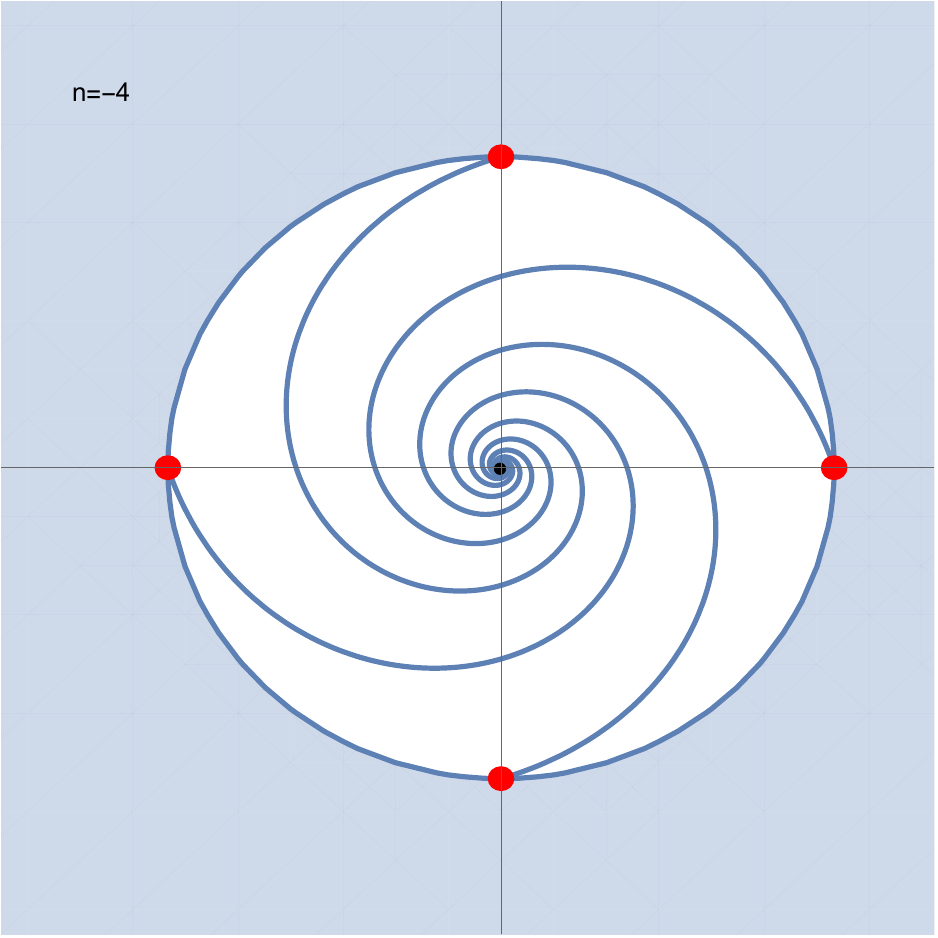}
\includegraphics[width=0.30\textwidth]{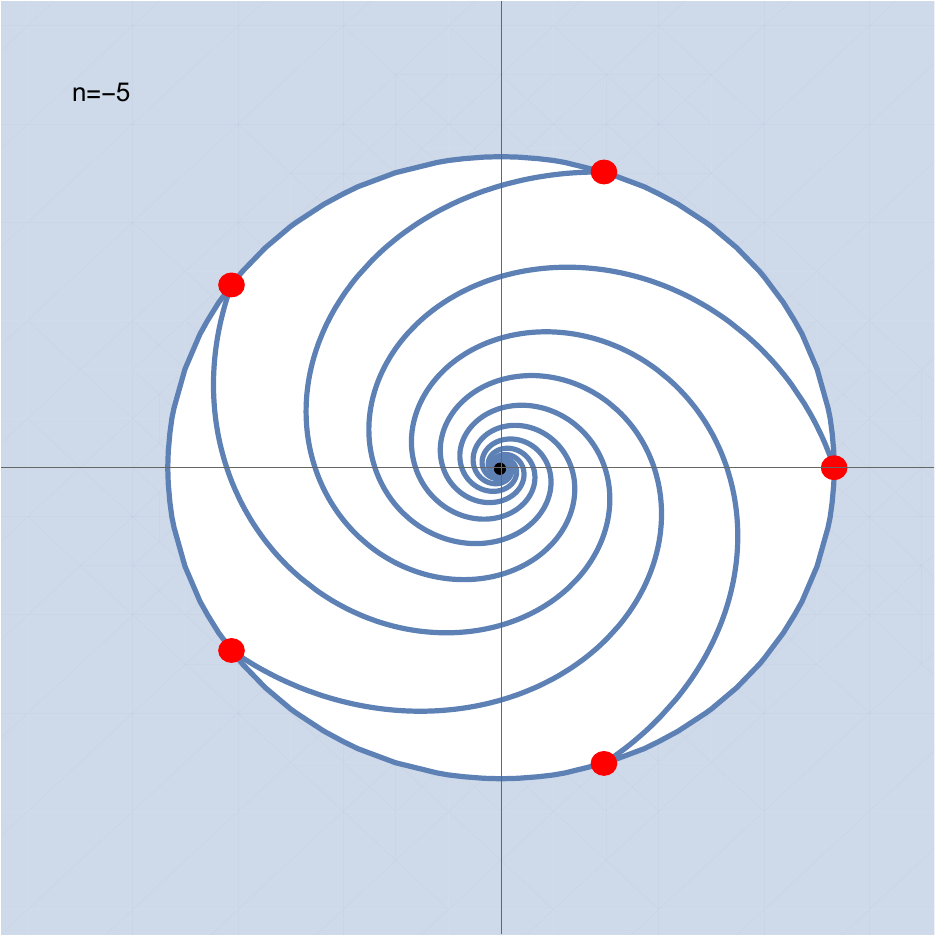}
\includegraphics[width=0.30\textwidth]{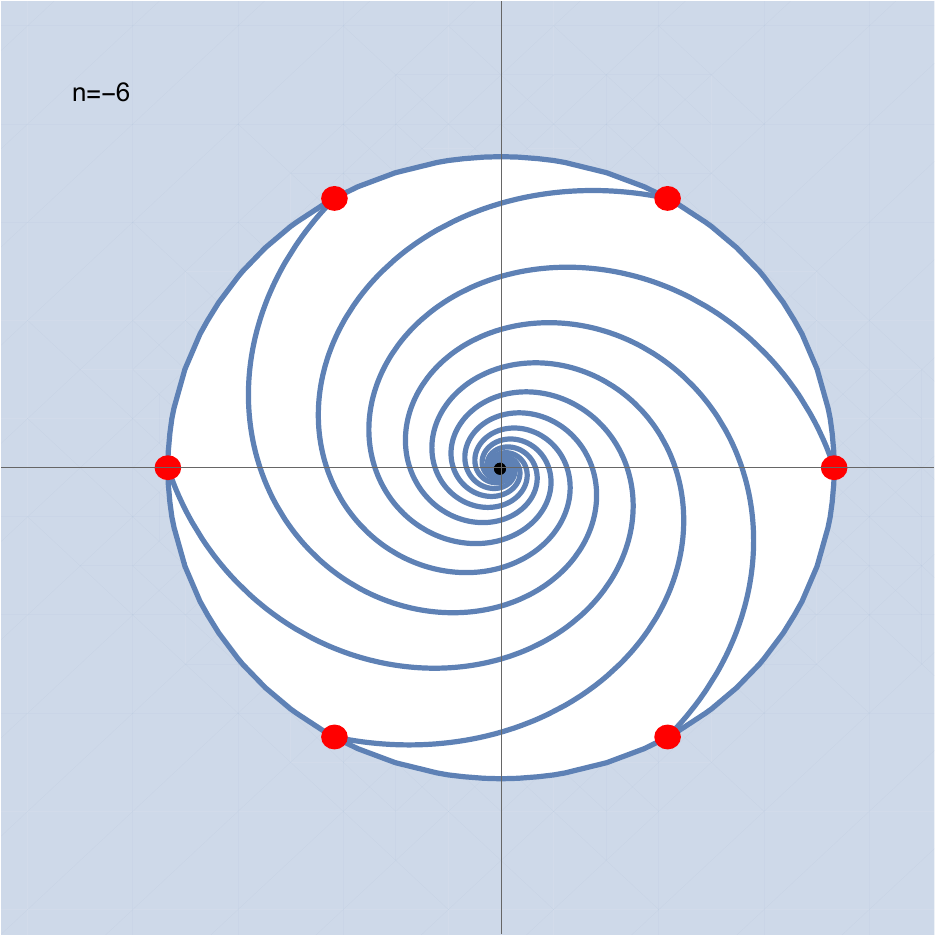}
    \caption{Vortex fans $\CF_n$ for generic $\vartheta$.}
\label{fig:vortex-fans}
\end{center}
\end{figure}

It will be useful to organize the spirals \eqref{eq:vortex-line-kn} into families with the same vorticity $n$.
For each $n$ there are $|n|$ points on the unit circle where trajectories of type $(ii,n)$ can pass, labeled by 
\be
	k = 0,1,\dots |n|-1\,.
\ee
The collection of $(ii,n)$ trajectories forms a fan of spirals, that we call the \emph{$n$-th vortex fan}
\be
	\CF_n = \bigcup_{k=0}^{|n|-1} \CS_{k/n}\,.
\ee
Examples are shown in Figure \ref{fig:vortex-fans}.

\begin{figure}[h!]
   \includegraphics[width=0.6\textwidth]{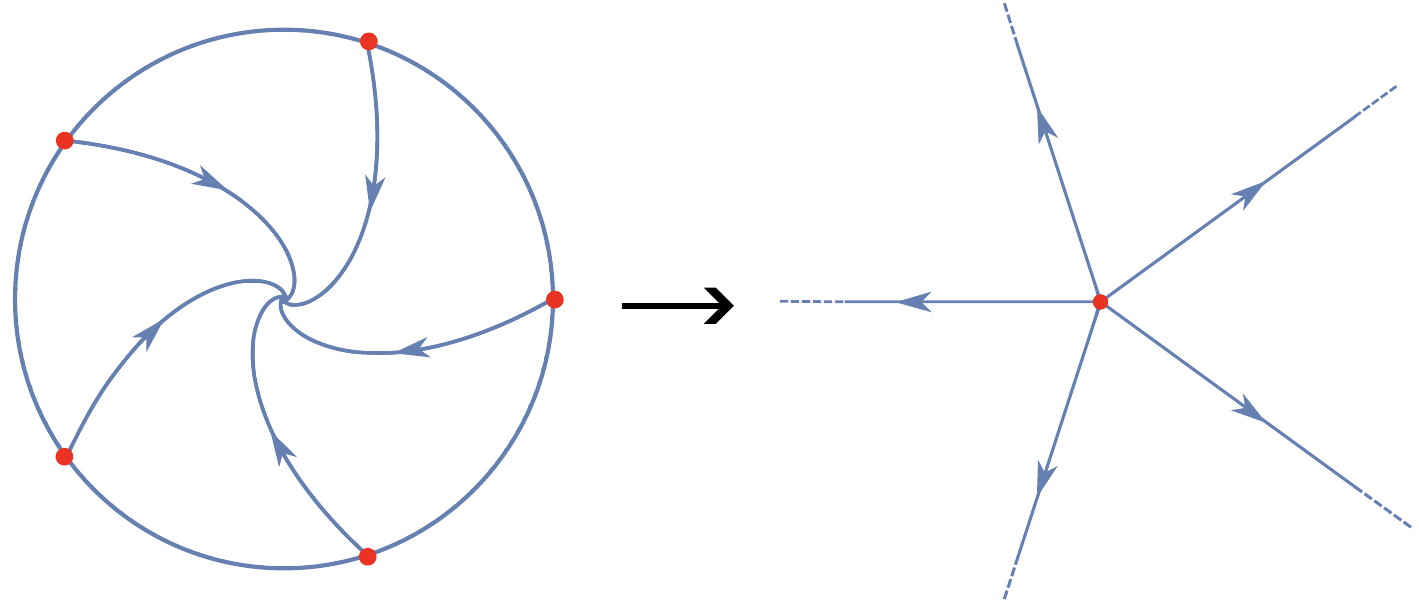}
   \centering
   \caption{Unspiralling coordinate transformation for $\CF_5$.}
   \label{spiralcoordinatetransfimg}
\end{figure}

Applying $\unspmap^{\vartheta}$ from \eqref{unspiraltransformation} to a vortex fan maps spirals to dual trajectories in the $\xunsp$-plane 
\be
	\tilde \CS_{k/n} := 
	\unspmap^{\vartheta} \left( \CS_{k/n} \right)\,,
\ee
which are straight radial lines, see Figure \ref{spiralcoordinatetransfimg}
\be\label{eq:n-fan-lines-w}
	\tilde \CS_{k/n}\,:\ \ \xunsp(t) = \frac{t}{2\pi n}\, e^{-\frac{2\pi i \,k}{n}} \,.
\ee
The dual vortex fan $\tilde\CF_n$ is therefore \emph{stable}, i.e. its image is independent of $\vartheta$. This is thanks to the fact that basepoints for $\CS_{k/n}$ are on the unit circle, which maps to $\xunsp=0$.

\begin{remark}\label{rmk:S-vs-E}
It is important to distinguish between $\CS_{k/n}$ and actual $\CE$-walls of the exponential networks.
Vortex fans $\CF_n$ represent possible shapes of $\CE$-walls of types $(ii,n)$, for any value of $\vartheta$. 
Instead actual $\CE$-walls carry additional information, the soliton data $\mu(a)$. 
In general the soliton data of an $\CE$-wall with shape $\CS_{k/n}$ is \emph{not} constant along the trajectory, due to 3d BPS wall-crossing. 
If soliton data vanishes the $\CE$-wall is simply absent, and more generally its soliton content. 
Although $\CF_n$ does not encode soliton data, it provides a useful parametrization of $\CE$-walls of types $(ii,n)$, which can then be used to study the BPS spectrum with additional information supplied by actual exponential networks. 
\end{remark}

\begin{figure}[h!]
\begin{center}
 \boxed{\includegraphics[width=0.3\textwidth]{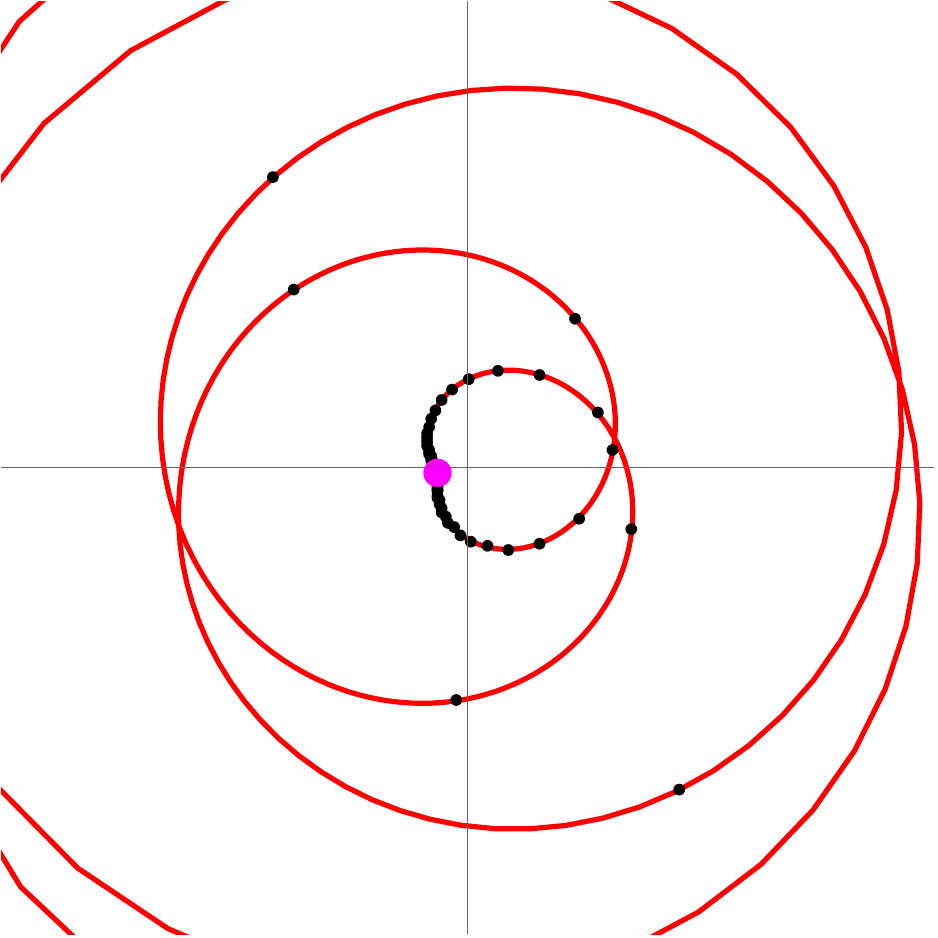}}
 \boxed{\includegraphics[width=0.3\textwidth]{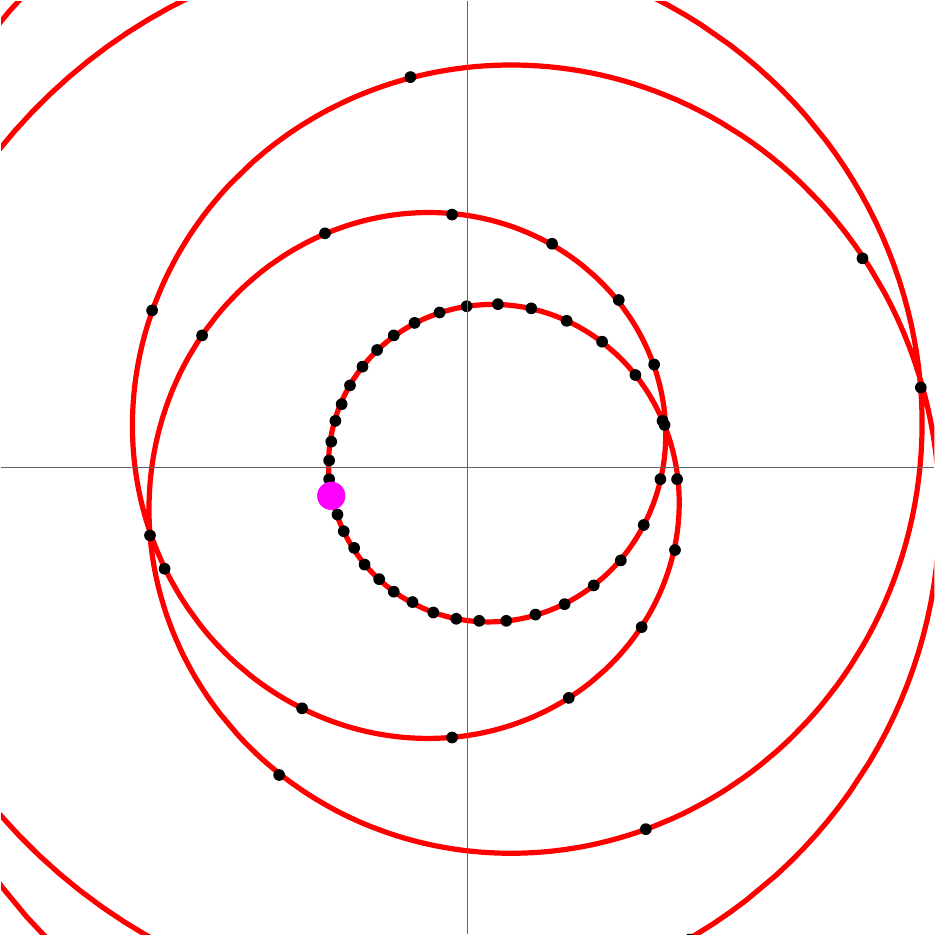}}\\
 \boxed{\includegraphics[width=0.3\textwidth]{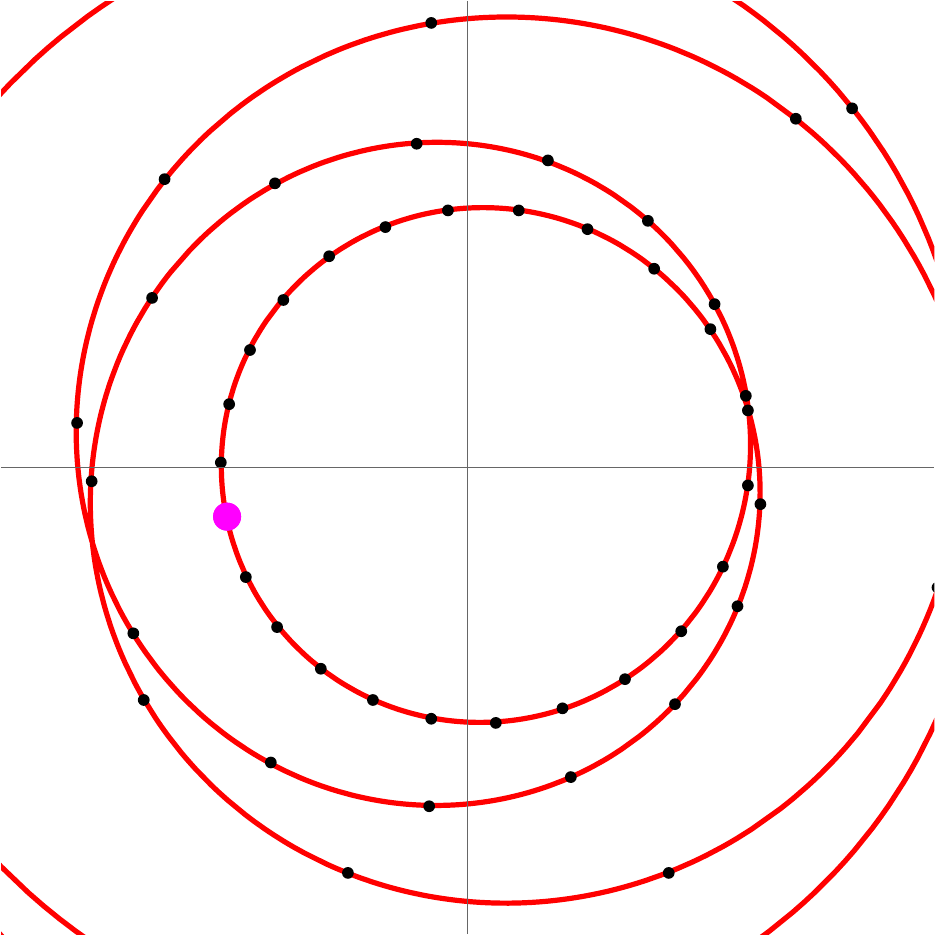}}
 \boxed{\includegraphics[width=0.3\textwidth]{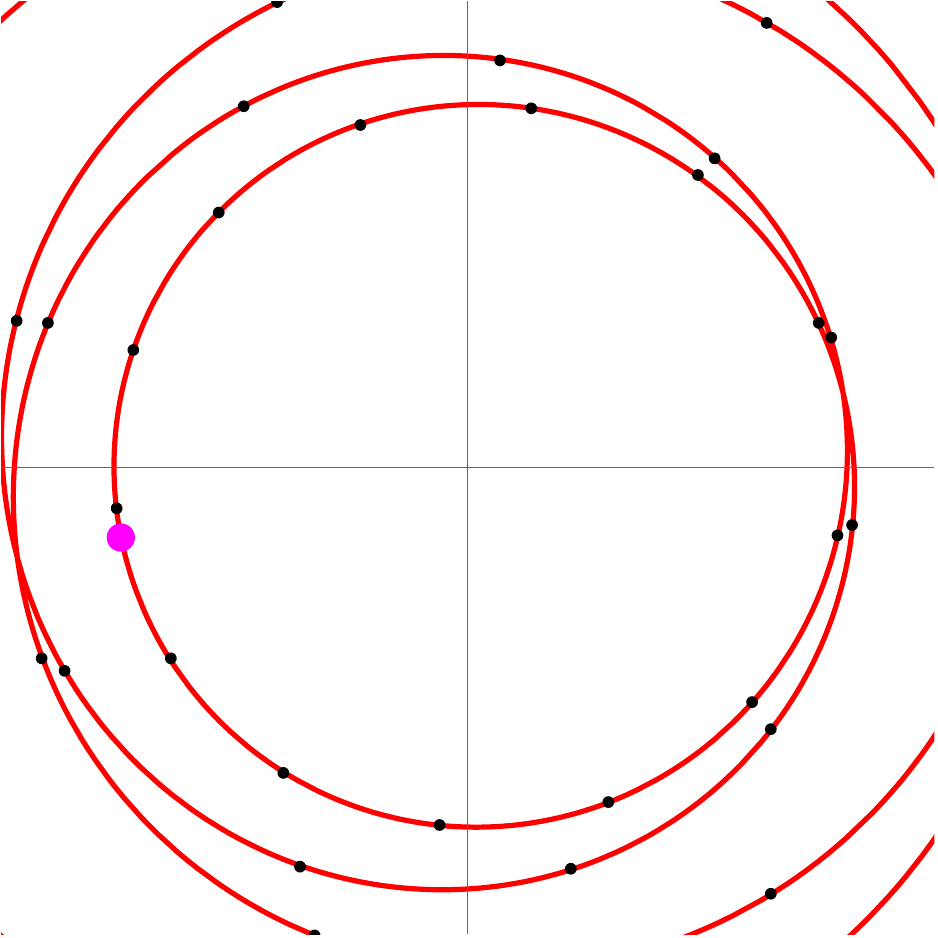}}
   \caption{Orbit of the theory point traced by $\phi^\vartheta(x_\theory)$ in the $\xunsp$-plane, for  $x_\theory=10^{-s}(5+i)$ with $s=1, 2, 3, 4$.}
   \label{fig:theory-orbit}
\end{center}
\end{figure}

To compute the BPS spectrum at a theory point $x_\theory$ we proceed as follows. 
As in the case $f=0$, consider the orbit of $\xunsp_\theory(\vartheta):= \unspmap^\vartheta(x_\theory)$ as $\vartheta$ varies.
The generic orbit is shown in Figure \ref{fig:theory-orbit} for $\vartheta\in(0,\pi)$, the interval of interest.
Marked points on the orbit correspond to equal spacings in $\vartheta$, and the distinguished point in magenta is the position at $\vartheta=\pi/2$. 

From \eqref{unspiraltransformation} we can see that the radius of the orbit is minimal for $\vartheta=\pi/2$, and runs to infinity when $\vartheta \to 0,\pi$.
Therefore, in the limit $x_\theory \to 0$ the orbit moves towards infinity in the $\xunsp$ plane and has approximately the following shape
\be\label{eq:orbit-minimal-point}
	\xunsp_\theory(\vartheta) \approx \frac{\log |x_\theory|}{\sin\vartheta}\exp\left(\frac{i\log|x_\theory|}{\tan \vartheta}\right)\,.
\ee
The speed of the angular motion is controlled by $(\log|x_\theory|)/{\tan \vartheta}$ which goes to zero for $\vartheta=\pi/2$, but spirals faster and faster as $\vartheta\to 0$ or $\pi$.

\begin{remark}\label{rmk:w-theory-radius}
In the limit $x_\theory\to 0$ the
minimal radius of the orbit, which occurs for $\vartheta=\pi/2$
\be
	|w_\theory|_{\rm min} 
	= -\log |x_\theory|\,,
\ee
grows infinitely large.
\end{remark}

Whenever the orbit $\xunsp_\theory$ crosses one of the radial lines with rational slope $\tilde\CS_{k/n}$, there is a possibility that some BPS states with charge $(ii,n,k)$ are present in the spectrum of $x_\theory$.
However the CFIV index $\mu(a)$ can only be determined by studying the exponential network, since it depends on the location of the point of intersection between the orbit and $\tilde\CS_{k/n}$, see Remark \ref{rmk:S-vs-E}. 
One should therefore study the image of the exponential network $\CW(\vartheta)$ under the map $\unspmap^\vartheta$, we turn to this next.

%%%%%%%%%%%%%%%%%%%%%%%%%%%%%%
\subsection{The warped network $\tilde{\mathcal W}(\vartheta)$}

We now study the image of the exponential network $\CW(\vartheta)$ under the map \eqref{eq:unspiralling-map}
\be
	\tilde\CW(\vartheta) := \unspmap^\vartheta(\CW(\vartheta))\,.
\ee
Since $\CE$-walls of type $(ii,n)$ lie along $n$-vortex fans $\CF_n$, we can conclude that their image must lie on the dual fan $\tilde\CF_n$, i.e. the collection of radial lines with $\vartheta$-independent rational slopes in \eqref{eq:n-fan-lines-w}. 

Since $(ii,n)$ $\CE$-walls are generated at intersections of $(ij,n)$ and $(ji,m-n)$ $\CE$-walls, it is natural to ask whether also the latter, and therefore the (image of the) \emph{entire network}, might also be stable.
From the plots of $\unspmap^\vartheta(\CW(\vartheta))$ shown in Table \ref{comptable}, it can be deduced that the network is indeed nearly stable as $\vartheta$ varies.%
\footnote{{{\color{blue}}I think we should add the $(ii,n)$ trajectories to the figures, in some distinguished color or thickened, since they are relevant for this discussion.}}
More precisely, while $\tilde \CS_{k/n}$ are completely fixed, the intersections of $(ij,n)$ and $(ji,m-n)$ $\CE$-walls where $(ii,n)$-walls are generated can slide up and down $\tilde \CS_{k/n}$.

\begin{table}
    \begin{tblr}{
        colspec = {c X[c,h]X[c,h]},
        stretch = 0,
        rowsep = 6pt,
        hlines = {black, 0.5pt},
        vlines = {black, 0.5pt},
    }
        $\frac{\vartheta\text{ (radians)}}{\pi}$ & $\net(\vartheta)$ & $\unspnet(\vartheta):=\unspmap^\vartheta(\net(\vartheta))$\\
	$0.232$ & \includegraphics[width=0.4\textwidth]{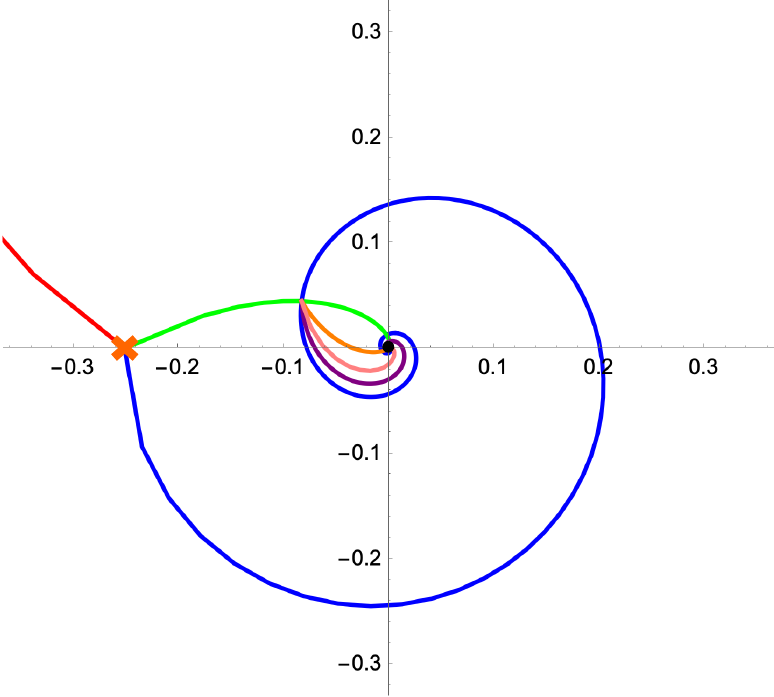}& \includegraphics[width=0.4\textwidth]{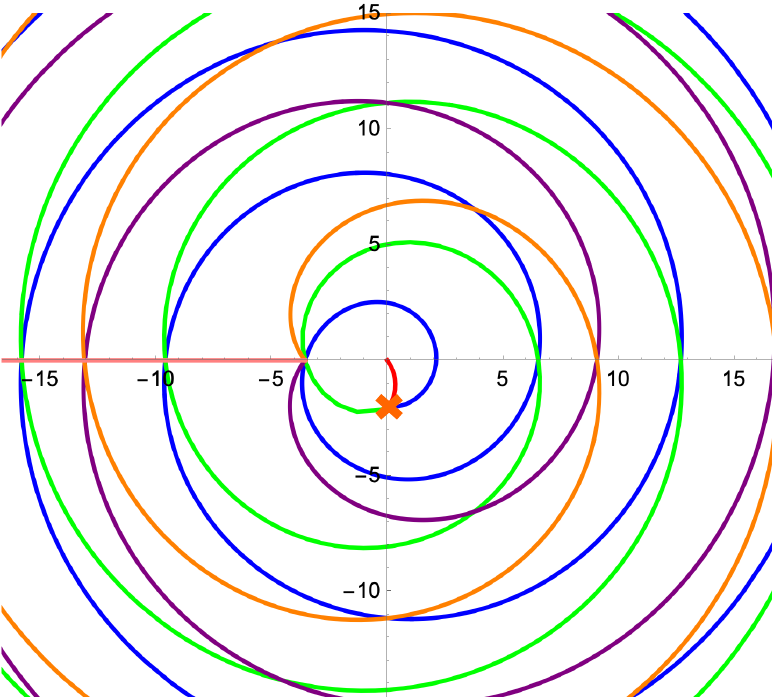}\\
        $0.107$ & \includegraphics[width=0.4\textwidth]{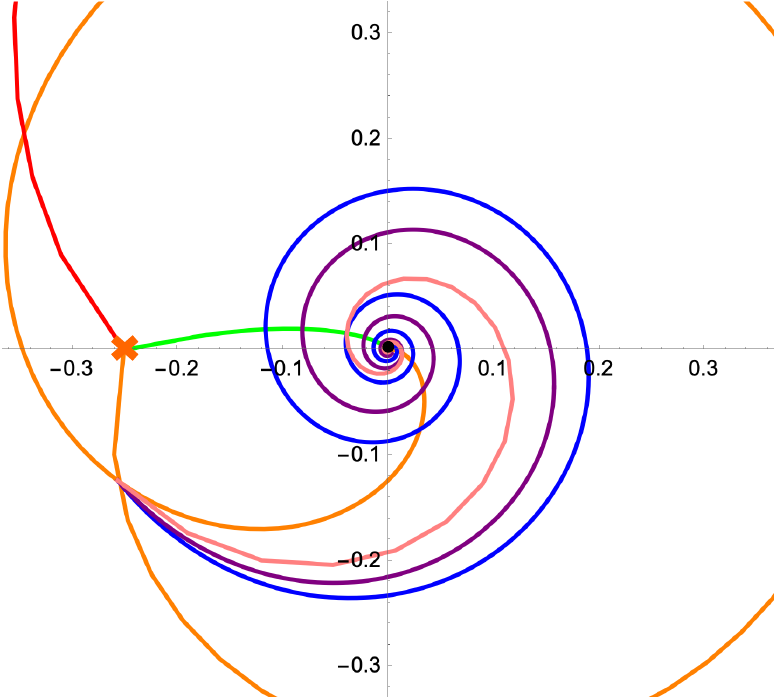}& \includegraphics[width=0.4\textwidth]{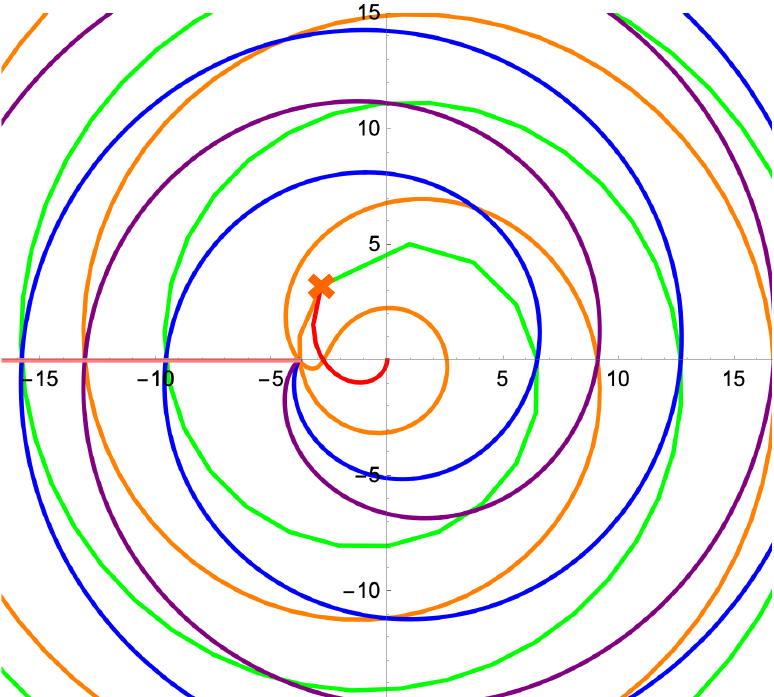}\\
    \end{tblr}
     \caption{Comparison of $\net_\vartheta$ and $\unspnet_\vartheta$ for different topologies of the network. Here, we only show the descendant trajectories corresponding to the first intersection point for simplicity. The colouring scheme is chosen to show that the ``untangled'' network has $\vartheta$ independent asymptotics.}
    \label{comptable}
\end{table}

However, another important phenomenon occurs as we vary $\vartheta$, as can be seen by
comparing the networks in Table \ref{comptable}. 
The image of the branch point 
\begin{equation}\label{branchpointunspiralled}
	\sqrtbpw(\vartheta) := \phi^\vartheta(\sqrtbp) = \frac{ \sqrtbp^{i \cot\vartheta}\log |\sqrtbp|}{ \sin \vartheta}\,,
\end{equation}
moves in the $\xunsp$ plane as we vary $\vartheta$, and for a certain value $\vartheta = \vartheta_\orcross$ it crosses one of the $\CE$-walls. 
Of course, this phenomenon is also visible in the original coordinates, as Table \ref{comptable} also shows. This topological transition of the network may cause, in principle, the BPS spectrum on $\CE$-walls near $x=0$ to jump.
As emphasized in remark \eqref{rmk:w-theory-radius}, when $x_\theory$ is small its dual orbit $\xunsp_\theory(\vartheta)$ becomes large. The BPS spectrum for the theory therefore depends on the values of CFIV indices on the $\CE$-walls in the region of large $\xunsp$, and these may be affected by the topological transition at $\vartheta_\orcross$.
Moreover, the transition in question is only the first of an infinite sequence happening as we vary $\vartheta$.

\begin{figure}[h!]
   \centering
   \begin{subfigure}{.5\textwidth}
       \centering
\includegraphics[width=0.9\textwidth]{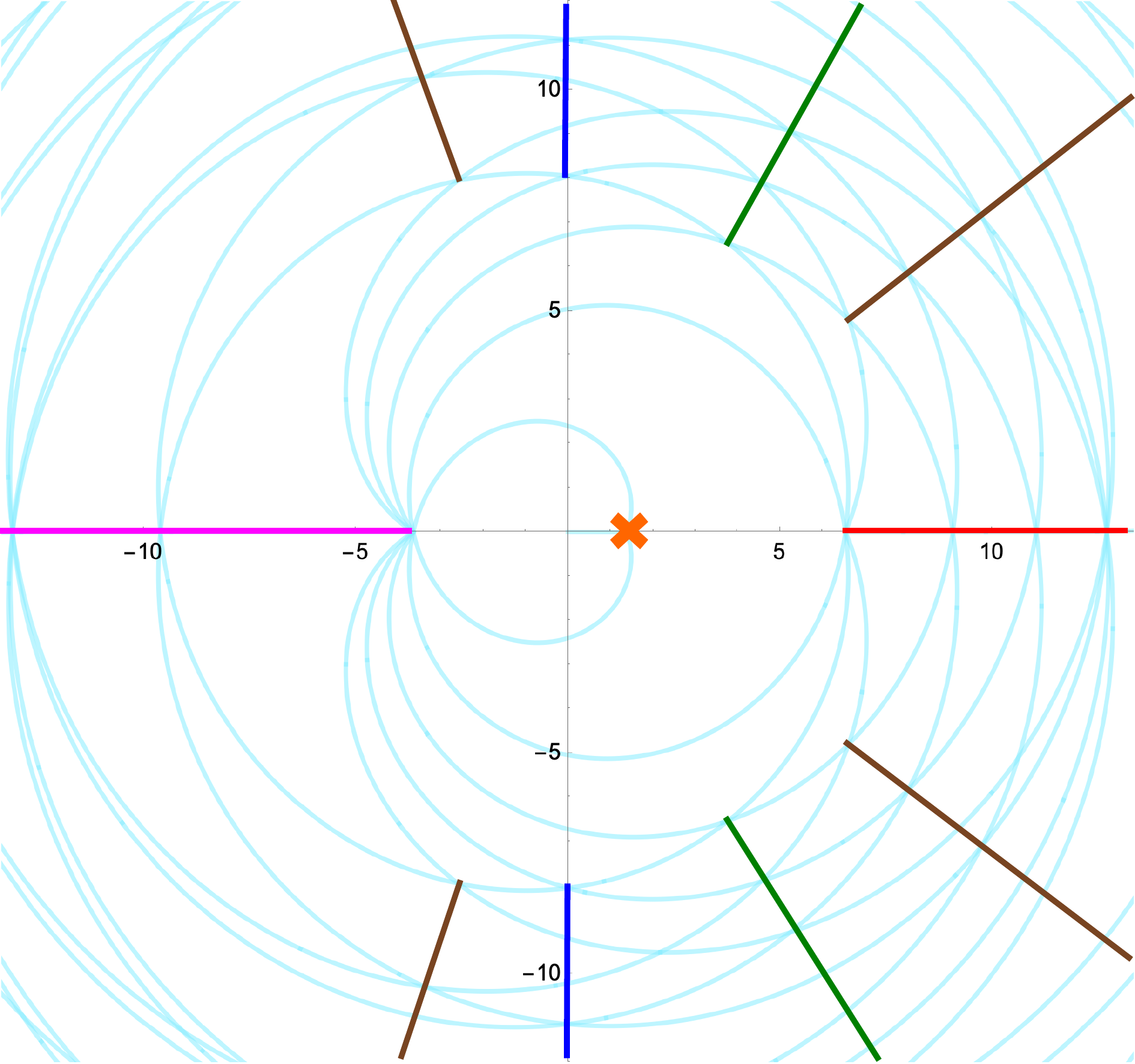}
       \caption{$\unspnet(\frac{\pi}{2})$}
       \label{ppiby2}
   \end{subfigure}%
   \begin{subfigure}{.5\textwidth}
       \centering
       \includegraphics[width=0.9\textwidth]{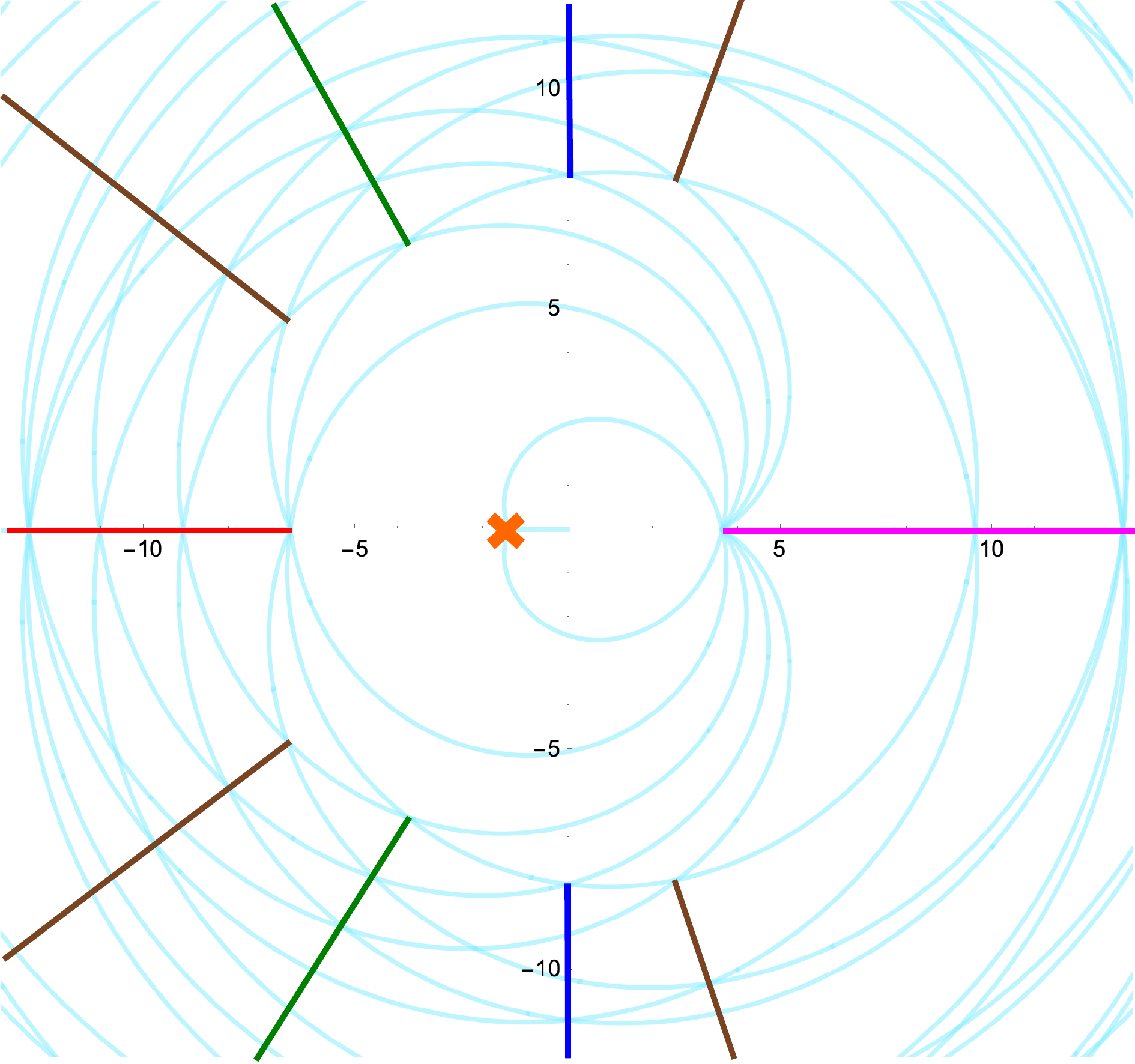}
       \caption{$\unspnet(-\frac{\pi}{2})$}
       \label{mpiby2}
   \end{subfigure}
   \centering
   \caption{Here we show the untangled networks for phases $\pm \frac{\pi}{2}$ which are relevant for computing the BPS spectrum as $x_{\rm{theory}}\to 0$. The translucent blue coloured trajectories show the $ij$ trajectories, while the rest show the $ii$ trajectories. Note that we only show here some descendents for simplicity. The colours of $ii$ trajectories represent the different fan trajectories $\tilde \CS_{k}$ (pink), $\tilde \CS_{k/2}$ (red), $\tilde \CS_{k/3}$ (green), $\tilde \CS_{k/4}$ (blue) and $\tilde \CS_{k/5}$ (brown). In case of common trajectories between fans, we use the colour for the ``lowest'' fan. To get higher order fans, we would have to plot more trajectories.}
   \label{fig:W-pi-over-2}
\end{figure}

Happily, we can deal with this issue in a very simple way, thanks to our previous observations.
Recall that in the limit $x_\theory\to 0$ the phases of $(++,n,k)$ BPS states tend to collimate around $\vartheta_{n,k=0} \approx \pm\pi/2$, see Remark \ref{rmk:phase-stab}.
Therefore, for the purpose of studying Conjecture \ref{conj:disks-from-iin}, which concerns BPS states of $x_\theory\to 0$, we only need to consider the exponential network at a specific phase, namely $\CW(\pi/2)$. We show this and its image $\tilde \CW(\pi/2)$ in Figure \ref{fig:W-pi-over-2}.
Working with this network we can obtain the spectrum for $x_\theory\to 0$ at arbitrary order in $(n,k)$ by going close enough to the puncture.

%%%%%%%%%%%%%%%%%%%%%%%%%%%%%%%%
\subsection{Large radius stabilization of $(++,n,k)$ CFIV indices}

We will now determine the spectrum of BPS vortices in vacua $(++)$ with vorticity $n$ and Kaluza-Klein momentum $k$.
Since we are interested in the limit of large and positive FI coupling, we consider $x_\theory\to 0$. 
This causes all BPS states, up to a certain mass, to have phase near $\vartheta_{n,0}\approx\pi/2$, see \eqref{eq:central-charge-nonlinear-framing}. 
As explained in Section \ref{eq:Open-GW-from ii} this corresponds to sending the size of a BPS vortex to zero, thereby effectively decompactifying the cylinder.
Consider the warped network $\tilde \CW(\vartheta)$ and recall from Remark \ref{rmk:w-theory-radius} that when $x_\theory$ is small, the dual theory point $\xunsp_\theory$ moves on an orbit of large radius as we vary $\vartheta$. 
The problem of computing BPS states at $x_\theory$ therefore translates into the study of intersections of the theory orbit with the dual vortex fan at large $\xunsp$, for values of $\vartheta$ in a range near $\vartheta_{n,0}\approx \pi/2$.

\subsubsection{Kaluza-Klein zero modes}
We first focus on vortices with $k=0$, i.e. no KK momentum, as these are directly relevant to Conjecture~\ref{conj:disks-from-iin}.
For this purpose we only need to study the trajectory $\tilde\CS_{0}$ of the dual vortex fans. 
In fact, since this line corresponds to $\tilde \CS_{n,k}$ for $k=0$ and \emph{any} value of $n<0$, 
it is the unique trajectory that belongs to \emph{all} vortex fans $\{\tilde\CF_n\}_{n\leq -1}$.

According to \eqref{eq:n-fan-lines-w}, $\tilde\CS_{0}$ is the line running along the negative real axis of the $\xunsp$-plane. This coincides with the image of the theory point through $\unspmap^\vartheta$ for $\vartheta = \vartheta_{n,k=0}$
\be
	\arg \xunsp_\theory(\vartheta_{n,0}) = \pi \,.
\ee

Our next task is to determine the CFIV indices for the soliton data of the $\CE$-walls that map to $\tilde \CS_0$.
The main challenge is presented by the fact that $\CE$-walls of type $(++,n)$ keep being generated as $x(t)$ from moves closer to zero (which happens for $t\to\infty$ from \eqref{eq:vortex-line-kn}), and existing $(ii,n)$ trajectories often run into triple intersections with $(ij,m)$ and $(ji,l)$ trajectories, causing further jumps of CFIV indices of already-existing $\CE$-walls. See Figure \ref{fig:W-pi-over-2}.

\begin{figure}[h!]
    \includegraphics[width=1\textwidth]{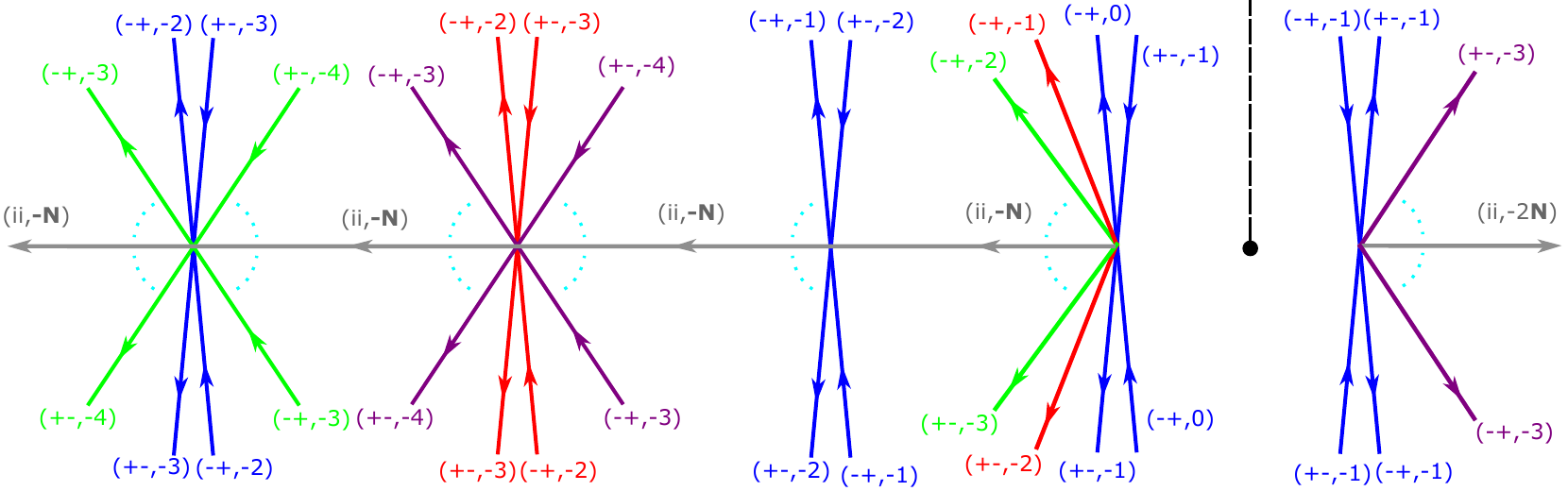}
    \centering
    \caption{Here we show the intersections and trajectories relevant for computing soliton data of the $(++,-n)$ line in $\unspnet$ up to $n\leq 6$. Similar coloured $ij$  trajectories are the same trajectories upto some winding that is not shown here (see Fig \ref{ppiby2} for more details), $ii$ lines are showed in gray, and the logarithmic branch cut is shown as a black dashed line. The labels represent the $(ij,n)$ branch labels for these trajectories. }
    \label{relevantstuff}
 \end{figure}

When this line begins\footnote{Here we fix $n=-1$ for the parametrization with $t$ of $\tilde\CS_0$.} 
at $t=0$ (corresponding to $\xunsp=0$) there is no $\CE$-wall supported on it, see Figure~\ref{fig:W-pi-over-2}. 
We can treat the absence of $\CE$-walls as a formal $(ii,n)$ wall with vanishing CFIV indices
\be
	\mu(a_{n,k=0}) = 0 \qquad n\leq 1,k\in \IZ  \,.
\ee
At a certain point $t_1>0$ along $\tilde\CS_0$ an intersection between $\CE$-walls of types $(+-,-1)$ and $(-+,0)$ generates $\CE_n$ walls of types $(++,n)$ for all $n \in -\mathbb{N}$.
This causes CFIV indices to jump 
\be
	\mu(a_{n,k}) = \delta_{k,0} \, r_{n}^{(1)}  \,,
\ee
where $r_{n}^{(1)}$ are determined by analyzing the soliton data at the intersection of $\CE$-walls, see Appendix \ref{app:computations}.
It will be convenient to encode this spectrum into a generating series\footnote{We omit the sum over $k$ because we are working with the line $\tilde S_{0}$ which only carries BPS states with $k=0$.}
\be
	\Xi[t] = \sum_{n\geq 1} r_{n}^{(1)} x^{n}\,. \qquad (t_1<t<t_2)
\ee
The trajectory $\tilde \CS_0$ proceeds and meets another intersection of $\CE$-walls at $t_2>t_1$. 
This is an intersection of $\CE$-walls of types $(+-,-2)$ and $(-+,-1)$. Since $-2-1=-3$, at this point the CFIV indices that change are those of type $(++,-n-3)$ with $n \geq 0$, giving us the generating series
\be
	\Xi[t] = \sum_{n\geq 1} r_{n}^{(1)} x^{n}
	+  \sum_{n\geq 3} r_{n}^{(2)} x^{n}
	\qquad (t_2<t<t_3) \,.
\ee
At $t=t_2$ the soliton data of the incoming trajectory induces a jump also in the soliton data of incoming $(+-,-2)$ and $(-+,-1)$ trajectories. Moreover, these trajectories are themselves those that generated the line $\tilde\CS_{-1/2}$ (with a shift of log index due to crossing the log cut), see Figure \ref{relevantstuff}.

Further down the line, intersections of trajectories continue to generate corrections to $\Xi[t]$ with jumps at MS walls $t_1<t_2<t_3<\dots$. We collect a complete analysis of these transitions and of the soliton data in Appendix \ref{app:computations}. What we find is the following changes in the generating series 
\be\label{eq:Xi-table}
\begin{array}{c|c}
	& \Xi[t]\\
	\hline
	0<t<t_1 & 0 \\
	t_1<t<t_2 & -x +\frac{1}{2}x^2 -\frac{1}{3} x^3 + \frac{1}{4} x^4 -\frac{1}{5} x^5 +\frac{1}{6} x^6 +\dots \\
	t_2<t<t_3 & -x +\frac{1}{2}x^2 -\frac{4}{3} x^3 + \frac{9}{4} x^4 -\frac{16}{5} x^5 +\frac{14}{3} x^6 +\dots \\
	t_3<t<t_4 & -x +\frac{1}{2}x^2 -\frac{4}{3} x^3 + \frac{9}{4} x^4 -\frac{21}{5} x^5 +\frac{26}{3} x^6 +\dots \\
	t_4<t<t_5 & -x +\frac{1}{2}x^2 -\frac{4}{3} x^3 + \frac{9}{4} x^4 -\frac{26}{5} x^5 +\frac{38}{3} x^6 +\dots 
\end{array}
\ee
The series $\Xi$ stabilizes as $t \to \infty$ to
\be
\begin{split}
	\Xi
	& = -x+\frac{x^2}{2}-\frac{4 x^3}{3}+\frac{9 x^4}{4}-\frac{26 x^5}{5}+\frac{38 x^6}{3}-\frac{246 x^7}{7}+\frac{809 x^8}{8}+O\left(x^9\right)
	\\
	& = - \sum_{\ell\geq 1} \fn_\ell \log(1-x^\ell)
\end{split}
\ee
with
\be
	\fn_\ell = -1, 1, -1, 2, -5, 13, -35, 100,\dots
\ee
These are precisely the integers obtained from the 
decomposition \eqref{eq:intro-conj-2} of $W_\vortex$ (also see \cite[p.30]{Aganagic:2001nx} with $p=-1$)
\be
\begin{split}
	W_\vortex 
	& = \int^x \log y_+\, d\log x \\
	& = 
	-\Li_2\left(\frac{1}{2} \left(1-\sqrt{1+4 x}\right)\right)
	+\frac{1}{2} \log ^2\left(\frac{1}{2} \left(1+\sqrt{1+4 x}\right)\right)
	\\
	& = - \sum_{\ell\geq 1} \fn_\ell \Li_2(x^\ell)\,.
\end{split}
\ee
This matching of integers provide a partial but highly nontrivial evidence for Conjecture \ref{conj:disks-from-iin} for $f=-1$.

\begin{remark}
The genus zero LMOV invariants $\fn_\ell$ for the toric brane with framing $f=-1$
seem to admit the following closed form expression 
\begin{equation}
    \fn_\ell = \frac{1}{2\ell^2}\sum_{d|\ell}(-1)^{d} {2d \choose d} \mu_M\left(\frac{\ell}{d}\right)
\end{equation}
where $\mu_M(n)$ is the M\"obius function. This formula is given in \cite{oeis}.
Notably, it is unclear from this expression whether the $\fn_\ell$ are integers for any value of $\ell\in \IN$, see e.g. \cite{mathoverflow}. However, it is known since \cite{Ekholm:2018eee} that the theory under study arises in the context of the knots-quivers correspondence, and in this case LMOV invariants are integer combinations of motivic Donaldson-Thomas invariants of symmetric quivers, which are known to be integers.
\end{remark}

\subsubsection{Boosted vortices}\label{sec:higher-KK-modes}
Finally we consider BPS states with nonzero KK charge, namely $\mu_{++,-n,k}$ with $k\neq 0$.
We will give a formula for these based on the CFIV indices with $k=0$, and hence in terms of $\fn_\ell$. For simplicity, we will replace $\mu_{++,-n,k}$ with $\mu_{n,k}$ in the discussion below.
 
Consider a vortex configuration with vorticity $n$ and KK momentum $k=0$.
Recall from Section \ref{sec:moduli-space-decomposition} that these vortex configurations 
are labeled by rectangular partitions, where a vortex with vorticity $n$ corresponds to a collection of $d$ copies of single-center vortices with vorticities $n_i=n/d$.
The contributions for each $d$ (dividing $n$) depend on the theory under consideration, i.e. on the choice of CS coupling $\kappa$ (or framing $f$), since the moduli space of single-center vortices with flux $n/d$ changes with $\kappa$.

Next let us consider boosting a vortex configuration labeled by a rectangular partition $\lambda=(n/d)^{d}$. As emphasized in Remark~\ref{rmk:vortex-partitions}, a Lorentz boost along the KK circle assigns to each single-center vortex the same amount of KK momentum. 
Therefore, different partitions contributing to $\CM_{n,0}$ in \eqref{eq:mod-space-decomp} get boosted in different ways. Nevertheless, a boost does not affect neither the internal degrees of freedom $\CI_{n/d}$ nor the degrees of freedom parameterizing the positions of vortices relative to their center of mass i.e. $\CH_d$. This leads to a simple prediction for the moduli spaces of boosted vortices $\CM_{n,k}$ in terms of the building blocks of $\CM_{n,0}$.

As an illustration consider the moduli space of vortices with vorticity 2
\be
	\CM_{2,0} \approx \left(\CI_{2} \times \CH_1 \right) \cup\left(\CI_{1} \times \CH_2 \right)\,.
\ee
Upon boosting by one unit of KK momentum, the single-center configuration in the first component contributes to $\CM_{2,1}$ while the two-center configuration in the second component will contribute to $\CM_{2,2}$, which also gets contributions from the single-center ($\CI_2$) configuration with 
\emph{two} units of KK momentum
\be\label{eq:n=2-boosted}
	\CM_{2,1} \approx  \left(\CI_{2} \times \CH_1 \right)\,,
	\qquad
	\CM_{2,2} \approx \left(\CI_{2} \times \CH_1 \right) \cup \left(\CI_{1} \times \CH_2 \right) \,.
\ee

In general, the moduli space $\CM_{n,k}$ of $(ii)$ kinky vortices with vorticity $n$ and $k$ units of KK momentum gets contributions from $(\CI_{n'} \times \CH_{d})$ if and only if
\be
    n'd = n \hspace{0.3cm} \text{and} \hspace{0.3cm} d|k.
\ee
The above condition implies $d$ divides both $n$ and $k$, allowing us to write an equivalent condition to the above:
\begin{equation}
    d|\operatorname{GCF}(n,k) \hspace{0.3cm} \text{and} \hspace{0.3cm} n' = \frac{n}{d}.
\end{equation}
Here $\operatorname{GCF}(n,k)$ denotes the greatest common factor of $n$ and $k$.
Using the above, we can write the decomposition of $\CM_{n,k}$ as
\begin{equation}
    \CM_{n,k} \approx \bigcup_{d|\operatorname{GCF}(n,k)} \CI_{n/d}\times \CH_{d}.
\end{equation}
Taking the Euler characteristic on both sides, we get the following prediction for CFIV indices with nonzero $k$, in terms of those with $k=0$, or in terms of $\fn_\ell$
\begin{equation}
    \mu_{n,k} = \sum_{d|\operatorname{GCF}(n,k)} \fn_{n/d} \frac{1}{d} 
\end{equation}
This formula agrees with our computations from exponential networks directly, both in the case of zero framing and in the case of framing $-1$. We discuss each in turn.

\paragraph{Framing $0$.}
Recall from \eqref{eq:W-vortex-f=0} that in the case of zero framing the only nonzero LMOV invariant is $\fn_1=1$. This immediately implies that $\mu_n$ only gets partitions from $n$-center configurations with single-center vorticities $n_i=1$
\be
	\CM_{n,0} \approx \CI_{1} \times \CH_n \qquad (f=0)\,.
\ee 
Upon boosting by $k$ units of KK momentum, the component contributes to
\be
	\CM_{n,nk}\approx \CI_{1} \times \CH_n  \qquad (f=0)\,.
\ee
Taking the Euler characteristic recovers \eqref{eq:r-n-k-f0}, stating that for $f=0$ the CFIV index is independent of $k$.

\paragraph{Framing $-1$.}
For the case of framing $-1$ we computed a few CFIV indices with nonzero value of the KK momentum from Exponential Networks.
These are the following
\begin{center}
\begin{tabular}{llll}
	$\mu_{1,1}  = -1$\,,
	\\
	$\mu_{2,1}  = 1$\,,
	\qquad&
	$\mu_{2,2} = \frac{1}{2}$ \,,
	\\
	$\mu_{3,1} = -1$ \,,
	\qquad&
	$\mu_{3,2} = -1$ \,,
	\qquad&
	$\mu_{3,3} = -\frac{4}{3}$ \,,
	\\
	$\mu_{4,1}  = 2$\,,
	\qquad&
	$\mu_{4,2} =  \frac{5}{2}$\,,
	\qquad&
	$\mu_{4,3} = 2$\,,
	\qquad&
	$\mu_{4,4} = \frac{9}{4}$\,.
\end{tabular}
\end{center}
We now explain how these numbers are predicted by the picture of moduli spaces of boosted vortices.
\begin{itemize}
    \item A KK boost by one unit of KK momentum on a vortex with vorticity $n=1$ gives the moduli space $\CM_{1,1}\approx \CM_{1,0}$. Taking the Euler characteristic gives therefore
    \be
	   \mu_{1,1} = \fn_1 = -1 = \mu_{1,0} \,.
    \ee
    \item A KK boost by one or two units of KK momentum on vortices with vorticity $n=2$ gives the two moduli spaces \eqref{eq:n=2-boosted}. Taking the Euler characteristic gives
    \be
	   \mu_{2,1} = \fn_2 = 1\,,
	   \qquad
	   \mu_{2,2} = \fn_2 + \fn_1 \frac{1}{2} = \frac{1}{2} = \mu_{2,0} \,.
    \ee
    \item KK boosts with up to three units of KK momentum on vortices with vorticity $n=3$ give the three moduli spaces
    \be\label{eq:n=3-boosted}
	   \CM_{3,1} \approx \CM_{3,2} \approx  \CI_{3} \times \CH_1 \,,
	   \qquad
	   \CM_{3,3} \approx \left(\CI_{3} \times \CH_1 \right) \cup \left(\CI_{1} \times \CH_3 \right)  \,.
    \ee 
    Here for the case of $3$ units of KK momentum the moduli space gets two contributions: either a single center with vorticity $3$ boosted by $3$ units of KK momentum, or a $3$-center configuration of single-vorticity boosted by one unit of KK momentum each.
    Taking the Euler characteristic, we get
    \be
        \mu_{3,1} = \mu_{3,2} = \fn_3 = -1 \,, \qquad   \mu_{3,3} = \fn_3 + \fn_1 \frac{1}{3} = -\frac{4}{3} =\mu_{3,0} \,.
    \ee
    \item KK boosts with up to four units of KK momentum on vortices with vorticity $n=4$ give the four moduli spaces
    \be
    \begin{gathered}
    \CM_{4,1} \approx \CM_{4,3} \approx  \CI_{4} \times \CH_1 \,,
	\qquad
	\CM_{4,2} \approx \left(\CI_{4} \times \CH_1 \right) \cup \left(\CI_{2} \times \CH_2 \right) \,,\\
    \qquad
    \CM_{4,4} \approx \left(\CI_4 \times \CH_1\right) \cup \left(\CI_2 \times \CH_2\right) \cup \left(\CI_1 \times \CH_4\right)\,.
    \end{gathered}
    \ee 
    Here for the case of $2$ units of KK momentum the moduli space gets two contributions: either a single center with vorticity $4$ boosted by $2$ units of KK momentum, or a $2$-center configuration of vorticity $2$ boosted by one unit of KK momentum each.
    Similarly, for the case of $4$ units of KK momentum there are three contributions corresponding to a single center, to two centers, and to four centers with respectively $4,2,1$ units of KK momentum on each center.
    Taking the Euler characteristic, we get
    \be
        \begin{gathered}
            \mu_{4,1} = \mu_{4,3} = \fn_4 = 2 \,, \qquad 
            \mu_{4,2} = \fn_4 +  \fn_2 \frac{1}{2} = \frac{5}{2} \,, \qquad \\ \mu_{4,4} = \fn_4 +  \fn_2 \frac{1}{2} +  \fn_{1}\frac{1}{4} = \frac{9}{4}.
        \end{gathered}
    \ee
\end{itemize}
This physical interpretation of CFIV indices as Euler characteristics of moduli spaces provides good control over boosted CFIV indices in terms of unboosted CFIV indices. Morover, the fact that the above calculation via moduli spaces matches with our calculations from exponential networks is again a highly non-trivial verification of Conjecture \ref{conj:disks-from-iin} and of the moduli space decomposition \eqref{eq:mod-space-decomp} that follows from it.

\appendix

\section{Analysis of soliton data}\label{app:computations}

\begin{figure}[h!]
       \centering
           \includegraphics[width=0.5\linewidth]{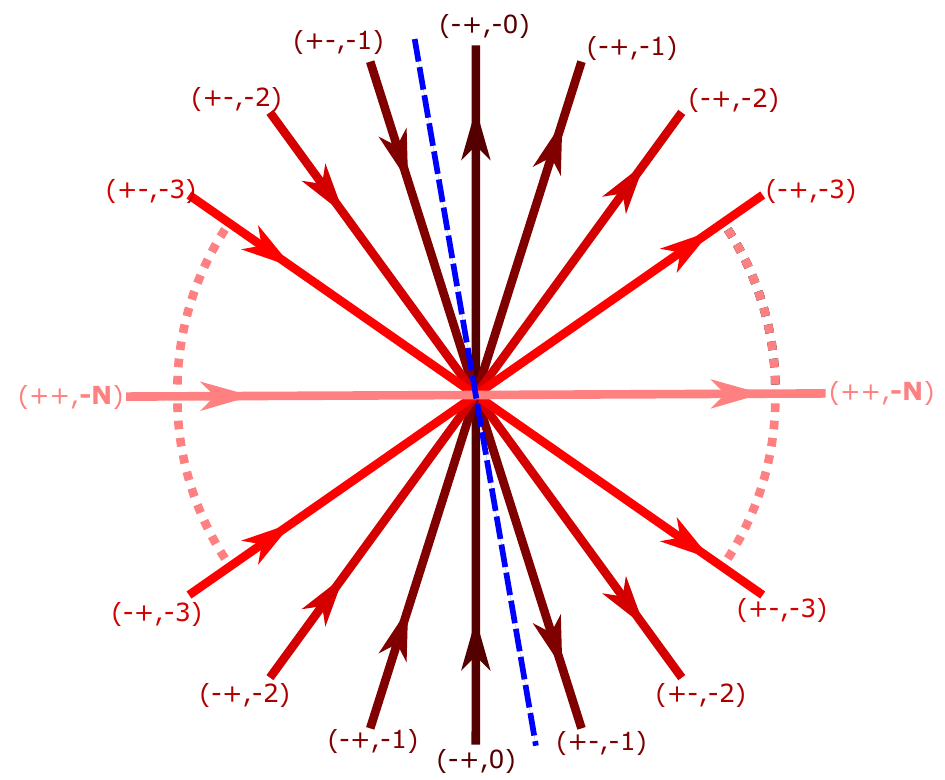}
       \caption{The 
        generic type of intersection occuring in our network, with the incoming and outgoing trajectories separated by the dashed blue line.}
       \label{genericintersections}
    \end{figure}

We start by computing the soliton data for the $(++,-\mathbb{N},0)$ line, that is $\tilde\CS_{0}$, which collects information about all BPS states with zero units of Kaluza-Klein momentum. 
From now on we will label charges of BPS states zero-modes as $(++,-\mathbb{N})$ for succintness. 
To that end, we recall that soliton data jumps at intersections which generically look as shown in Fig \ref{genericintersections}. Collection of these intersection points at different phases forms MS walls.

These jumps at intersections are determined by solving the twisted homotopy invariance equation (sometimes also called the flatness condition) around them. Hence, we start by constructing this equation:
\begin{itemize}
    \item We start by defining a representation of the Stoke's matrices for $\CE$-walls. For each wall corresponding to soliton data of type $(ij,n)$, we have
 \begin{equation}
        \begin{split}
            &S_{++,n}[\xi] =\exp\begin{pmatrix}
                \mu_{++,n}\xi^n & 0\\
                0 & 0
            \end{pmatrix}, \hspace{0.2cm}
            S_{+-,n}[\xi] =\exp\begin{pmatrix}
                    0 & 0\\
                \mu_{+-,n}\xi^n  & 0
            \end{pmatrix}, 
            \\& S_{-+,n}[\xi] =\exp\begin{pmatrix}
                0& \mu_{-+,n}\xi^n\\
                  0 & 0
            \end{pmatrix}, \hspace{0.2cm}
            S_{--,n}[\xi] =\exp\begin{pmatrix}
                0 & 0 \\
                0 & \mu_{--,n}\xi^n
            \end{pmatrix},
        \end{split} 
        \label{StokesMatrices}
    \end{equation}
    with $\mu_{ij,n}$ being the soliton data of the $\CE$-walls shown in Fig \ref{genericintersections}.
    \item Next, we arrange all trajectories on the basis of incoming and outgoing trajectories, and construct two paths $\rho$ and $\rho'$ that close. This is shown Fig \ref{io}.
    \begin{figure}[h!]
       \centering
           \includegraphics[width=0.3\linewidth]{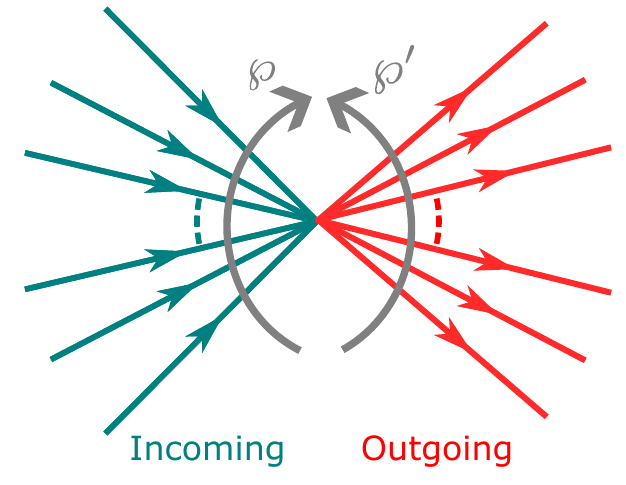}
       \caption{Homotopy paths around a generic intersection point}
       \label{io}
    \end{figure}
    \item Using the above representation of Stoke's matrices and the two paths $\rho$ and $\rho'$, we construct the twisted homotopy invariance equation as
    \begin{equation}
        F(\rho)\equiv \vec{\prod_{\alpha \in \rho}} S_{\alpha} = \vec{\prod_{\alpha \in \rho'}} S_{\alpha}^{-1} \equiv F(\rho')\,.
        \label{eqn:homtopinv}
    \end{equation}
\end{itemize}
Eq \eqref{eqn:homtopinv} can now used to find the outgoing soliton data, ie. $\mu_{ij,n}$'s at all intersection points recursively (as described in \cite{Gaiotto:2009hg}) in terms of the incoming soliton data.

We do this to find jumps in the soliton data for the $(++,-n)$ $\CE$-wall with $n \in \mathbb{N}$. Due to finiteness of paper, we only show the explicit calculation steps of soliton data for $n \leq 6$. This is the situation shown in Fig \ref{relevantstuff}, with the $\CE$-wall of interest shown in grey and going towards the left. 
\begin{itemize}
    \item The rightmost intersection is a part of the $(++,-2\mathbb{N})$ line, however its $ij$ trajectories as shown in purple in Fig \ref{relevantstuff} can be seen to intersect the $(++,-\mathbb{N})$ line in a way that contributes to its soliton data at $n=6$ (observe the third intersection from the left in Fig \ref{relevantstuff}). 
    
    Looking at our universal network Fig \ref{ppiby2}, we realise that the incoming trajectories (as shown in blue here) are the initial trajectories which have wound around the origin once, ie. an odd number of times. This also means that both the initial trajectories have crossed the logarithmic shift. Combining these facts, we see that the non-zero incoming soliton data is
    \begin{equation}
        \mu_{-+,-1}^{\incoming} = \mu_{+-,-1}^{\incoming} = -1. 
    \end{equation}
    Substituting this into the twisted homotopy invariance Eq \eqref{eqn:homtopinv}, we get the outgoing soliton data as
    \begin{equation}
        \begin{split}
            &\mu_{-+,1-2n}^{\outgoing} = \mu_{+-,1-2n}^{\outgoing} = (-1)^n \AND
            \\& \mu_{++,-2n}^{\outgoing} = -\mu_{--,-2n}^{\outgoing} = -\frac{(-1)^n}{n},
        \end{split}
        \label{out1}
    \end{equation}
    for all $n \in \mathbb{N}$, with all the rest being $0$. For our purposes, we only need $(+-,-3)$ and $(-+,-3)$ trajectories generated here (the rest contribute to soliton data with $n>6$ on the $(++,-\mathbb{N})$ line).
    \item For the second intersection from the right, the incoming trajectories are just the initial trajectories, with one of them having a logarithmic shift. Therefore, the non-zero incoming CFIV indices are
    \begin{equation}
        \mu_{-+,0}^{\incoming} = \mu_{+-,-1}^{\incoming} = 1.
    \end{equation}
    Substituting this in Eq \eqref{eqn:homtopinv}, we get the outgoing soliton data
    \begin{equation}
        \begin{split}
            &\mu_{-+,1-n}^{\outgoing} = \mu_{+-,-n}^{\outgoing} = (-1)^{n+1} \AND
            \\& \mu_{++,-n}^{\outgoing} = -\mu_{--,-n}^{\outgoing} = \frac{(-1)^n}{n},
        \end{split}
        \label{out2}
    \end{equation}
    for all $n \in \mathbb{N}$, with all the rest being $0$. It is at this point that the $(++,-\mathbb{N})$ line has been generated.
    \item For the third intersection from the right, the initial trajectories intersect at the $(++,-\mathbb{N})$ line, leading to jumps in the soliton data. Before reaching this point, the initial trajectories have wound around once, hence we have to add the winding factor \cite{Gaiotto:2012rg}. This gives us the non-zero incoming soliton data as
    \begin{equation}
        \begin{split}
            &\mu_{+-,-2}^{\incoming} = \mu_{-+,-1}^{\incoming} = -1 \AND
            \\&\mu_{++,-n}^{\incoming} = -\mu_{--,-n}^{\incoming} = \frac{(-1)^n}{n}.
        \end{split}
    \end{equation}
    Using this in Eq \eqref{eqn:homtopinv}, we get the outgoing soliton data and show them in Table \ref{out3}.
    \begin{table}[h!]
        \centering
        \begin{tblr}{
            colspec = {cccc},
            stretch = 0,
            rowsep = 6pt,
            hlines = {black, 0.5pt},
            vlines = {black, 0.5pt},
        }
        $n$ & $-+$ & $+-$ & $++$ & $--$\\
        $1$ & $-1$ & $0$ & $-1$ & $1$\\
        $2$ & $2$ & $-1$ & $\frac{1}{2}$ & $-\frac{1}{2}$ \\
        $3$ & $-3$ & $2$ & $-\frac{4}{3}$ & $\frac{4}{3}$\\
        $4$ & $5$ & $-3$ & $\frac{9}{4}$ & $-\frac{9}{4}$\\
        $5$ & $-9$ & $5$ & $-\frac{16}{5}$ & $\frac{16}{5}$\\
        $6$ & $16$ & $-9$ & $\frac{14}{3}$ & $-\frac{14}{3}$\\
        \end{tblr}
        \caption{Outgoing soliton data $\mu_{ij,-n}^{\outgoing}$ at the second intersection point along the $(++,-\mathbb{N})$ line}
        \label{out3}
    \end{table}
    We remark that this is the last intersection along this $ii$ line where we can determine all the soliton data by a finite number of incoming trajectories. After this, using the trajectories shown in Fig \ref{relevantstuff} we will only be able to determine soliton data upto $n=6$, and beyond this we will have to use incoming $ij$ trajectories with higher $n$.
    \item For the fourth intersection from the right in Fig \ref{relevantstuff}, we see an interesting phenomena: two pairs of $(+-)$ and $(-+)$ trajectories intersect at precisely the same point on the $(++,-\mathbb{N})$ line. The incoming $ii$ soliton data is taken exactly from Table \ref{out3}, while the incoming $ij$ soliton data is taken from Eqs \eqref{out1} and \eqref{out2}, and then corrected with the winding factor. This gives us the non-zero incoming $ij$ soliton data as
    \begin{equation}
        \mu_{+-,-3}^{\incoming} = \mu_{-+,-2}^{\incoming} =-\mu_{+-,-4}^{\incoming} = -\mu_{-+,-3}^{\incoming}= 1.
    \end{equation}
    Using Eq \eqref{eqn:homtopinv}, we get the outgoing non-zero soliton data and show them in Table \ref{out4}.
    \begin{table}[h!]
        \centering
        \begin{tblr}{
            colspec = {cccc},
            stretch = 0,
            rowsep = 6pt,
            hlines = {black, 0.5pt},
            vlines = {black, 0.5pt},
        }
        $n$ & $-+$ & $+-$ & $++$ & $--$\\
        $1$ & $0$ & $0$ & $-1$ & $1$\\
        $2$ & $1$ & $0$ & $\frac{1}{2}$ & $-\frac{1}{2}$ \\
        $3$ & $-3$ & $1$ & $-\frac{4}{3}$ & $\frac{4}{3}$\\
        $4$ & $5$ & $-3$ & $\frac{9}{4}$ & $-\frac{9}{4}$\\
        $5$ & $-9$ & $5$ & $-\frac{21}{5}$ & $\frac{21}{5}$\\
        $6$ & $19$ & $-9$ & $\frac{26}{3}$ & $-\frac{26}{3}$\\
        \end{tblr}
        \caption{Outgoing soliton data $\mu_{ij,-n}^{\outgoing}$ at the third intersection point along the $(++,-\mathbb{N})$ line}
        \label{out4}
    \end{table}
    \item Finally, in the last case of Fig \ref{relevantstuff} we again see multiple pairs of $+-$ and $-+$ trajectories intersecting. The initial blue trajectory has now wound twice, so the winding factor goes away and we get the original soliton data as given in Eq \eqref{out2}. However, the green trajectory has wound once, hence we have to correct its soliton data as given in Eq \eqref{out2} with the winding factor. This gives us the non-zero incoming $ij$ soliton data as
    \begin{equation}
          \mu_{+-,-3}^{\incoming} = \mu_{-+,-2}^{\incoming} = -\mu_{+-,-4}^{\incoming} = -\mu_{-+,-3}^{\incoming} = 1,
    \end{equation} 
    and the non-zero incoming $ii$ soliton data is shown in Table \ref{out4}. Using Eq \eqref{eqn:homtopinv}, we get the outgoing soliton data as shown in Table \ref{out5}.
    \begin{table}[h!]
        \centering
        \begin{tblr}{
            colspec = {cccc},
            stretch = 0,
            rowsep = 6pt,
            hlines = {black, 0.5pt},
            vlines = {black, 0.5pt},
        }
        $n$ & $-+$ & $+-$ & $++$ & $--$\\
        $1$ & $0$ & $0$ & $-1$ & $1$\\
        $2$ & $1$ & $0$ & $\frac{1}{2}$ & $-\frac{1}{2}$ \\
        $3$ & $-3$ & $1$ & $-\frac{4}{3}$ & $\frac{4}{3}$\\
        $4$ & $5$ & $-3$ & $\frac{9}{4}$ & $-\frac{9}{4}$\\
        $5$ & $-9$ & $5$ & $-\frac{26}{5}$ & $\frac{26}{5}$\\
        $6$ & $19$ & $-9$ & $\frac{38}{3}$ & $-\frac{38}{3}$\\
        \end{tblr}
        \caption{Outgoing soliton data $\mu_{ij,-n}^{\outgoing}$ at the fourth intersection point along the $(++,-\mathbb{N})$ line}
        \label{out5}
    \end{table}
    We remark that this is the final correction to soliton data with $n \leq 6$ along the $(++,-\mathbb{N})$ line. This is because all the other intersections on the line are of type $(-+,-n_1)$ and $(+-,-n_2)$ such that $n_1+n_2 > 6$. One can say that the soliton data ``stabilises'' after a certain number of intersections, or equivalently, we say that the soliton data stabilises as we go down the $(++,-\mathbb{N})$ line, in the order of increasing $n$, ie. higher $n$ soliton data gets stabilised after lower $n$ soliton data.
\end{itemize}
In this way, we have found the soliton data explicitly for the full $(++,-\mathbb{N})$ line in our universal network $\unspnet$ upto $n=6$.

\bibliographystyle{unsrt}
\bibliography{bibliography.bib}

\begin{thebibliography}{10}

\bibitem{blr2018}
Sibasish Banerjee, Pietro Longhi, and Mauricio Romo.
\newblock {Exploring 5d BPS Spectra with Exponential Networks}.
\newblock {\em Annales Henri Poincare}, 20(12):4055--4162, 2019.

\bibitem{Cecotti:1992rm}
Sergio Cecotti and Cumrun Vafa.
\newblock {On classification of N=2 supersymmetric theories}.
\newblock {\em Commun. Math. Phys.}, 158:569--644, 1993.

\bibitem{Cecotti:1992qh}
Sergio Cecotti, Paul Fendley, Kenneth~A. Intriligator, and Cumrun Vafa.
\newblock {A New supersymmetric index}.
\newblock {\em Nucl. Phys. B}, 386:405--452, 1992.

\bibitem{Cecotti:1991me}
Sergio Cecotti and Cumrun Vafa.
\newblock {Topological antitopological fusion}.
\newblock {\em Nucl. Phys. B}, 367:359--461, 1991.

\bibitem{Cecotti:2013mba}
Sergio Cecotti, Davide Gaiotto, and Cumrun Vafa.
\newblock {$tt^*$ geometry in 3 and 4 dimensions}.
\newblock {\em JHEP}, 05:055, 2014.

\bibitem{Klemm:1996bj}
Albrecht Klemm, Wolfgang Lerche, Peter Mayr, Cumrun Vafa, and Nicholas~P.
  Warner.
\newblock {Selfdual strings and N=2 supersymmetric field theory}.
\newblock {\em Nucl. Phys. B}, 477:746--766, 1996.

\bibitem{Gaiotto:2012rg}
Davide Gaiotto, Gregory~W. Moore, and Andrew Neitzke.
\newblock {Spectral networks}.
\newblock {\em Annales Henri Poincare}, 14:1643--1731, 2013.

\bibitem{Eager:2016yxd}
Richard Eager, Sam~Alexandre Selmani, and Johannes Walcher.
\newblock {Exponential Networks and Representations of Quivers}.
\newblock {\em JHEP}, 08:063, 2017.

\bibitem{GL-to-appear}
Kunal Gupta and Pietro Longhi.
\newblock {To appear}.

\bibitem{Bullimore:2018yyb}
Mathew Bullimore and Andrea Ferrari.
\newblock {Twisted Hilbert Spaces of 3d Supersymmetric Gauge Theories}.
\newblock {\em JHEP}, 08:018, 2018.

\bibitem{Banerjee:2024smk}
Sibasish Banerjee, Mauricio Romo, Raphael Senghaas, and Johannes Walcher.
\newblock {Exponential Networks for Linear Partitions}.
\newblock 3 2024.

\bibitem{Ooguri:1999bv}
Hirosi Ooguri and Cumrun Vafa.
\newblock {Knot invariants and topological strings}.
\newblock {\em Nucl. Phys. B}, 577:419--438, 2000.

\bibitem{Dimofte:2010tz}
Tudor Dimofte, Sergei Gukov, and Lotte Hollands.
\newblock {Vortex Counting and Lagrangian 3-manifolds}.
\newblock {\em Lett. Math. Phys.}, 98:225--287, 2011.

\bibitem{Aganagic:2000gs}
Mina Aganagic and Cumrun Vafa.
\newblock {Mirror symmetry, D-branes and counting holomorphic discs}.
\newblock 12 2000.

\bibitem{Aganagic:2001nx}
Mina Aganagic, Albrecht Klemm, and Cumrun Vafa.
\newblock {Disk instantons, mirror symmetry and the duality web}.
\newblock {\em Z. Naturforsch. A}, 57:1--28, 2002.

\bibitem{Aganagic:2013jpa}
Mina Aganagic, Tobias Ekholm, Lenhard Ng, and Cumrun Vafa.
\newblock {Topological Strings, D-Model, and Knot Contact Homology}.
\newblock {\em Adv. Theor. Math. Phys.}, 18(4):827--956, 2014.

\bibitem{Kucharski:2017poe}
Piotr Kucharski, Markus Reineke, Marko Stosic, and Piotr Su\l{}kowski.
\newblock {BPS states, knots and quivers}.
\newblock {\em Phys. Rev. D}, 96(12):121902, 2017.

\bibitem{Ekholm:2018eee}
Tobias Ekholm, Piotr Kucharski, and Pietro Longhi.
\newblock {Physics and geometry of knots-quivers correspondence}.
\newblock {\em Commun. Math. Phys.}, 379(2):361--415, 2020.

\bibitem{Intriligator:2013lca}
Kenneth Intriligator and Nathan Seiberg.
\newblock {Aspects of 3d N=2 Chern-Simons-Matter Theories}.
\newblock {\em JHEP}, 07:079, 2013.

\bibitem{Dunne:1998qy}
Gerald~V. Dunne.
\newblock {Aspects of Chern-Simons theory}.
\newblock In {\em {Les Houches Summer School in Theoretical Physics, Session
  69: Topological Aspects of Low-dimensional Systems}}, 7 1998.

\bibitem{Dimofte:2011jd}
Tudor Dimofte and Sergei Gukov.
\newblock {Chern-Simons Theory and S-duality}.
\newblock {\em JHEP}, 05:109, 2013.

\bibitem{Yoshida:2014ssa}
Yutaka Yoshida and Katsuyuki Sugiyama.
\newblock {Localization of three-dimensional $\mathcal{N}=2$ supersymmetric
  theories on $S^1 \times D^2$}.
\newblock {\em PTEP}, 2020(11):113B02, 2020.

\bibitem{Beem:2012mb}
Christopher Beem, Tudor Dimofte, and Sara Pasquetti.
\newblock {Holomorphic Blocks in Three Dimensions}.
\newblock {\em JHEP}, 12:177, 2014.

\bibitem{Dimofte:2011ju}
Tudor Dimofte, Davide Gaiotto, and Sergei Gukov.
\newblock {Gauge Theories Labelled by Three-Manifolds}.
\newblock {\em Commun. Math. Phys.}, 325:367--419, 2014.

\bibitem{Terashima:2011qi}
Yuji Terashima and Masahito Yamazaki.
\newblock {SL(2,R) Chern-Simons, Liouville, and Gauge Theory on Duality Walls}.
\newblock {\em JHEP}, 08:135, 2011.

\bibitem{Gaiotto:2009hg}
Davide Gaiotto, Gregory~W. Moore, and Andrew Neitzke.
\newblock {Wall-crossing, Hitchin systems, and the WKB approximation}.
\newblock {\em Adv. Math.}, 234:239--403, 2013.

\bibitem{Kontsevich:2008fj}
Maxim Kontsevich and Yan Soibelman.
\newblock {Stability structures, motivic Donaldson-Thomas invariants and
  cluster transformations}.
\newblock 11 2008.

\bibitem{Gaiotto:2011tf}
Davide Gaiotto, Gregory~W. Moore, and Andrew Neitzke.
\newblock {Wall-Crossing in Coupled 2d-4d Systems}.
\newblock {\em JHEP}, 12:082, 2012.

\bibitem{Banerjee:2022oed}
Sibasish Banerjee, Pietro Longhi, and Mauricio Romo.
\newblock {A-branes, Foliations and Localization}.
\newblock {\em Annales Henri Poincare}, 24(4):1077--1136, 2023.

\bibitem{Grassi:2022zuk}
Alba Grassi, Qianyu Hao, and Andrew Neitzke.
\newblock {Exponential Networks, WKB and Topological String}.
\newblock {\em SIGMA}, 19:064, 2023.

\bibitem{Alim:2022oll}
Murad Alim, Lotte Hollands, and Iv\'an Tulli.
\newblock {Quantum Curves, Resurgence and Exact WKB}.
\newblock {\em SIGMA}, 19:009, 2023.

\bibitem{Labastida:2000zp}
J.~M.~F. Labastida and Marcos Marino.
\newblock {Polynomial invariants for torus knots and topological strings}.
\newblock {\em Commun. Math. Phys.}, 217:423--449, 2001.

\bibitem{Labastida:2000yw}
J.~M.~F. Labastida, Marcos Marino, and Cumrun Vafa.
\newblock {Knots, links and branes at large N}.
\newblock {\em JHEP}, 11:007, 2000.

\bibitem{Galakhov:2014xba}
Dmitry Galakhov, Pietro Longhi, and Gregory~W. Moore.
\newblock {Spectral Networks with Spin}.
\newblock {\em Commun. Math. Phys.}, 340(1):171--232, 2015.

\bibitem{Coman:2020qgf}
Ioana Coman, Pietro Longhi, and J\"org Teschner.
\newblock {From quantum curves to topological string partition functions II}.
\newblock 4 2020.

\bibitem{oeis}
Online~Encyclopedia of~Integer~Sequences.
\newblock {Sequence A131868 \url{https://oeis.org/A131868}}.

\bibitem{mathoverflow}
M.~Kontsevich R.~Stanley Max~Alekseyev, O.~Gorodetsky.
\newblock
  {\url{https://mathoverflow.net/questions/195339/a-congruence-involving-binomial-coefficients}},
  2015.

\end{thebibliography}

\end{document}